\documentclass[traditabstract, twocolumns]{aa}

\usepackage{graphicx}
\usepackage{epstopdf}
\usepackage{times,epsfig}
\usepackage{deluxetable}
\usepackage{longtable}
\usepackage{natbib}
\usepackage{amsmath}
\usepackage{amssymb}
\usepackage{mathrsfs}
\DeclareMathOperator\erf{erf}
\bibpunct{(}{)}{;}{a}{}{,}
\usepackage[english]{babel}
\usepackage{lscape}
\usepackage{url}

\authorrunning{A. Nyholm et al.}
\titlerunning{Type~IIn SN light-curve properties}

\begin{document}

\defcitealias{ackermann15}{A15}
\defcitealias{arcavi10}{A10}
\defcitealias{ofek14prec}{O14a}
\defcitealias{ofek14rise}{O14c}

\title{Type~IIn supernova light-curve properties measured from an\\untargeted survey sample\thanks{Full Table~\ref{tab:phot} is only available at the CDS via anonymous ftp to \protect\url{cdsarc.u-strasbg.fr} (\protect\url{130.79.128.5}) or via \protect\url{http://cdsarc. u-strasbg.fr/viz-bin/cat/J/A+A/637/A73}. The classification spectra and the photometry is available via the Weizmann Interactive Supernova Data Repository (WISeREP) at \protect\url{https://wiserep.weizmann.ac.il}. Based on observations made with the Palomar Transient Factory and intermediate Palomar Transient Factory surveys.}}

\author{A.~Nyholm\inst{1}
\and J.~Sollerman\inst{1}
\and L.~Tartaglia\inst{1}
\and F.~Taddia\inst{1}
\and C.~Fremling\inst{2}
\and N.~Blagorodnova\inst{3}
\and A.~V.~Filippenko\inst{4,5}
\and A.~Gal-Yam\inst{6}
\and D.~A.~Howell\inst{7,8}
\and E.~Karamehmetoglu\inst{1}
\and S.~R.~Kulkarni\inst{2}
\and R.~Laher\inst{9}
\and G.~Leloudas\inst{10}
\and F.~Masci\inst{9}
\and M.~M.~Kasliwal\inst{2}
\and K.~Mor\r{a}\inst{11}
\and T.~J.~Moriya\inst{12}
\and E.~O.~Ofek\inst{13}
\and S.~Papadogiannakis\inst{11}
\and R.~Quimby\inst{14,15}
\and U.~Rebbapragada\inst{16}
\and S.~Schulze\inst{6}
}

\institute{Department of Astronomy and the Oskar Klein Centre, Stockholm University, AlbaNova, 10691 Stockholm, Sweden\\(\email{anders.nyholm@astro.su.se})
\and Division of Physics, Mathematics, and Astronomy, California Institute of Technology, Pasadena, CA 91125, USA
\and Department of Astrophysics/IMAPP, Radboud University, Houtlaan 4, 6525 XZ Nijmegen, The Netherlands
\and Department of Astronomy, University of California, Berkeley, CA 94720-3411, USA
\and Miller Senior Fellow, Miller Institute for Basic Research in Science, University of California, Berkeley, CA 94720, USA
\and Department of Particle Physics and Astrophysics, Weizmann Institute of Science, 234 Herzl St., Rehovot, 76100, Israel
\and Las Cumbres Observatory, 6740 Cortona Drive, Suite 102, Goleta, CA 93117-5575, USA
\and Department of Physics, University of California, Santa Barbara, CA 93106-9530, USA
\and IPAC, California Institute of Technology, 1200 E. California Blvd, Pasadena, CA 91125, USA
\and DTU Space, National Space Institute, Technical University of Denmark, Elektrovej 327, 2800 Kgs. Lyngby, Denmark
\and Department of Physics and the Oskar Klein Centre, Stockholm University, AlbaNova, 10691 Stockholm, Sweden
\and National Astronomical Observatory of Japan, 2-21-1 Osawa, Mitaka, Tokyo 181-8588, Japan
\and Benoziyo Center for Astrophysics, Weizmann Institute of Science, 76100 Rehovot, Israel
\and Department of Astronomy/Mount Laguna Observatory, San Diego State University, 5500 Campanile Drive, San Diego, CA 92812-1221, USA
\and Kavli Institute for the Physics and Mathematics of the Universe (WPI), The University of Tokyo Institutes for Advanced Study, The University of Tokyo, Kashiwa, Chiba 277-8583, Japan
\and Jet Propulsion Laboratory, California Institute of Technology, Pasadena, CA 91109, USA}

\date{Received; accepted}

\abstract{The evolution of a Type~IIn supernova (SN~IIn) is governed by the interaction between the SN ejecta and a hydrogen-rich circumstellar medium. The SNe~IIn thus allow us to probe the late-time mass-loss history of their progenitor stars. We present a sample of SNe~IIn from the untargeted, magnitude-limited surveys of the Palomar Transient Factory (PTF) and its successor, the intermediate PTF (iPTF). To date, statistics on SN~IIn optical light-curve properties have generally been based on small ($\lesssim 10$ SNe) samples from targeted SN surveys. The SNe~IIn found and followed by the PTF/iPTF were used to select a sample of 42 events with useful constraints on the rise times as well as with available post-peak photometry. The sample SNe were discovered in 2009--2016 and have at least one low-resolution classification spectrum, as well as photometry from the P48 and P60 telescopes at Palomar Observatory. We study the light-curve properties of these SNe~IIn using spline fits (for the peak and the declining portion) and template matching (for the rising portion). We study the peak-magnitude distribution, rise times, decline rates, colour evolution, host galaxies, and K-corrections of the SNe in our sample. We find that the typical rise times are divided into fast and slow risers at $20\pm6$~d and $50\pm11$~d, respectively. The decline rates are possibly divided into two clusters (with slopes $0.013 \pm 0.006$\,mag\,d$^{-1}$ and $0.040\pm0.010$\,mag\,d$^{-1}$), but this division has weak statistical significance. We find no significant correlation between the peak luminosity of SNe~IIn and their rise times, but the more luminous SNe~IIn are generally found to be more long-lasting. Slowly rising SNe~IIn are generally found to decline slowly. The SNe in our sample were hosted by galaxies of absolute magnitude $-22 \lesssim M_g \lesssim -13$\,mag. The K-corrections at light-curve peak of the SNe~IIn in our sample are found to be within 0.2\,mag for the observer's frame $r$-band, for SNe at redshifts $z < 0.25$. By applying K-corrections and also including ostensibly "superluminous" SNe~IIn, we find that the peak magnitudes are $M_{\rm peak}^{r} = -19.18\pm1.32$\,mag. We conclude that the occurrence of conspicuous light-curve bumps in SNe~IIn, such as in iPTF13z, are limited to $1.4^{+14.6}_{-1.0} ~\%$ of the SNe~IIn. We also investigate a possible sub-type of SNe~IIn with a fast rise to a $\gtrsim 50$\,d plateau followed by a slow, linear decline.}


\maketitle

\section{Introduction}
A supernova (SN) that has a spectrum showing Balmer emission lines with narrow or intermediate-width central components and broad line wings is classified as a Type~IIn SN. This SN classification (where "II" indicates presence of hydrogen and "n" stands for narrow) was proposed by \citet{schlegel90} and reached wider use via the review by \citet{filippenko97}. It has been shown in observations and modelling \citep[e.g.][]{chevalier94,smith17hsn,dessart16} that the SN~IIn spectral signature arises from the interaction of SN ejecta with a hydrogen-rich circumstellar medium (CSM), where shocks form when the ejecta sweep up the CSM, converting the kinetic energy of the ejecta into radiated energy.

An SN~IIn spectrum is produced in a process involving the SN environment, and shows little direct signatures of the explosion itself. Many different scenarios can therefore potentially give rise to SN~IIn spectral signatures, making it difficult to tell whether a core-collapse (CC) SN explosion or a violent, but nondisruptive, stellar outburst sends the ejecta off \citep[e.g.][]{dessart09}. Whereas the ejecta-CSM interaction makes SNe~IIn useful probes of progenitor mass-loss histories, SNe~IIn are also challenging to understand since the nature of the underlying energy source remains elusive. The spectroscopic criteria defining the SN~IIn classification makes them spectroscopically somewhat similar, but their peak absolute magnitudes vary greatly, as do their light-curve shapes and the presence of undulations and bumps in their light-curves. The wide range of light-curve properties shown by SNe~IIn indicates significant variety in both CSM and ejecta properties \citep[e.g.][]{moriya14}, suggesting that different mechanisms and progenitor channels lead to the SNe~IIn we observe \citep{smith14}.

SNe~IIn are intrinsically rare. A volume-limited SN sample from the targeted Lick Observatory Supernova Search \citep[LOSS;][]{li11} shows that $\sim 7$\% of all CC~SNe are SNe~IIn. Their diverse properties call for large samples in order to improve our understanding of this SN type. However, the current SN~IIn literature is skewed towards a small number of well-studied events, which are often nearby (at redshifts $z \lesssim 0.02$) or characterised by some unusual property. Our aim in this paper is to study a sample of SNe~IIn from the untargeted Palomar Transient Factory (PTF; \citealp{law09}) survey and its successor, the intermediate PTF \citep[iPTF;][]{kulkarni13}, with emphasis on the SN~IIn optical light-curves.

Among the SN~IIn samples presented in the literature are works concerning SN data releases and analyses \citep[e.g.][]{kiewe12,taddia13,taddia15}, and also environmental \citep[e.g.][]{kelly12,habergham14,taddia15,galbany18,kuncarayakti18} and SN rate studies \citep{li11}, as well as precursor outburst analyses \citep[e.g.][]{ofek14prec,bilinski15}. In Table~\ref{tab:IInsamples} we present a summary of the SN~IIn samples in the literature concerning optical light-curve properties.

The decline behaviour after peak brightness differs considerably among SNe~IIn. Among slowly declining SNe~IIn, SN~1988Z has become a benchmark event in the literature \citep{stritzinger12, habergham14, turatto93}. Other SNe~IIn decline considerably faster, with SN~1998S \citep{liu00, fassia00} commonly given as a specimen. Apart from decline rates, light-curve shapes can also distinguish SNe~IIn. SN~1994W \citep{sollerman98} often exemplifies SNe~IIn with plateaus in their light-curves followed by a sharp decline in brightness, sometimes called SNe~IIn-P \citep{mauerhan13}. Some SNe~IIn exhibit distinct episodes of re-brightening ("bumps"), thus breaking their decline in brightness \citep[e.g.][]{stritzinger12,graham14,martin15,nyholm17}.

An important finding facilitated by modern surveys, such as the PTF, is that some SNe~IIn have precursor outbursts in the years before the main SN outburst. Sample studies \citep{ofek14prec,bilinski15} give inconclusive results regarding whether precursor events are common (possibly owing to the different methods used in the respective studies). Precursor outbursts suggest connections between SNe~IIn and phenomena such as luminous blue variable (LBV; \citealp{humphreys94}) outbursts \citep{smith11_lbv} as well as other so called SN impostors \citep{vandyk00,vandyk12}.

The SN~IIn samples in the literature have uneven photometric coverage of the SNe during the rising phase of their light-curves. A number of the benchmark SN~IIn events (e.g. SN~1988Z, \citealp{stathakis91}; SN~1995N, \citealp{fransson02}; SN~2005ip and SN~2006jd, \citealp{stritzinger12}) lack well-determined rise times to peak. The literature sample of light-curves shown by \citet[][their Fig.~10]{taddia15} also indicates the lack of pre-peak photometry for a number of otherwise well-characterised SNe~IIn. The existence of a correlation between rise time and peak luminosity for SNe~IIn has been investigated \citep{ofek14_10jl, ofek14rise, moriyamaeda14} and suggested to exist \citep{ofek14rise}. To further study this possible correlation, more knowledge of SNe~IIn rise times is needed.

By using a sample from the high-cadence and untargeted PTF/iPTF search to study light-curve rise times, peak absolute magnitudes, and light-curve slopes during the decline phase, we can characterise the SN~IIn type. The untargeted nature of the PTF/iPTF survey provides a sample which is not biased towards SNe in luminous, metal-rich, high-star-formation-rate host galaxies. The high cadence of PTF/iPTF also allows us to put tighter constraints on the rise times of our SNe.

In Sect.~\ref{sec:obs}, we present the SN search and follow-up observations done by the PTF/iPTF and the telescopes and instruments used, as well as the reduction methods adopted to prepare the photometric data for analysis. We discuss the selection of our SN~IIn sample as well as its properties in Sect.~\ref{sec:sample}, including a discussion of distances and foreground extinction of the SNe. In Sect.~\ref{sec:analysis}, we measure the light-curve peak magnitudes and peak epochs (Sect.~\ref{sec:peak}) and the decline rates of the SNe (Sect.~\ref{sec:decline}). We also study the light-curve rise times (Sect.~\ref{sec:risetimes}), as well as possible correlations between these light-curve-shape parameters (Sect.~\ref{sec:corr}). Colour curves are examined using our optical photometry (Sect.~\ref{sec:colour}). The luminosity function is examined (Sect.~\ref{sec:lumfunc}). A simple study of the host galaxies of our sample SNe is done, using Sloan Digital Sky Survey (SDSS) and Panoramic Survey Telescope and Rapid Response System (Pan-STARRS) photometry for the hosts (Sect.~\ref{sec:host}). In Sect.~\ref{sec:discu}, we discuss our results and highlight some sample SNe with interesting properties, and estimate the fraction of SNe~IIn having light-curve bumps. We summarise our conclusions in Sect.~\ref{sec:concl}. In Appendix \ref{sec:bigfig}, we show compilations of spectra and light-curves. In Appendix \ref{sec:kcor}, we use the spectra of our SNe to study their K-corrections.

\section{Observations and data reduction \label{sec:obs}}
The PTF and iPTF surveys did untargeted searches for astronomical transients during 2009--2017, using the 1.2-m Samuel Oschin telescope (known as P48) at Palomar Observatory. P48 equipped with the CFH12K mosaic CCD gave a 7.26\,deg$^2$ field of view \citep{law09}. Images were taken through either a Mould $R$-band filter \citep{law09,ofek12,laher14} or an SDSS $g$-band filter \citep{fukugita96}, giving a limiting magnitude of $R \approx 20.5$ or $g \approx 21$ (respectively) under typical conditions. After image processing and an automatic initial selection of candidates \citep{cao16}, a human scanner vetted the images and flagged interesting candidates for follow-up photometry and spectroscopy on other telescopes. All SNe in the sample presented in this paper were found independently with the P48. Whereas several different pipelines were used for photometry in P48 images during the course of PTF/iPTF, all SN photometry from P48 presented in this paper has been extracted with the PTFIDE \citep[PTF Image Differencing and Extraction;][]{masci17} pipeline.

An important instrument for photometric follow-up observations during PTF/iPTF was the 1.52-m telescope (known as P60) at Palomar Observatory. The telescope was used for imaging with the GRBCam camera \citep{cenko06} or the Spectral Energy Distribution Machine (SEDM) integral field spectrograph \citep{blagorodnova18} using SDSS $gri$ filters. A typical strategy with the P60 was to obtain SDSS $gri$ images of a SN field with a cadence of 1\,d or 3\,d (for young SNe) and a cadence of 6\,d for older SNe. Other cadences and filters (for example SDSS $z$ and Johnson $B$) were also sometimes used. For this paper, SN photometry of the images from P60 (both GRBCam and SEDM) was done with the FPipe pipeline \citep{fremling16}. All magnitudes reported here for SNe in our sample (Table~\ref{tab:phot}) are AB magnitudes from PSF photometry on host-subtracted images.

Classification spectra of identified transients were obtained continuously by members of the PTF/iPTF collaboration, using different telescopes with apertures ranging from 2 to 10 m, and their spectrographs. While some SNe were monitored extensively with spectroscopy, the majority of the SNe in our sample only have $\sim 1$ spectrum. The classification spectra used in this work are shown in Fig.~\ref{fig:class_spec} and are listed in Table~\ref{tab:speclog}.

\begin{figure}
   \centering
    \includegraphics[width=\linewidth]{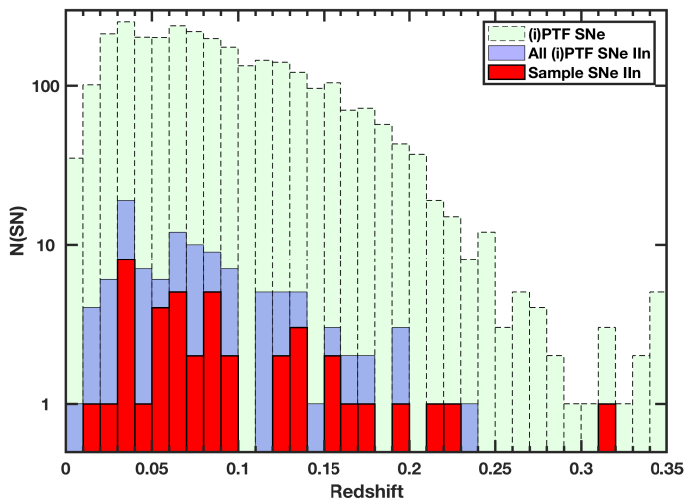}
    \caption{Histogram of redshifts for the 42 SNe~IIn included in our sample, shown in red. The redshifts of the full PTF/iPTF sample of 111 SNe~IIn is shown in blue. For comparison, the redshifts of the full PTF/iPTF yield of spectroscopically classified SNe (for $z < 0.35$) is shown in green.\label{fig:redshifts}}
\end{figure}

\begin{figure*}
   \centering
  \includegraphics[width=17.5cm]{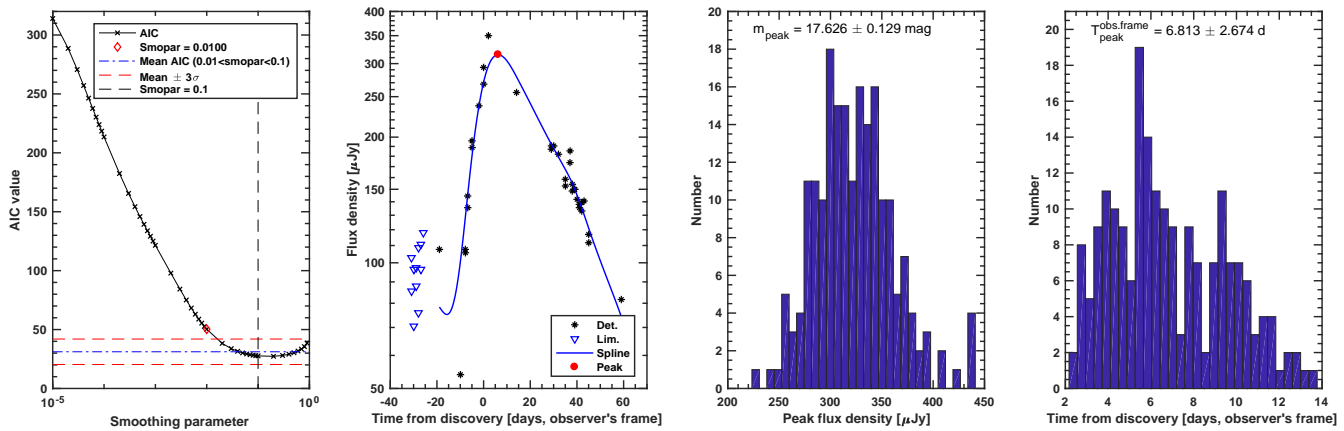}
    \caption{\textit{Left to right:} AIC values as a function of the smoothing parameter of the attempted spline fits, light-curve (in flux density), peak flux density (Monte Carlo histogram), and peak time (Monte Carlo histogram) example for PTF10achk.}
    \label{fig:AICpanels}
\end{figure*}

\section{The sample\label{sec:sample}}
A total of 3018 confirmed SNe was found by PTF/iPTF in the years 2009--2017, all spectroscopically classified by the collaboration using template-matching software like SNID \citep{blondin07}, Gelato \citep{harutyunyan08}, or Superfit \citep{howell05}. Classifications of the PTF/iPTF SNe were done by a joint effort of members of the PTF/iPTF collaboration, involving discussions of ambiguous cases.

Among the SNe found and spectroscopically classified by PTF/iPTF during 2009--2017, a total of 111 are SNe~IIn. These SNe~IIn are located at declinations $-23\degr < \delta < +75\degr$ and in the redshift interval $0.0071 < z < 0.31$. The mean $z$ of the total SN~IIn sample is 0.083 and the median $z$ is 0.07. For our SN~IIn sample, we made a selection based on criteria related to the availability of photometry for the rising and declining portions of the SN light-curves. Selection of our final sample of SNe~IIn was conducted according to the following steps.

To obtain photometry of even quality from the P48 images, the PTFIDE pipeline was run on all available $R$ and $g$ images from P48 for the locations of the 111 SNe in the initial sample to get forced template-subtracted photometry \citep[using the procedure from][]{masci15}. For the light-curves thus obtained, we applied the constraint that a photometric upper limit must occur less than 40\,d before the epoch of the PTF/iPTF discovery of the SN in order to allow a useful constraint on the rise time. This decreased the sample from 111 to 55 events. Including also inspection of the preliminary reductions of the P60 photometry, only SNe~IIn with detections past 40\,d after discovery were kept in the sample. After final refinement using the fully reduced P60 photometry, we were left with a sample of 42 SNe~IIn with satisfactory light-curve coverage. The time intervals considered in the selection were in the observer's frame. Our SN sample selection process is summarised in Table~\ref{tab:IInselection}. The classification spectra in the region around H$\alpha$ for these 42 SNe~IIn are provided in Fig.~\ref{fig:class_spec}. Basic properties of the SNe in the sample are summarised in Table~\ref{tab:IInsummary}. In the cases when the PTF/iPTF discovery of a SN constituted an independent discovery of a SN already found (or later found) by others, a reference is given in Table~\ref{tab:IInsummary}. The 42 light-curves are shown in Fig.~\ref{fig:LCs}. The SNe in the sample are covered in $BgRriz$ photometry (but not all SNe are covered in all photometric bands). The photometry presented here (Table~\ref{tab:phot}) for the 42 SNe is available via the VizieR database \citep{ochsenbein00} and the classification spectra and the photometry is available via WISeREP \citep{yaron12}.

Table~\ref{tab:IInsummary} shows that most of the SNe~IIn in our sample have been mentioned in the literature, mostly in three sample studies \citep{ackermann15, ofek14rise, ofek14prec} as well as in Circulars and other non-refereed sources. These sample studies mainly concerned themselves with $R$-band photometry \citep{ofek14rise, ofek14prec} or searches for $\gamma$-ray emission from the positions of the SNe \citep{ackermann15}, and did not publish full light-curves. PTF10tel ($=$ SN~2010mc) and its precursor activity is the topic of a paper by \citet{ofek13} and PTF12glz was studied by \citet{soumagnac19}. Early-time spectra of PTF10gvf and PTF10tel were studied in the spectral sample paper by \citet{khazov16}. For consistency, we will refer to the SNe by their PTF/iPTF names, even in the cases when International Astronomical Union names exist.

The superluminous SNe (SLSNe) IIn deserve special attention. A $M~<~-21$ mag limit has commonly been used to define SLSNe \citep{galyam12,galyam19} but it is not clear whether SNe~IIn fulfilling this criterion constitute a separate population of events \citep[e.g.][]{moriya18,galyam19}. During the time when the SNe~IIn in our sample were found and classified, the $M < -21$ mag limit was in common use and the 111 SNe Type~IIn that we used to construct our final sample are therefore typically fainter at peak than this limit.

Not all SLSNe Type~II display narrow Balmer lines in their spectra \citep{inserra18}. In the full PTF/iPTF catalogue of spectroscopically classified SNe a total of 15 SLSNe~II are listed. For 7 of the SNe, the spectra are either noisy or contaminated by host galaxy emission lines and make a SLSN~Type~IIn classification difficult. The spectrum of PTF12gwu has weak Balmer lines \citep{perley16}, but possibly broad. The remaining 7 of the SLSN~Type~II display a narrow H$\alpha$ emission component ostensibly making them SLSNe Type~IIn. Applying our criteria (see above) requiring a maximum gap of 40\,d between discovery epoch and last pre-discovery upper limit, leave 6 SLSNe Type~IIn. Also requiring photometry past 40\,days after discovery leaves 5 SLSNe~IIn: PTF10heh, PTF12mkp, PTF12mue, iPTF13duv and iPTF13dol. The PTF/iPTF photometry shows that these 5 SLSNe Type~IIn had light-curve peaks at $\sim -21$ mag. The photometry for two of these (PTF12mue and iPTF13duv) is too sparse to allow us to further characterise their light-curves.

We therefore used the sample of 111 PTF/iPTF SNe~IIn (with its inherent use of the $M < -21$ mag SLSN criterion) when selecting our SN IIn sample. For discussing whether the SLSNe Type~IIn constitute a separate population of SNe, we include the additional 3 SLSNe~IIn (PTF10heh, PTF12mkp and iPTF13dol\footnote{\object{iPTF13dol} was found by iPTF on 2013 September 29 at $\alpha = ~\rm 22^h$$\rm 22^m$$07\fs27$, $\delta =$ $12\degr 30\arcmin 39\farcs9$ (J2000.0) and $z = 0.225$.}) in the peak-magnitude distribution and in our luminosity function study in Sect.~\ref{sec:lumfunc}. For comparison purposes, we also include \object{PTF12mkp} when discussing the duration-luminosity phase space in Sect.~\ref{sec:model}. PTF10heh and PTF12mkp are already presented in the literature \citep{perley16}. A study dedicated to SLSNe ~II from PTF is ongoing (Leloudas et al., in prep.).

In this paper, our SN distance estimates are based on the SN redshifts, assuming a $\Lambda$CDM cosmology and using the cosmological parameters $H_0 = 70.0$\,km\,s$^{-1}$\,Mpc$^{-1}$, $\Omega_{\Lambda} = 0.721$, and $\Omega_{M} = 0.279$ \citep{hinshaw13}. We use a method based on the NASA/IPAC Extragalactic Database (NED) routine (following \citealp{mould00a,mould00b}) to compensate for Virgo, Great Attractor, and Shapley cluster infall in our distance estimates. The redshifts of the sample SNe were measured by fitting Gaussian profiles to the H$\alpha$ emission lines of their spectra. The histogram of the redshifts is shown in Fig.~\ref{fig:redshifts}.

Milky Way (MW) extinction values from \citet{SF11} were obtained via NED. Extinctions for different photometric bands were calculated using the \texttt{extinction} function by \citet{ofek14code} based on \citet{cardelli89} and assuming an absorption-to-reddening ratio $R_V = 3.1$. If SN spectra showed a clear \ion{Na}{I\,D} absorption doublet, host-galaxy extinction was computed using \citet[][their Eq.~1]{taubenberger06}. Such indications of host-galaxy extinction were seen only in the spectra of PTF09tm ($E(B-V) = 0.16$\,mag), PTF11qnf ($E(B-V) = 0.72$\,mag), and PTF12frn ($E(B-V) = 0.57$\,mag). We remind the reader that given the quality and resolution of our classification spectra, we are not very sensitive to this method. The effects of these assumptions on our results are discussed in Sect.~\ref{sec:errors}.

\section{Analysis\label{sec:analysis}}
In the following analysis we characterise the main light-curve parameters, that is peak absolute magnitude and its epoch, light-curve decline rates, and rise times. We also investigate possible correlations between these properties. We describe the optical colours and the host-galaxy properties. Finally, we compare our observations to simple models in order to derive information on the CSM and on the SN progenitor scenarios. In our analysis, we apply corrections to the following SNe: for PTF10acsq, a correction (found by interpolation on P60 $r$-band detections) of $0.07$\,mag was added to the P48 $R$-band; by similar interpolation on P60 $g$-band detections, for iPTF15aym $0.72$\,mag was added to the P48 $g$-band, and for iPTF15eqr $0.19$\,mag was added (by interpolation on P60 $r$) to the P48 $R$-band.

\subsection{Peak magnitudes and peak times\label{sec:peak}}

\begin{figure*}
    \centering
    \includegraphics[width=14cm]{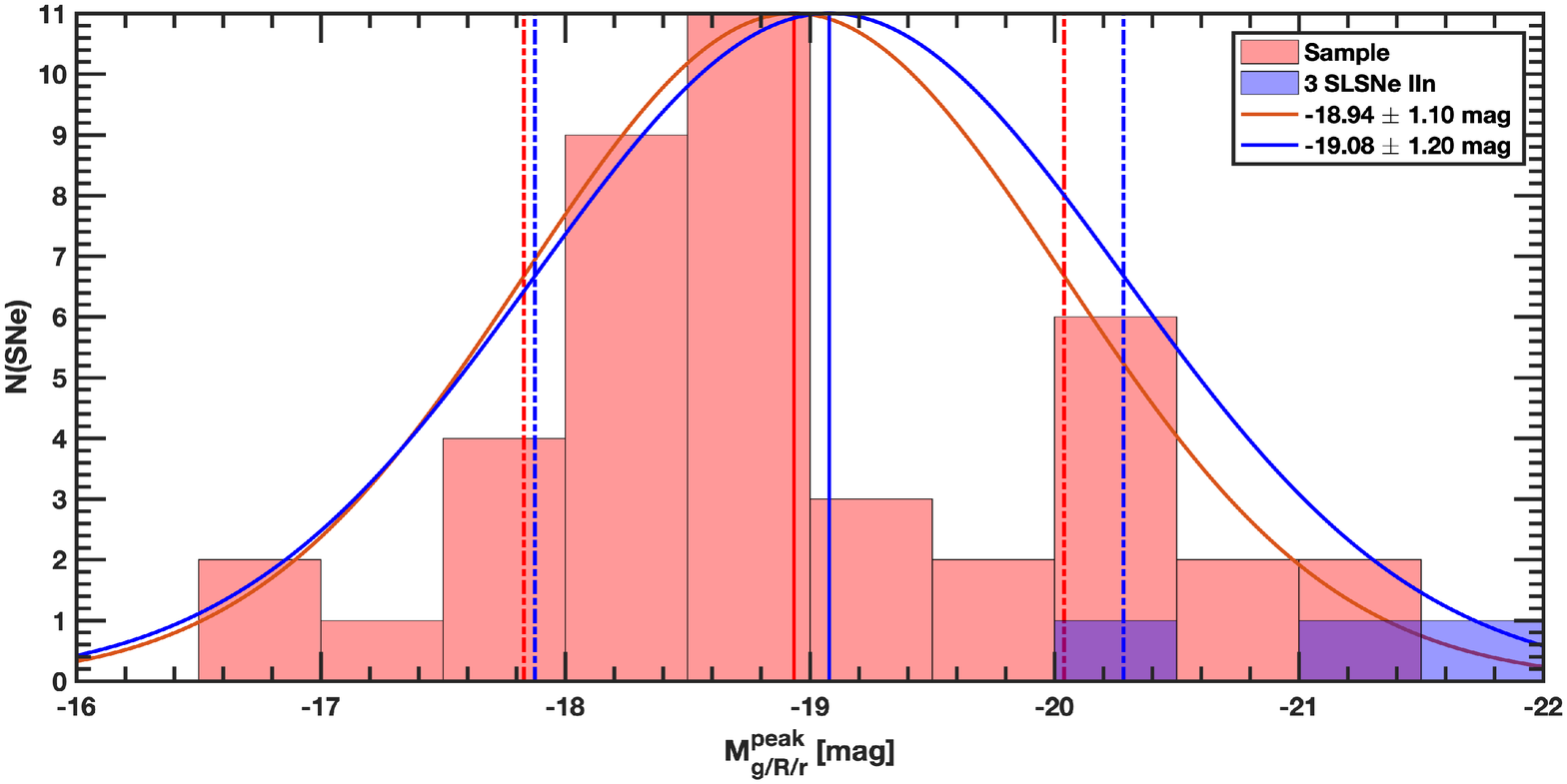}\\
    \includegraphics[width=14cm]{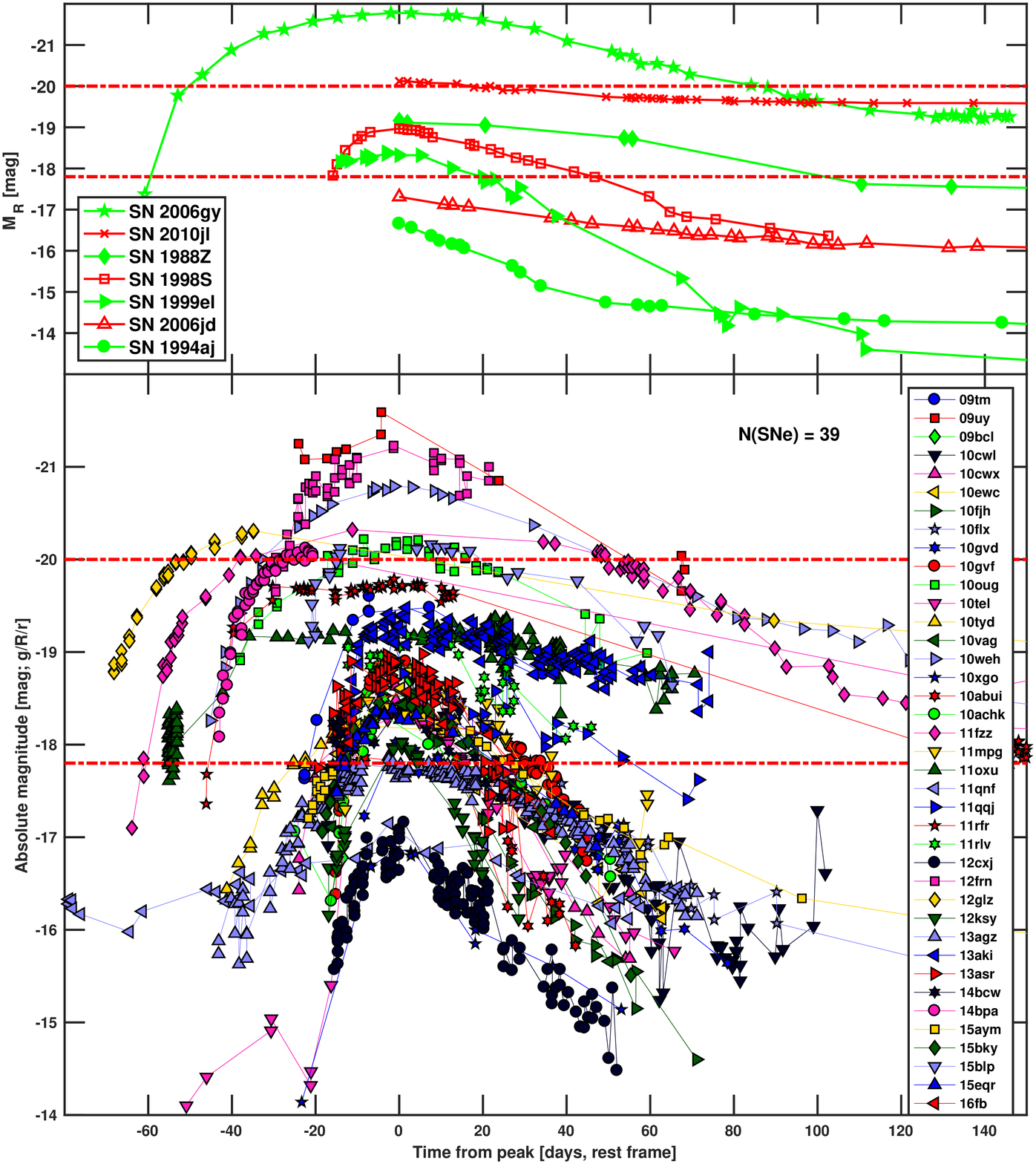}
     \caption{\textit{Top panel:} histogram for the peak absolute magnitudes of our 42+3 SNe~IIn, overplotted with the best-fit Gaussian distributions as identified with GMM. The Gaussian describing the distribution for the sample of 42 SNe is shown in red, the Gaussian for the sample 42 SNe~IIn + 3 SLSNe~IIn is shown in blue. Dash-dot lines show the $1\sigma$ values for the respective Gaussians. \textit{Central panel:} Some well-studied SNe~IIn from the literature, showing photometry of \object{SN 1988Z} \citep{aretxaga99}, \object{SN 1994aj} \citep{benentti98}, \object{SN 1998S} \citep{liu00}, \object{SN 1999el} \citep{dicarlo02}, SN~2006gy \citep{smith07phot06gy}, \object{SN 2006jd} \citep{stritzinger12}, and \object{SN 2010jl} \citep{fransson14}, with $-17.8$\,mag and $-20.0$\,mag ($1\sigma$ values from the sample distribution) shown as dash-dot red lines. \textit{Bottom panel:} Absolute magnitudes of the $\rm N(SNe) = 39$ SNe~IIn with determined peak epochs as a function of time relative to light-curve peak (with sample $1\sigma$ values shown as dashed red lines). No K-corrections were applied in these plots.}
       \label{fig:peakmags}
 \end{figure*}

The differences in cadence, photometric errors, as well as intrinsic light-curve shapes make the use of a single analytical template difficult for characterising the light-curves. Instead, we use cubic smoothing splines (CSS; \citealp{deboor01}) to fit the light-curves and to find peak-magnitude values and times of peak for our sample SNe. Before fitting, we convert the photometry from AB magnitudes to flux densities (using the \texttt{convert\_flux} function by \citealp{ofek14code}).

The fitting of a CSS to a set of data is determined by the smoothing parameter $s$, where the fit approaches a least-squares (LSQ) fit of a straight line as $s \rightarrow 0$, whereas the fit approaches a cubic-spline interpolant between each data point as $s \rightarrow 1$. To objectively choose an $s$ value for each light curve, we proceed in the following way.

For a grid of 45 $s$ values with decadal spacing in the interval $10^{-5} \leq s \leq 0.9$, a CSS fit was made for each $s$ value. For each CSS fit, the number of degrees of freedom (d.o.f.) was computed as the number of observations minus the number of parameters of the CSS fit, and then used to calculate the $\chi^2$ values of each CSS fit (per d.o.f., i.e., the reduced $\chi^2$ values, $\chi^2_{\rm {d.o.f.}}$).

For selecting the $s$ value, we use the Akaike Information Criterion (AIC; see, e.g. Sect. 2 of \citealp{davis07}) as our statistic. The AIC punishes models of high orders (i.e., "overparametrised" models). As we desire a CSS fit capturing the large-scale behaviour of the light-curve (while not being sensitive to smaller fluctuations), we choose to consider values of $s < 0.1$. We generally select the smoothing parameter $s$ giving the smallest AIC value for $s < 0.1$. If no minimum of the AIC value is found for $s < 0.1$, an $s$ value is selected corresponding to the AIC value 3$\sigma$ above the mean AIC value for the interval $0.01 < s < 0.1$.

Armed with our selected $s$ parameter for each SN, Monte Carlo (MC) experiments were done to estimate the uncertainties of the maximum flux density and the peak epoch. For each MC run, pseudo data points were generated for each epoch in the light curve, drawn from Gaussian distributions around the original data points (from within the $1\sigma$ photometric error bars). For the $s$ value found above, CSS fits were made to each such pseudo light curve, and maximum flux density and peak epoch were determined for each of them. The mean values of the highest flux density and its epoch are used as the final values, and their standard deviations are used as their 1$\sigma$ uncertainties. This process is illustrated in Fig.~\ref{fig:AICpanels}. The CSS fits are shown in Fig.~\ref{fig:lcfit_allIIn}, along with the measured peak of each light curve. The peak magnitudes are presented in Table~\ref{tab:peakmag_time}, where the absolute magnitudes are corrected for extinction. In our later analysis of decline rates and rise times, the SNe with peak time error $> 10$\,d (marked with red crosses in Fig.~\ref{fig:lcfit_allIIn}) are excluded.

To evaluate if one or more groupings can be distinguished among the SNe when considering a given property (such as peak absolute magnitude), we use Gaussian mixture models (GMM) following \citet{papadogiannakis19}. The method works by applying an increasing number of Gaussian distributions to the data, using the AIC for each step to evaluate the significance of each combination of Gaussians. The best combination of Gaussians is the one giving the lowest value of AIC. Given our small sample we also tried the Bayesian information criterion (BIC), which is considered to rule out over-parametrised models in a way stronger than the AIC \citep{liddle04}. Following \citet[][their Sect.~4.1]{sollerman09} we consider models separated by $\Delta$IC $\geq 6$ from the surrounding models as significant. Owing to our relatively small sample size (42 SNe), we limit ourselves to examining 1 to 3 combined Gaussians.

Applying GMM to the distribution of peak absolute magnitudes, the AIC would prefer a combination of 3 Gaussians describing the distribution. The BIC favours a single Gaussian describing the peak absolute magnitude distribution. For simplicity, we consider this single Gaussian (Fig.~\ref{fig:peakmags}) in our analysis.

The observed peak-magnitude distribution for the sample is shown in Fig.~\ref{fig:peakmags} along with the best-fit Gaussian distributions. The $g$-band photometry was used when measuring four of the SNe (see Table~\ref{tab:peakmag_time}). Of our sample SNe~IIn, the peak magnitudes within $1\sigma$ from the mean peak magnitudes range between $-17.8$ and $-20.0$\,mag (a factor of 8 in luminosity).

Comparing these peak-magnitude ranges to known SNe~IIn in the literature (central panel in Fig.~\ref{fig:peakmags}), we notice that many well-studied SNe~IIn peak within the $-17.8$ to $-20.0$\,mag range as found above. In the bottom panel of Fig.~\ref{fig:peakmags}, we show the 39 SNe with determined peak epoch and mark the $-17.8$ to $-20.0$\,mag range with horizontal dash-dot lines. The large total range of peak absolute magnitudes for the sample ($\sim 5$\,mag) is clearly shown. \citet{kiewe12} and \citet{taddia13} found the peak absolute magnitudes of the SNe~IIn they studied to be in the interval $-19 \lesssim M_B \lesssim -17$\,mag, consistent with the findings of other authors \citep{richardson02, li11, richardson14} and with our result.

\begin{figure*}
    \centering
      \includegraphics[width=14cm]{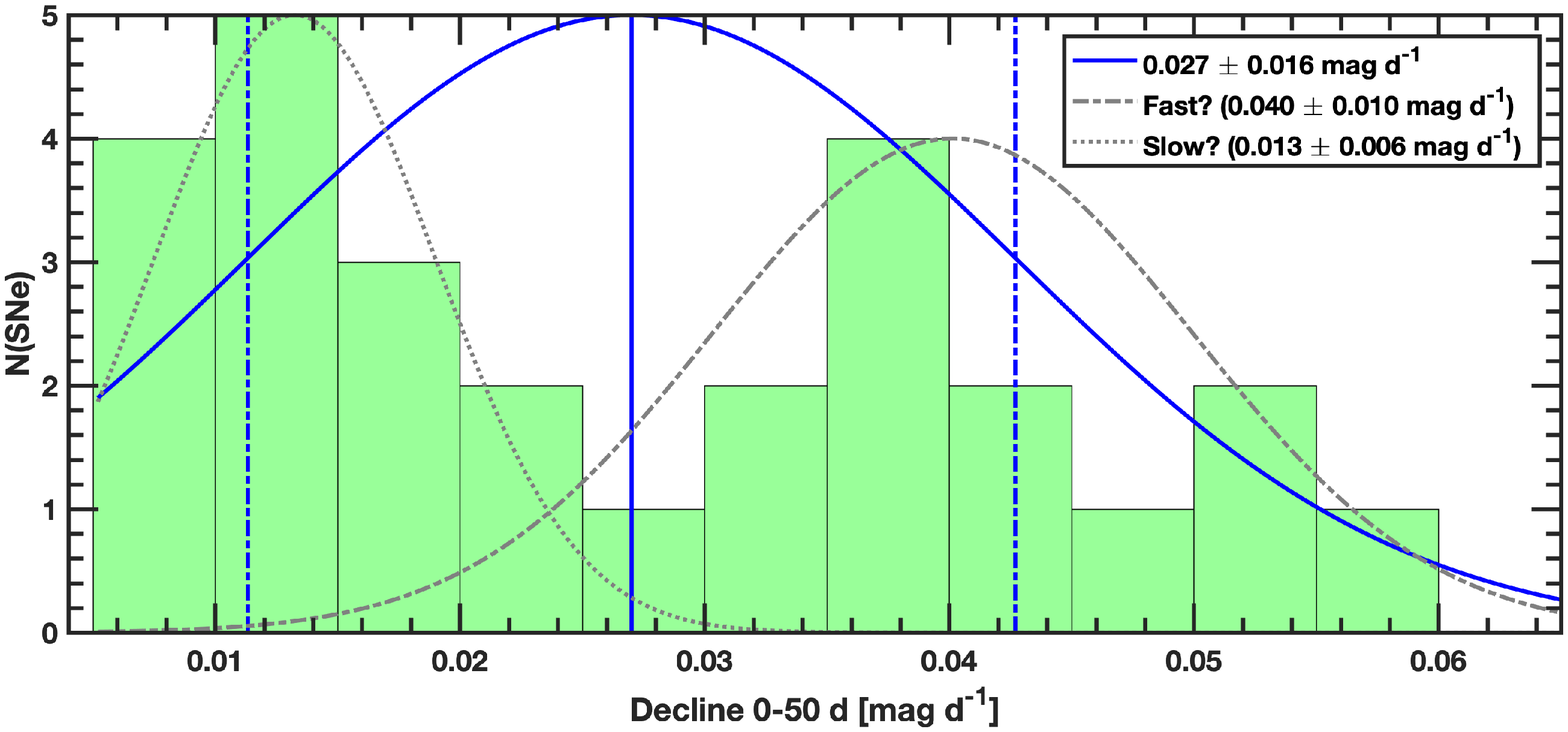}\\
      \includegraphics[width=14cm]{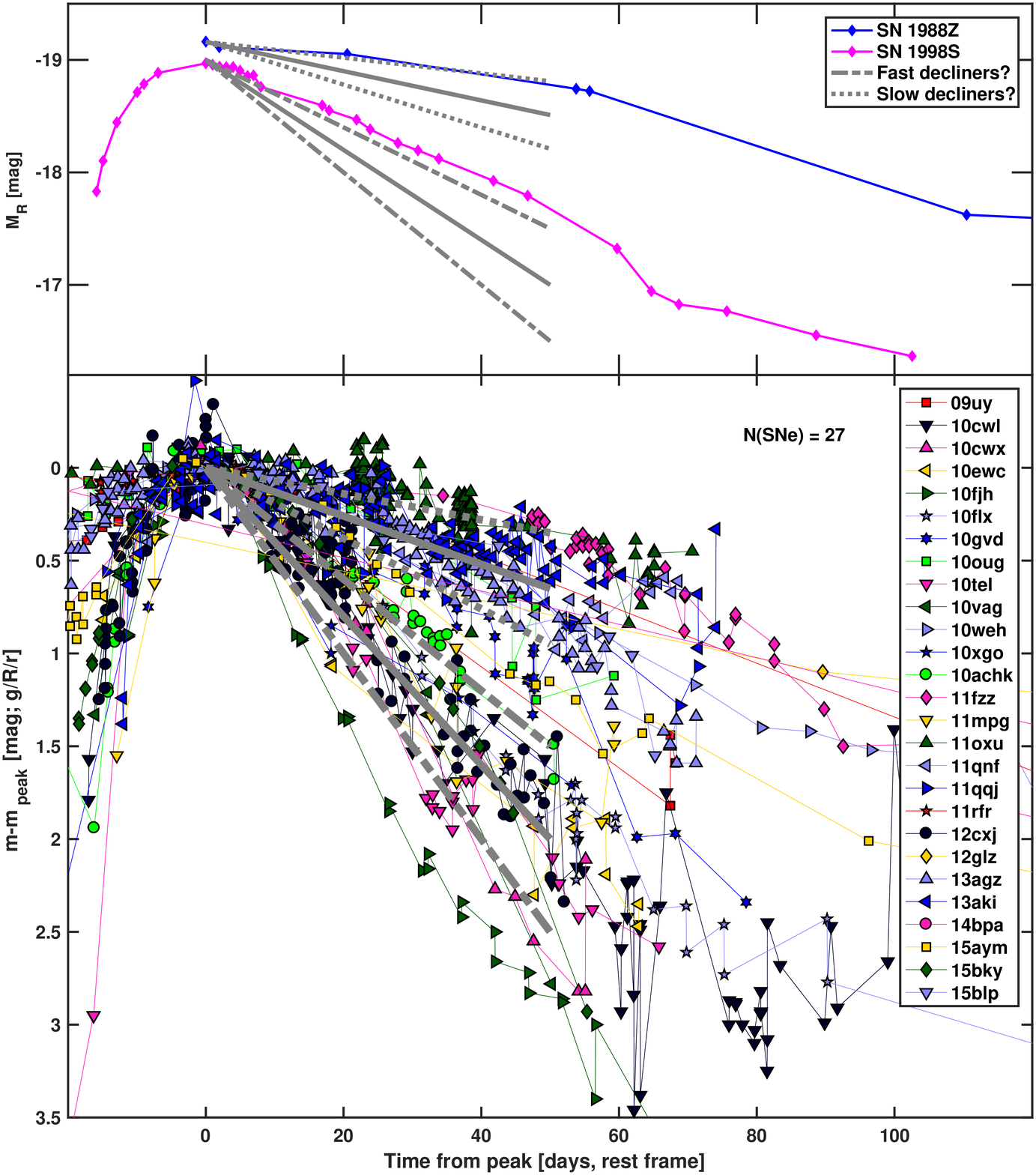}
     \caption{\textit{Top panel:} histogram for the decline rates between 0 and 50 d after peak brightness for 27 of our SNe~IIn. A single Gaussian (identified using GMM with BIC) describing the whole population is shown in blue, but our SNe~IIn might possibly divide into fast (grey dash-dot lines) and slow (grey dotted lines) decliners, with slopes $0.040 \pm 0.010$\,mag\,d$^{-1}$ and $0.013 \pm 0.006$\,mag\,d$^{-1}$, respectively (suggested by GMM with AIC). The respective mean decline rates are shown as solid grey lines. \textit{Central panel:} The two possible clusters of SNe~IIn (slow and fast decliners) compared to the prototypical slow and fast decliners from the literature (SN~1988Z, \citealp{aretxaga99}; SN~1998S, \citealp{liu00}). \textit{Bottom panel:} The light-curves of the $\rm N(SNe) = 27$ of our SNe~IIn with measured decline rates (between 0 and 50\,d), scaled to match at peak. The two different decline rates are overplotted. Fast (grey dash-dot lines) and slow (grey dotted lines) decliners are shown, with the respective mean decline rates marked with solid grey lines.} 
     \label{fig:compare_decline}
 \end{figure*}

\begin{figure*}
   \centering
    \includegraphics[scale=0.36]{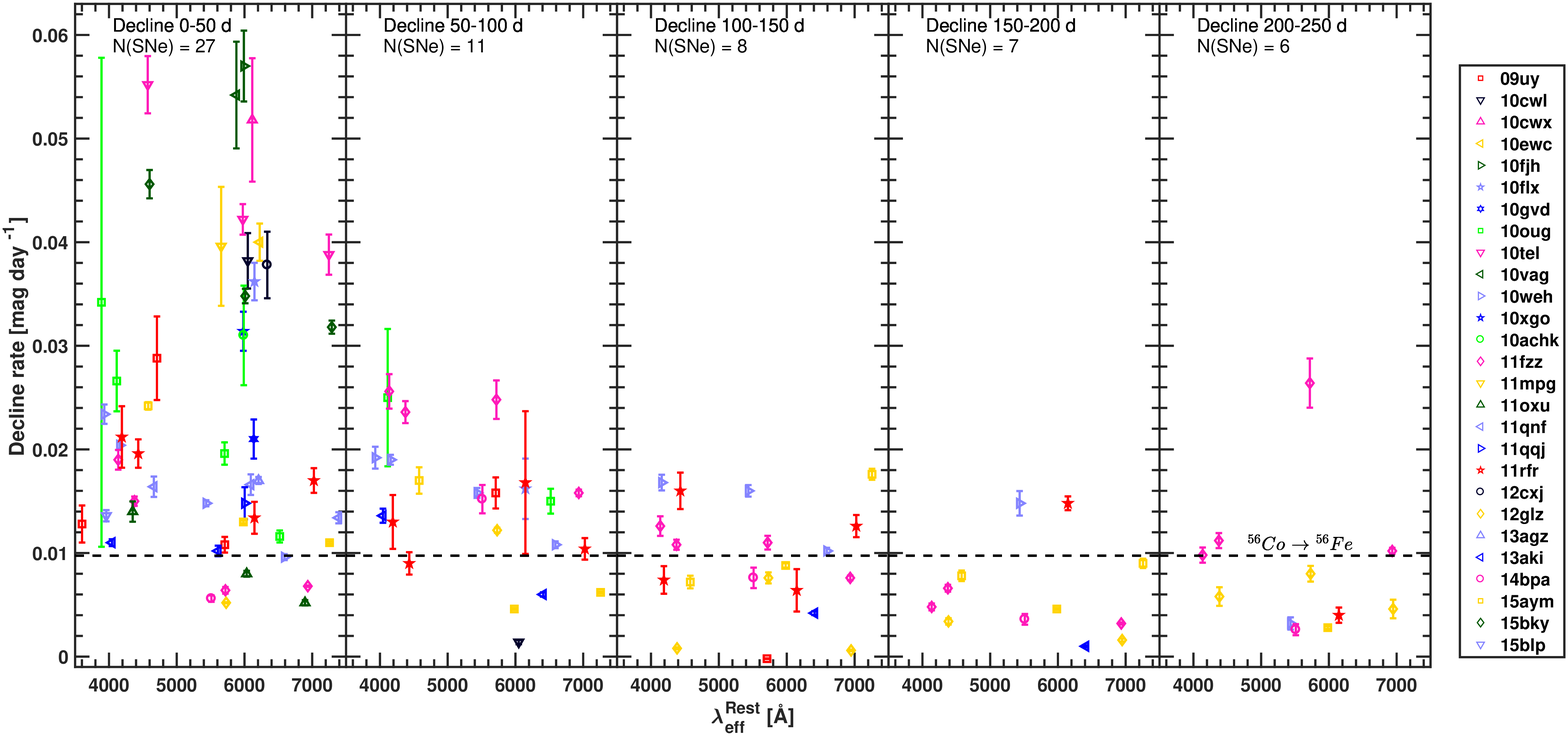}
    \caption{Decline rates 0--250 rest-frame days (past the peak time given in Table~\ref{tab:peakmag_time}) of 27 of the SNe in our sample. For reference, the decay rate 0.0097\,mag\,d$^{-1}$ of $\rm ^{56}Co$~$\rightarrow$~$\rm ^{56}Fe$ is shown as a dashed black line. The number of unique SNe in each panel is given as $\rm N(SNe)$.}
    \label{fig:declines}
\end{figure*}

\subsection{Light-curve decline rates\label{sec:decline}}
Among the first events classified as SNe~IIn it was noticed that some faded slowly compared to other CC~SNe. \object{SN 1988Z} \citep{stathakis91, aretxaga99} declined at $\sim 0.2$\,mag\,(100\,d)$^{-1}$ in the $R$ band past 100\,d after peak brightness. \object{SN 1998S} \citep{liu00, fassia00} instead had a fast initial decline of $\sim 2.5$\,mag\,(100\,d)$^{-1}$ in the $R$ band. Sample measurements of SN~IIn decline rates have been presented by \citet{taddia13} and \citet{kiewe12}.

In the MC experiments with the CSS fits (Sect.~\ref{sec:peak}), the flux density values at 50, 100, 150, 200, and 250 days past peak were determined (if occurring within the time range where the SN was detected) for calculation of decline rates in these time intervals. This is done for the photometry obtained in the $BgRri$ bands, in the cases where the coverage exceed 50\,d (rest frame) after the peak times given in Table~\ref{tab:peakmag_time}. The decline rates for 0--50\,d for the $R/r/g$ bands are provided in Table~\ref{tab:peakmag_time}. The distribution of the decline rates between 0 and 50\,d post-peak (for the 27 SNe possible to measure) is shown as a histogram in the top panel of Fig.~\ref{fig:compare_decline}.

A GMM test (Sect.~\ref{sec:peak}) gives inconclusive results, with the BIC significantly suggesting one Gaussian distribution of decline rates. Such a single distribution has $0.027\pm0.016$\,mag\,d$^{-1}$ and is overplotted in the top panel of Fig.~\ref{fig:compare_decline}. However, AIC suggests two Gaussian distributions, but with weak significance: one of slow decliners ($0.013\pm0.006$\,mag\,d$^{-1}$) and one of fast decliners ($0.040\pm0.010$\,mag\,d$^{-1}$). No K-corrections were applied when measuring these rates. Overplotting these decline rates on the actual SN light-curves (scaled to match at peak brightness) in the bottom panel of Fig.~\ref{fig:compare_decline} visualises this result. The central panel shows that most of our fast-declining SNe~IIn are actually declining faster than SN~1998S (often considered as a prototypical fast decliner). 

To allow a straightforward comparison between decline rates of SNe at different redshifts, in Fig.~\ref{fig:declines} we plot the decline rates as a function of the effective rest wavelength \citep[for the filter used and the redshift of the SN in question, as in for example][]{gonzalezgaitan15,taddia15_Ibc}. In the 0--50\,d  panel, we see the lack of slow decliners in the bluer bands. This is similar to what is seen in other SN types \citep[e.g. SNe~IIP;][]{valenti16} where also the bluer bands exhibit the fastest decline rates. For the panels showing times $> 50$\,d, the number of SNe gets smaller. This likely involves observational bias, since these SNe are becoming too faint for us to monitor. Some SNe~IIn \citep[like SN~2005ip; see][]{stritzinger12} had an initial rapid decline, followed by a long-lasting ($> 100$\,d) fainter plateau. Such faint and late plateaus would be missed in this study for many of our SNe, owing to their large distances. In Fig.~\ref{fig:declines}, the decline rate plotted for PTF11fzz in the $200-250$ d panel is illusive and caused by the photometric filter effect discussed in Sect.~\ref{sec:bumps}.

Although SNe~IIn are assumed to primarily be driven by circumstellar interaction (CSI), we can use the decline rates found here to evaluate if radioactive decay can be a possible energy source in some of the SNe in our sample. For CC~SNe such as SNe~IIP, the rate of decay for $\rm ^{56}Co$~$\rightarrow$~$\rm ^{56}Fe$ determine the light-curve slope around $\sim 200$\,d after light-curve peak. Figure~\ref{fig:declines} shows that some SNe around 200\,d after peak have decline rates consistent with radioactive decay. From Fig.~\ref{fig:LCs}, it can be seen that these SNe have absolute magnitudes of $M \approx -17$\ at this time. Calculations using \citet[][his Eq.~19]{nadyozhin94} indicate that the original $\rm ^{56}Ni$ mass necessary to drive such a light-curve is $\sim 1$\,M$_{\odot}$, which is at least an order of magnitude more than for normal SNe~IIP \citep[e.g.][]{hamuy03,rubin16,anderson19}. Most SNe in our sample indeed decline more slowly than the radioactive decay, and we consider radioactive decay as an unrealistic mechanism to explain the late-time luminosities and decay rates seen in our sample of SNe~IIn.

\subsection{Light-curve rise times\label{sec:risetimes}}
By fitting a function to the rising portion of a SN light curve, the starting time of the light-curve can be constrained, given the assumption that the unseen early portion of the rise is smooth and that the fitted function is consistent with the pre-discovery upper limits in the photometry. Functions commonly used for such fits are power-law functions \citep{gonzalezgaitan15, ofek14rise, cowen10} or exponential functions \citep{gonzalezgaitan15, ofek14rise, bazin09}, which may have little physical motivation but allow quantification of the SN light-curve rise.

In our work, we are primarily interested in finding a consistent way to measure the rise times of our sample SNe~IIn, without characterising the shape of the light-curve rise in detail for each SN. For this purpose, we use a power-law light-curve template based on a well-studied, nearby SN in our sample.

Based on the procedure in Sect.~\ref{sec:peak}, the SNe with peak time error $> 10$ days (PTF10qwu, PTF10acsq, iPTF13cuf) are excluded also in the rise time analysis. The available pre-peak photometry does not allow a reliable measurement of the rise time if the difference in magnitude covered during the rise is too small, or if the photometry is taken during a too short time interval. For the magnitude interval ($\Delta m$) and time interval ($\Delta t$) covered by $n$ pre-peak detections, we therefor exclude any SN having $\Delta m < 0.60$ mag and $n / \Delta t > 0.40$ $\rm d^{-1}$. This removes PTF10abui, PTF10flx, PTF11rlv, PTF11qqj, iPTF14bcw and iPTF16fb from the rise time analysis.

\begin{figure}
   \centering
    \includegraphics[width=\linewidth]{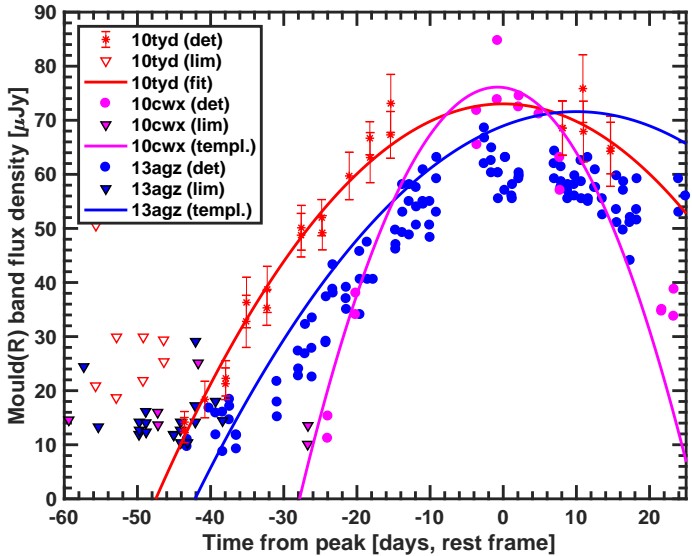}
    \caption{Light-curve ($R$-band) of PTF10tyd, with weighted fit of the $\propto t^2$ template shown along with the rise template candidates PTF10cwx and iPTF13agz. Scale, stretch, and shift parameters as found in Sect.~\ref{sec:risetimes} are applied.\label{fig:temp10tyd}}
\end{figure}

\begin{figure*}
    \centering
   \includegraphics[width=14cm]{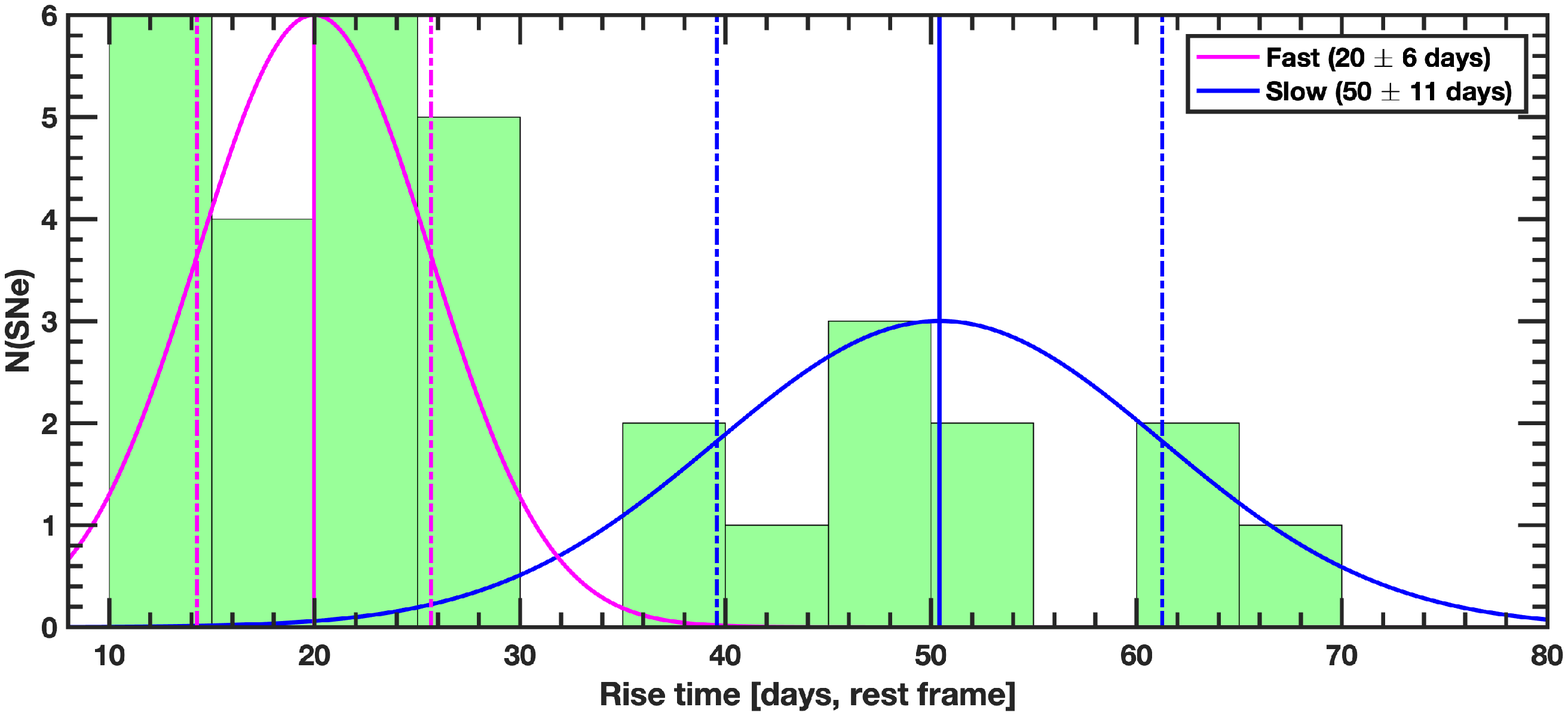}\\
    \includegraphics[width=14cm]{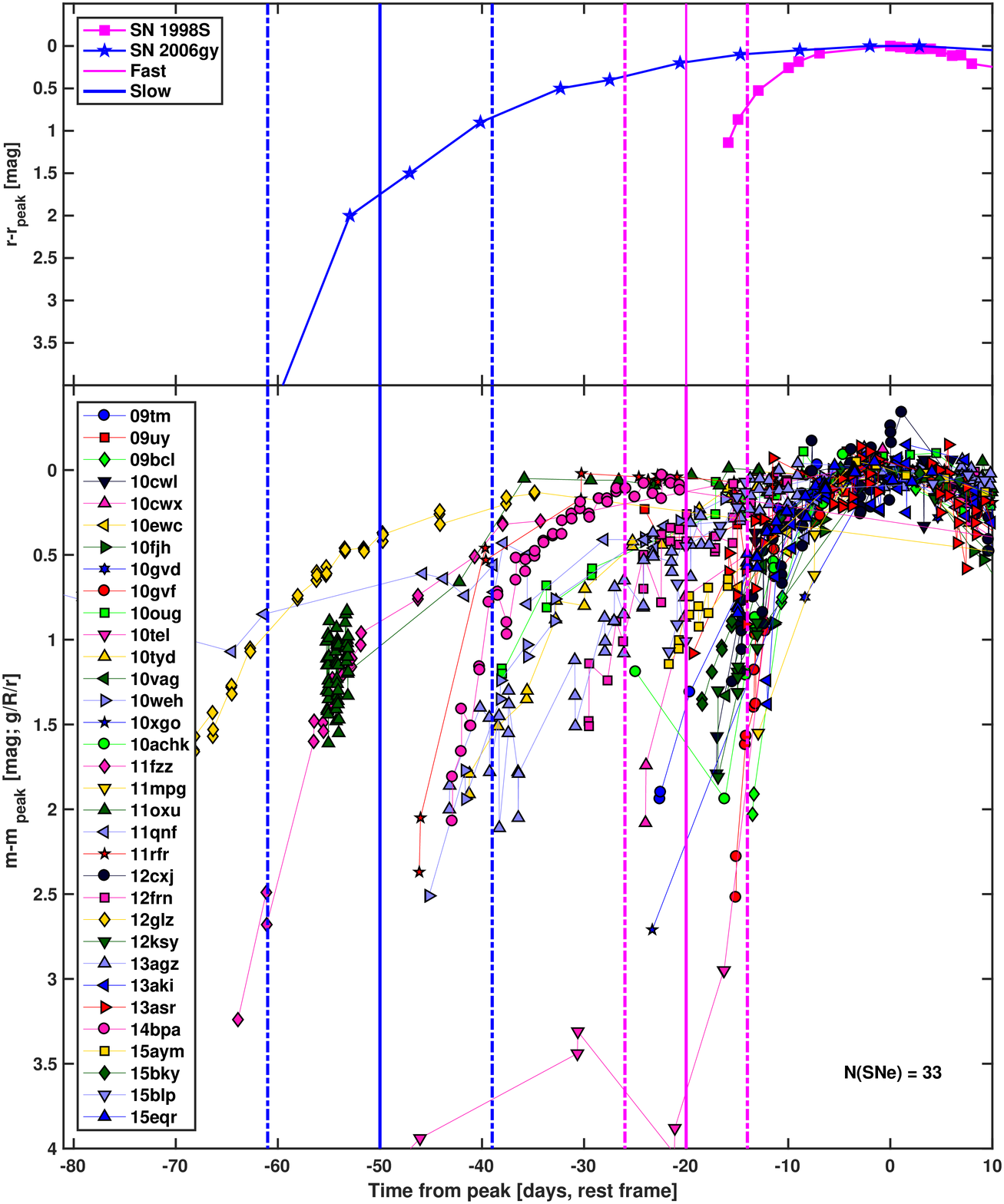}
     \caption{\textit{Top panel:} histogram for the rise times for 32 of our SNe~IIn. Our SNe~IIn are divided into fast (magenta lines) and slow (blue lines) risers, with rise times $20\pm6$\,d and $50\pm11$\,d, respectively (as identified with GMM). Dashed lines show $1\sigma$ of the Gaussians. For reasons discussed in Sect.~\ref{sec:impost}, we excluded PTF11qnf from the GMM analysis. \textit{Central panel:} The rise times of the two clusters of SNe~IIn (slow and fast risers) compared to two well-known slow and fast risers from the literature (SN 2006gy, \citealp{smith07phot06gy}; SN~1998S, \citealp{liu00}), with the rise-time ranges from the top panel overplotted. \textit{Bottom panel:} The light-curves of all the $\rm N(SNe) = 33$ SNe in our sample with well-determined peak times matched at peak brightness (here, we include PTF11qnf for completeness).}
     \label{fig:rise}
 \end{figure*}

To make a rising light-curve template for rise-time measurements, we considered SNe from our sample with $z < 0.075$, having the first detection within 4\,d of the last upper limit (in the observer's frame) and this first detection having $m_R > 21$\,mag. At $z = 0.075$, this means $M_R \gtrsim -16.5$\,mag. This gives the rising light-curve template candidates PTF10cwx, PTF10tel, PTF10tyd, PTF11qnf, and iPTF13agz. We choose PTF10tyd as our template owing to it having the lowest $\chi^2_{\rm {d.o.f.}}$ for a fitted $t^2$ power law. A least-squares fit to the PTF10tyd rising light-curve (flux density scale) gives the rest-frame template with the rise time $\Delta t = 47.5$\,d (Fig.~\ref{fig:temp10tyd}) for the power-law model

\begin{equation}
\label{eq:powerlaw}
L(t) = L_{\rm peak} \left[1 - \left(\frac{t - t_{\rm peak}}{\Delta t}\right)^2\right]
\end{equation}

\noindent
where $L_{\rm peak}$ is the peak luminosity from the fit, $\Delta t$ is the time from 0 to peak luminosity, and $t_{\rm peak}$ is the time of the peak luminosity \citep[][their Eq.~5]{ofek14rise}. This template is stretched, scaled, and shifted in order to fit the other SNe in our sample (see Fig.~\ref{fig:fitrise_all}). For demonstration, in Fig.~\ref{fig:temp10tyd} we also show the PTF10tyd template stretched, scaled, and shifted to the light-curves of template candidates PTF10cwx and iPTF13agz.

In Fig.~\ref{fig:fitrise_all} we have converted pre-peak AB magnitudes of the SN to flux density as in Sect.~\ref{sec:peak} and normalised the flux density values to peak brightness. We then stretch the light-curve data points in time by applying a multiplicative factor, scale their position in flux density (additively), and shift their position in time (additively) to minimise the distance (in a LSQ sense) of the light-curve data points to the power-law template. When this is done, the derived stretch value is used to compute the rise time of the input SN (as $t_{\rm rise} = \Delta t_{\rm PTF10tyd} / {\rm stretch}$). A MC experiment is done to generate new light-curve data points within the light-curve error bars. The light-curve stretch, scale, and shift operations are then repeated to estimate the uncertainty of the stretch value.

For the cases where the obtained $\propto t^2$ template fit ends up inconsistent (i.e., brighter) than an upper limit observed before the first detection, the time of that upper limit is assumed as an approximation of the "start of rise" epoch. In such cases, as an estimate of the uncertainty we assume the time between the upper limit and the first detection. For the special case of PTF10tel \citep{ofek13}, with its detected photometric activity before the start of rise, we adopt the explosion epoch given by \citet{ofek13} as the "start of rise" epoch. The derived rise times (in the rest frame) are presented in Table~\ref{tab:peakmag_time}.

We plot the histogram of the rise times in the top panel of Fig.~\ref{fig:rise}. The SN~IIn population is characterised by a large range of rise times, and a GMM test (Sect.~\ref{sec:peak}) shows a significant division (for both AIC and BIC) into two clusters: fast-rising SNe~IIn with rise times of $20\pm6$\,d (magenta lines in Fig.~\ref{fig:rise}) and slow-rising SNe~IIn with rise times of $50\pm11$\,d (blue lines). PTF11qnf, where we found a rise time to peak of 135\,d (Table~\ref{tab:peakmag_time}) and a faint ($\rm M_r \approx -17$\,mag) peak, is an outlier not included in this GMM rise-time analysis. PTF11qnf might have been an SN impostor (Sect.~\ref{sec:impost}).

The literature about SNe~IIn already showed examples of fast-rising SNe~IIn like SN~1998S and slow-rising SNe~IIn like the SLSN~Type~IIn \object{SN 2006gy}. These two SNe are displayed in the central panel of Fig.~\ref{fig:rise} and the rising portions of their light-curves are compared to the two rise-time ranges we found in our sample. We overplot all our SNe~IIn with measured rise times in the bottom panel of Fig.~\ref{fig:rise}, to further illustrate the variety of rise-time values and the relatively similar light-curve shapes. We particularly note the erratic light-curve of PTF11qnf as well as the precursor activity prior to the start of rise for PTF10tel (as analysed by \citealp{ofek13} and compared to \object{SN 2009ip} by, for example, \citealp{pastorello18} and \citealp{smith14_09ip}). In Sect.~\ref{sec:discu} we further discuss these events.

\subsection{Correlations between light-curve properties\label{sec:corr}}
\begin{figure*}
   \centering
    \includegraphics[width=9cm]{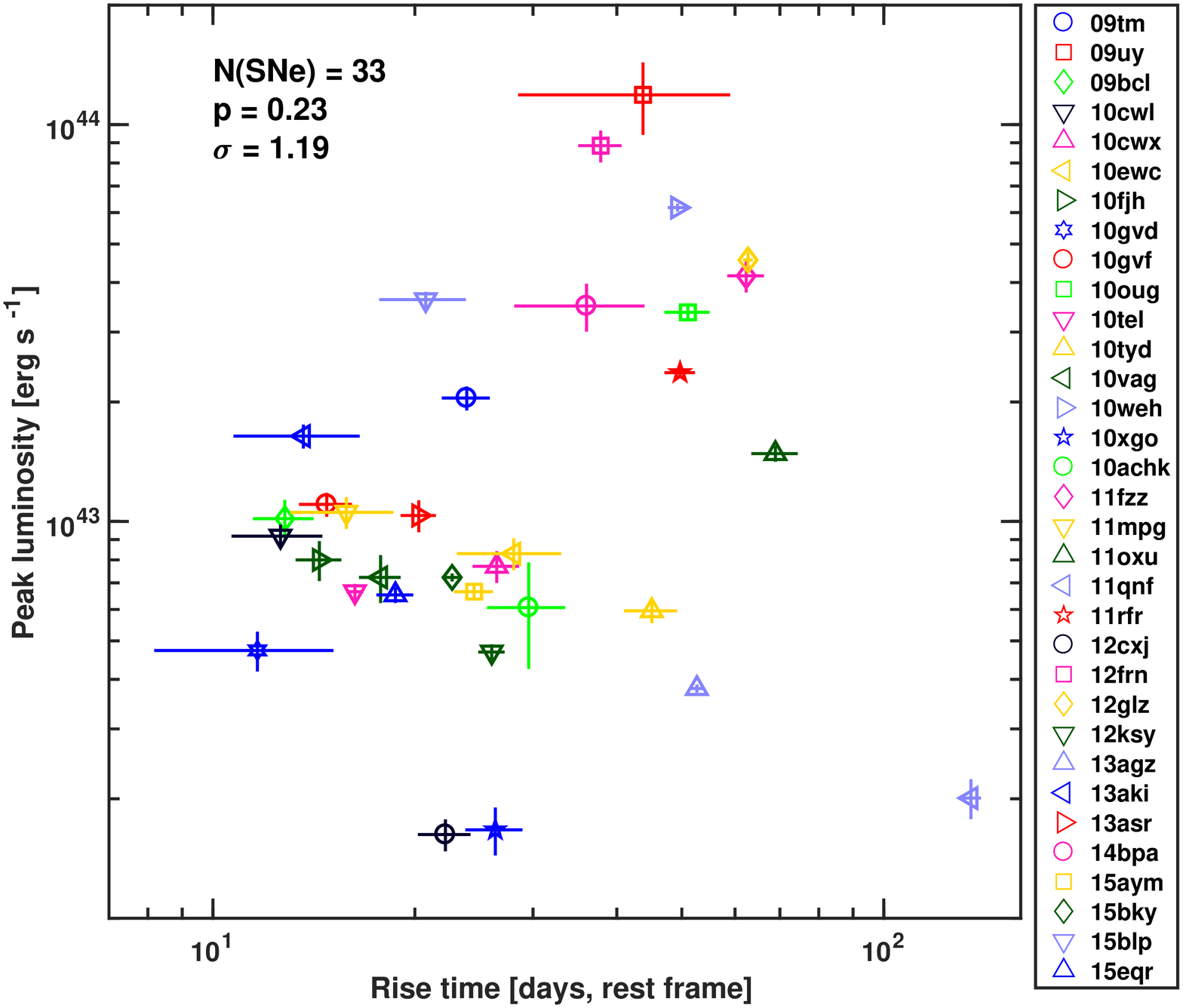}
    \caption{Rise times and peak luminosities for 33 of the SNe in the sample, using measurements described in Sect.~\ref{sec:risetimes}, following the investigation by \citet{ofek14rise}. The photometric band used for each SN is given in Table~\ref{tab:peakmag_time}. The Spearman correlation coefficient $p$, the corresponding significance $\sigma$ and the number of SNe, $\rm N(SNe)$, are shown.\label{fig:risetimes}}
\end{figure*}

Having measured the main light-curve-shape parameters, we proceed to investigate if these quantities are somehow correlated in our sample. \citet{ofek14_10jl} proposes a correlation between rise time and peak luminosity, based on the assumption of the SN shock breaking out in the CSM. This was studied observationally by \citet{ofek14rise}, inferring a possible correlation. From analytical models, \citet{moriyamaeda14} instead indicate that no strong dependence on rise time should be seen for the peak luminosity (shock breakout does not have to occur in the CSM of all SNe~IIn.). In Fig.~\ref{fig:risetimes} we use the peak luminosities obtained from absolute $R/r$-band magnitudes $M_r$ with

\begin{equation}
\label{eq:lum}
L(M_r) = (3.04 \times 10^{35}) \times 10^{(-0.4~M_r)}\,\rm erg\,s^{-1}
\end{equation}

\noindent
based on solar bolometric values and, following \citealp{ofek14prec} and \citealp{nyholm17}, neglecting bolometric corrections. A unweighted Spearman correlation test for the SNe plotted in Fig.~\ref{fig:risetimes} gives the correlation coefficient $p = 0.23$, corresponding to a significance of $\sim 1.19\sigma$. We report the significance for a given correlation coefficient $p$ as $\sqrt{2} \times \erf^{-1}({1-p})$, using a two-sided tail of the normal distribution. Excluding the possible SN impostor PTF11qnf (Sect.~\ref{sec:impost}) gives $p = 0.08$ ($\sim 1.8\sigma$). If a correlation between rise time and peak luminosity exists at all, it is weak.

Using a bootstrap technique, \citet{ofek14rise} claimed a significance of $\sim 2.5\sigma$ for their sample of 15 SNe~IIn (12 of which are in our sample). With our method, we find a significance of $\sim 1.75\sigma$ for this \citet{ofek14rise} sample. Finding a similar significance with our larger sample (32 SNe; excluding PTF11qnf) also suggests that the correlation, if any, is weak.

\begin{figure*}
   \centering
  \includegraphics[width=18cm,angle=0]{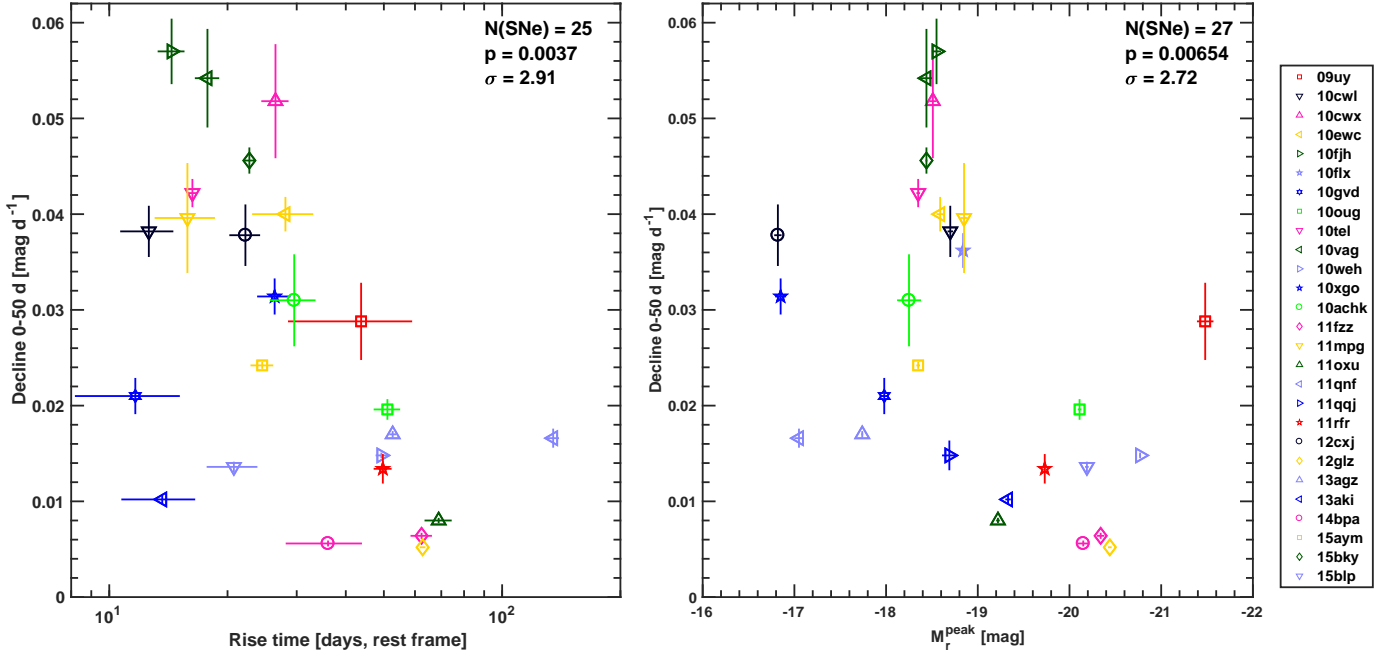}
    \caption{Examination of possible correlations between decline rates versus rise times and peak absolute magnitudes, respectively, of the sample SNe (Sect.~\ref{sec:corr}). Photometric bands are as specified in Table~\ref{tab:peakmag_time}. In each panel the Spearman correlation coefficient $p$, the corresponding significance $\sigma$ and the number of SNe, $\rm N(SNe)$, are shown.\label{fig:corr}}
\end{figure*}

In Fig.~\ref{fig:corr} (left panel) we find a correlation with $p = 0.0037$ ($\sim 2.9\sigma$) for decline rate versus rise time. This shows that slowly rising SNe~IIn are also slow decliners, and fast risers are fast decliners. The empty upper-right corner of the decline rate versus rise time plot of Fig.~\ref{fig:corr} shows that we do not see any SNe having both a slow rise and a fast decline. For decline rate versus peak absolute magnitude (Fig.~\ref{fig:corr}, right panel) we obtain $p = 0.0065$ ($\sim~2.7\sigma$), suggesting that the more luminous SNe~IIn decline slower. This is somewhat comparable to the Phillips relation \citep{pskovskii77,phillips93} between peak magnitude and decline rate for SNe~Ia, but less tight in the SN~IIn case. These correlations for SNe~IIn can be discerned in the bottom panel of Fig.~\ref{fig:peakmags}.

As a control, we investigate the presence of bias by looking for correlations between rise or decline rates and the peak \textit{apparent} magnitudes of the SNe~IIn. Such a correlation might indicate bias affecting the correlations discussed above. The decline rate versus peak apparent magnitude gives $p = 0.28$ ($\sim 1.1\sigma$). The rise time versus peak apparent magnitude gives $p = 0.52$ ($\sim 0.65\sigma$). The absence of a correlation in both cases suggests that the correlations with absolute magnitude could be real.

\subsection{Colour evolution\label{sec:colour}}
\begin{figure*}
   \centering
     \includegraphics[width=14cm,angle=0]{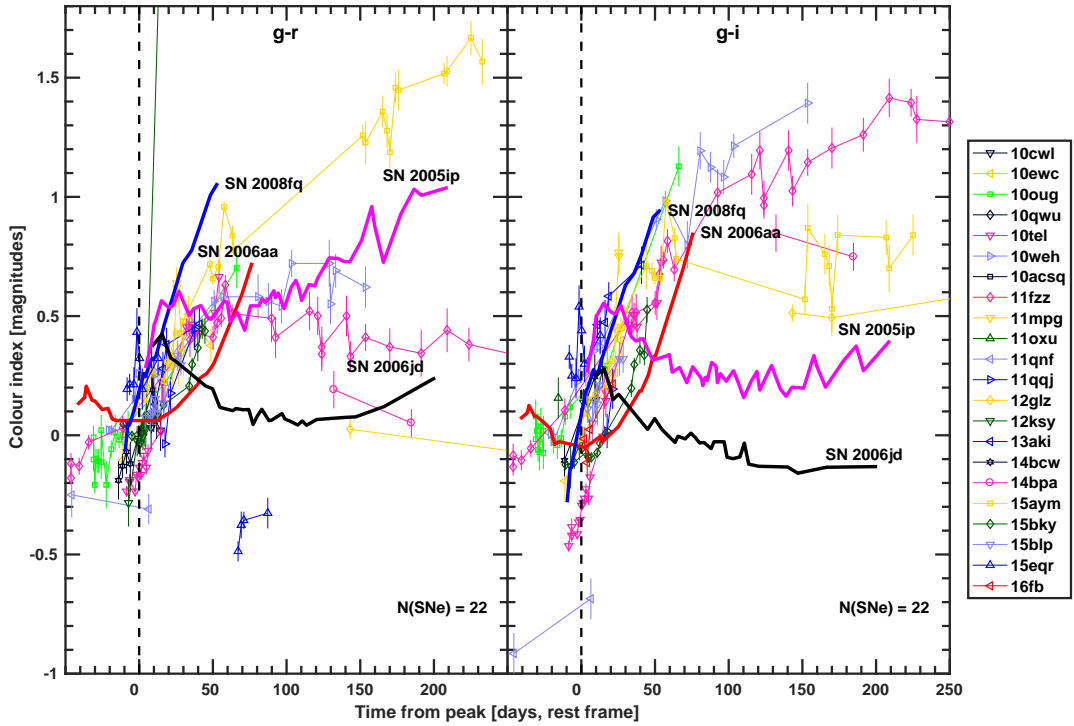}
    \caption{Evolution of $g-r$ and $g-i$ colours for 22 SNe in the sample. Host and MW extinction have been removed (Sect.~\ref{sec:sample}). The vertical black dashed line shows the light-curve peak epoch (Sect.~\ref{sec:peak}). Lines are drawn between data points to guide the eye. For clarity, colour data points with uncertainties $> 0.1$\,mag are not shown. For comparison, SNe~IIn 2005ip, 2006jd \citep{stritzinger12}, 2006aa, and 2008fq \citep{taddia13} have their colour curves shown with bold lines.\label{fig:colours}}
\end{figure*}

Only a small number of SN~IIn sample studies have considered the colour evolution of the SNe~IIn in a collective fashion. Compilations of colour evolution of SNe~IIn were presented by \citet{taddia13} and \citet{delarosa16}, based on their respective samples. In Fig.~\ref{fig:colours}, we present the $g-r$ and $g-i$ colour evolution for 22 of our SNe~IIn with respect to the light-curve peak epochs. The magnitudes used were corrected for extinction as described in Sect.~\ref{sec:sample}.

For comparison, we also plot the $g-r$ and $g-i$ colours of the slowly evolving SNe~IIn 2005ip and 2006jd (photometry from \citealp{stritzinger12}) and the rapidly evolving SNe~IIn 2006aa and 2008fq (photometry from \citealp{taddia13}). The colour curves for these four events from the literature have been computed using the MW and host extinctions and peak epochs given in the respective papers. The evolution of the $g-i$ colour is rather monotonic for the sample as a whole, albeit with some spread. The SNe 2006aa and 2008fq have $g-i$ colour evolution encompassing most of our sample SNe, while SNe~IIn with the slower $g-i$ colour evolution of SNe 2005ip and 2006jd are not seen among our 22 plotted SNe. The $g-i$ colour evolving towards the red for the majority of the SNe reflects their decreasing blackbody continuum temperatures. The $g-r$ colour evolution is likely affected by the evolution of the H$\alpha$ emission line (with a line centre within the SDSS $r$ filter for $z \lesssim 0.065$, where 7 of the 22 SNe in Fig.~\ref{fig:colours} are located).

The small spread in colours suggests that the colour corrections applied (Sect.~\ref{sec:sample}) are without large errors. Exceptions are the $g-r$ colour of iPTF15eqr and the $g-i$ colour of PTF11qnf. In the case of PTF11qnf, a possible SN impostor, its colour index and its host $E(B-V)$ is discussed in Sect.~\ref{sec:impost}.

Our analysis of rise and decline behaviour (Sect.~\ref{sec:corr}) as well as the luminosity function (Sect.~\ref{sec:lumfunc}) is based on $R/r$-band photometry as far as possible, but when the $R/r$-band is not available, we used the $g$-band instead. This is the case for four of our SNe (Table~\ref{tab:peakmag_time}). Via linear interpolation between epochs surrounding time of peak, we find the colours of each SN at peak. The $g-r$ colour index at peak of the SNe~IIn is $g-r = 0.06\pm0.21$\,mag and $g-i = 0.13\pm0.46$\,mag. This is consistent with $g-r \approx 0$\,mag and $g-i \approx 0$\,mag around peak brightness, and it indicates that our use of $g$-band for four SNe in our analysis should not affect our conclusions.

\subsection{Luminosity function\label{sec:lumfunc}}
The observed luminosity function (see Fig.~\ref{fig:peakmags}, top panel) of the SNe in our sample is affected by Malmquist bias \citep{malmquist22}. This means that SNe intrinsically luminous at peak (thus easier to detect in a magnitude-limited survey like PTF/iPTF) are over-represented compared to the intrinsically less luminous SNe. To reduce the impact of the Malmquist bias on our estimated luminosity function, we use the bootstrap method described by \citet{richardson14} as implemented by \citet{taddia19}.

\begin{figure}
 \centering
  \includegraphics[width=\linewidth]{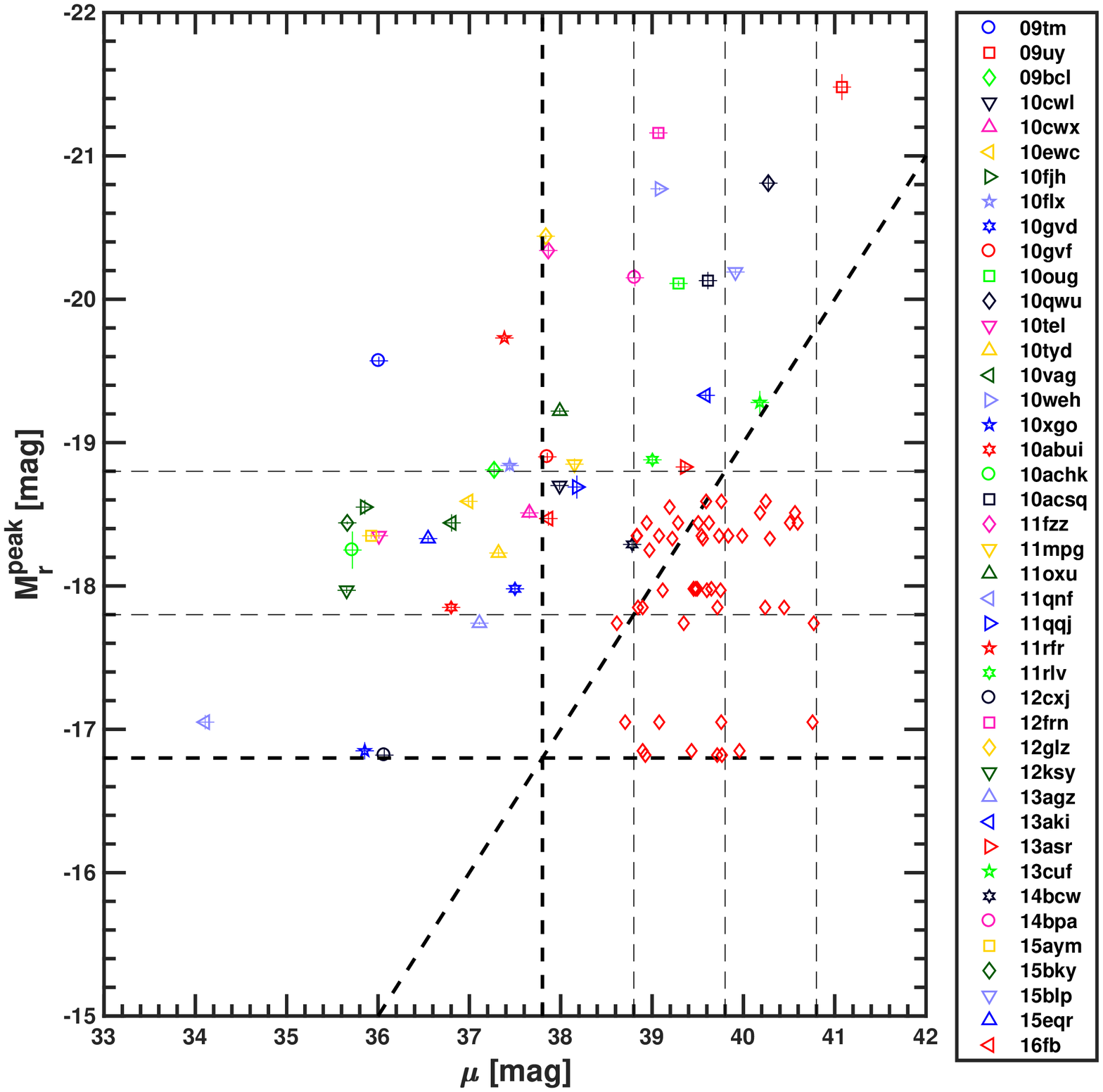}
  \includegraphics[width=\linewidth]{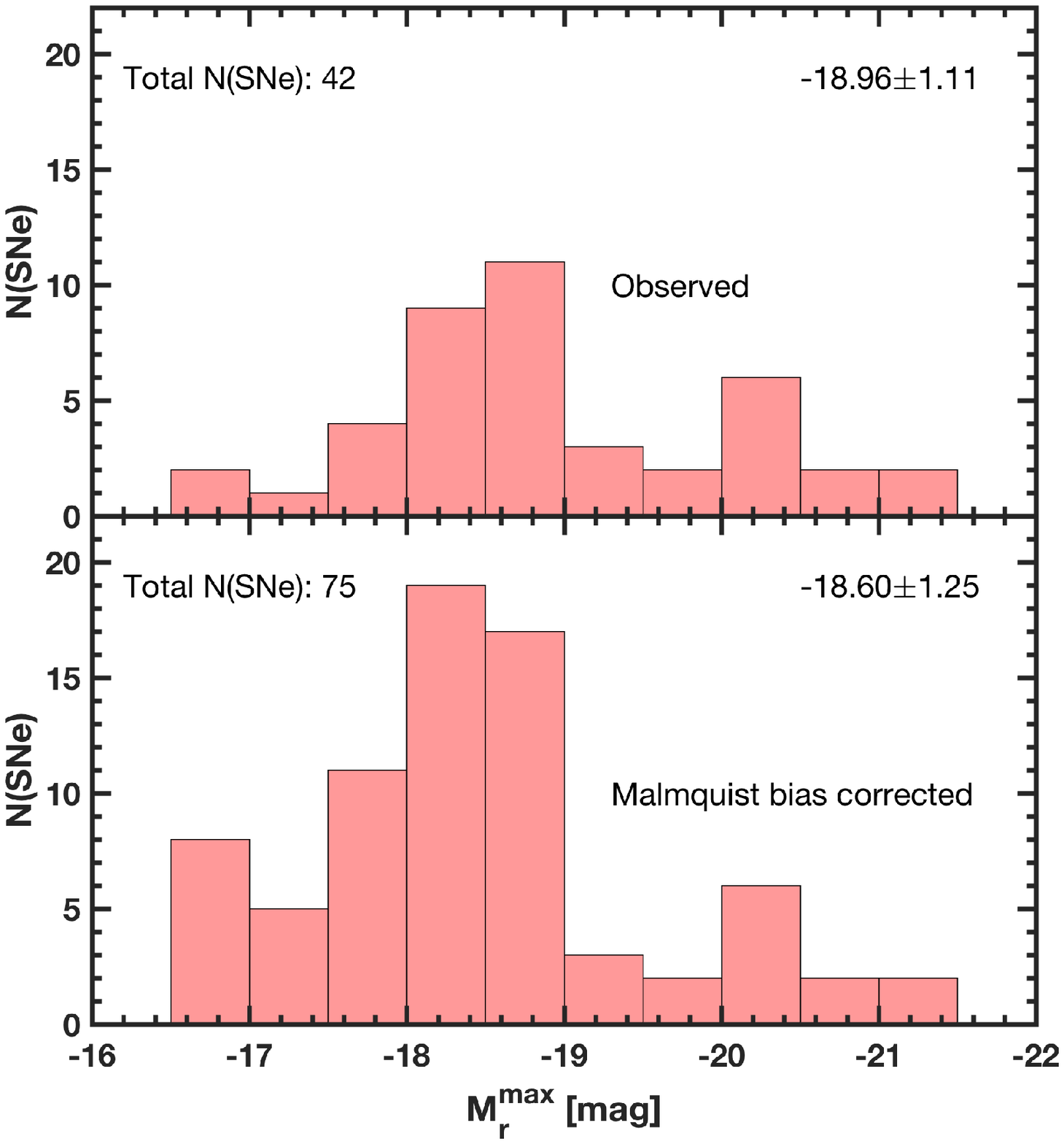}
  \caption{\textit{Top panel:} Peak absolute magnitude ($\rm M_{r}^{peak}$) as a function of distance modulus ($\rm \mu$) of the 42 SNe in the sample, shown with the randomly generated SNe~IIn (empty red diamonds) introduced to compensate for Malmquist bias. The limiting magnitude $m = 21$ typical for PTF/iPTF under favourable conditions is shown as a diagonal, dashed line. Vertical and horizontal dashed lines show $\rm M_{r}^{peak}$ and  $\rm \mu$ intervals described in Sect.~\ref{sec:lumfunc}. \textit{Bottom panel:} Histograms of the observed and bias-compensated luminosity distributions of the SNe~IIn at peak brightness. Bias compensation follows \citet{richardson14} and \citet{taddia19}. K-corrections are not taken into account in the results plotted here, but are studied in Sect.~\ref{sec:lumfunc}.\label{fig:malmquist}}
 \end{figure}

Figure~\ref{fig:malmquist} (top panel) shows the peak absolute magnitude as a function of distance modulus ($\mu$) for our 42 sample SNe. PTF12cxj with $M_{r}^{\rm max} = -16.82$\,mag is the intrinsically faintest SN in the sample and the dashed diagonal line represents the typical limiting magnitude of the PTF/iPTF survey under good conditions, $m_R \approx 21$\,mag. This indicates that our sample satisfactorily represents the SN~IIn population up to $\mu \approx 37.8$\,mag ($z \approx 0.08$). This is consistent with the sample redshift distribution shown in Fig.~\ref{fig:redshifts}. In this discussion, all magnitudes are assumed to be in the $r$ band (c.f.~Sect.~\ref{sec:colour}).

We consider the peak-magnitude distribution observed at $\mu < 37.8$\,mag as intrinsic in the intervals $-17.8 < M_{r}^{\rm max} < -16.8$\,mag and $-18.8 < M_{r}^{\rm max} < -17.8$\,mag, respectively, and randomly generate the peak magnitudes of the missing SNe with $-17.8 < M_{r}^{\rm max} < -16.8$\,mag in the interval $37.8 < \mu < 38.8$\,mag based on these intrinsic distributions. The same random generation of missing SNe is repeated for the more distant $\mu$ intervals and are shown as empty red diamond symbols in Fig.~\ref{fig:malmquist} (top panel). The observed luminosity function of the SN sample is $M_{\rm peak} = -18.96\pm1.11$\,mag whereas the Malmquist-bias-compensated luminosity function has $M_{\rm peak} = -18.60\pm1.25$\,mag ($1\sigma$ spread) as shown in Fig.~\ref{fig:malmquist}.

The sample of 42 SNe~IIn was selected from a body of SNe~IIn all deemed to have $M_{\rm peak} > -21$\,mag (Sect.~\ref{sec:sample}) and thus not being SLSNe according to the traditional \citep{galyam12} criterion. In the study by \citet{richardson14}, SNe~IIn with $M_{\rm peak} < -21$\,mag were also included, so to facilitate a comparison of results, we repeated the above Malmquist-bias correction for the 45 SNe~IIn encompassing the main sample of 42 SNe~IIn and the 3 SLSNe~IIn (PTF10heh, PTF12mkp, and iPTF13dol) introduced in Sect.~\ref{sec:sample}. The observed luminosity function of this extended SN sample has $M_{\rm peak} = -19.10\pm1.21$\,mag whereas its bias-corrected luminosity function has $M_{\rm peak} = -18.73\pm1.35$\,mag. \citet{richardson14} found a bias-corrected distribution of $M_{\rm peak} = -18.53\pm1.36$\,mag, similar to our result.

\citet{richardson14} applied a K-correction based on spectra of \object{SN 1995G} \citep[slowly declining;][]{pastorello02} and SN~1998S \citep[rapidly declining;][]{fassia01} in their calculations. When we include our K-corrections and their uncertainties (using Eqs. \ref{eq:kcorr_r} and \ref{eq:kcorr_r_err}; for any $z > 0.31$, we assume the K-correction for $z = 0.31$) the Malmquist-bias-compensated luminosity function becomes $M_{\rm peak} = -18.72\pm1.32$\,mag. Here, we assume that all the SNe were observed in the $R/r$-band. For the extended SN sample (including the 3 SLSNe~IIn) and including K-corrections, we find the Malmquist-bias-corrected luminosity function $M_{\rm peak} = -19.18\pm1.32$\,mag. The mean values of the derived K-corrected distributions are more luminous than the ones without K-correction by 0.12\,mag and 0.45\,mag, respectively, but consistent within 1$\sigma$.

\subsection{Host galaxies\label{sec:host}}
The host galaxies of our sample of SNe can be expected to have a wide range of absolute magnitudes, since the SNe were found in an untargeted search not giving priority to any special galaxy type. In earlier SN searches, nearby galaxies with lively star formation were targeted to increase the chance of finding SNe \citep{anderson13}. The sample of SNe~IIn whose environments were studied by \citet{taddia15} has a bias towards hosts being large spiral galaxies. Here, we examine whether our sample SNe are hosted by galaxies covering a different range of absolute magnitudes than the hosts in the \citet{taddia15} sample.

From a study of the properties of PTF/iPTF SN host galaxies (Schulze et al., in prep.), we collected $g$-band absolute magnitudes of the host galaxies of our 42 sample SNe. Schulze et al. obtained the total flux of a galaxy in the different SDSS filters via a curve-of-growth analysis. For galaxies outside the SDSS field, Pan-STARRS images were used instead. In the case of PTF09uy and PTF10qwu the measurements in \citet{perley16} were used. The absolute magnitudes were obtained by modelling their spectral energy distributions with \texttt{Le PHARE} \citep{arnouts99, ilbert06}. The host absolute magnitudes (largely consistent with Petrosian magnitudes, \citealp{petrosian76}) are given in Table~\ref{tab:IInsummary}.

\begin{figure}
\begin{center}
\includegraphics[width=\linewidth]{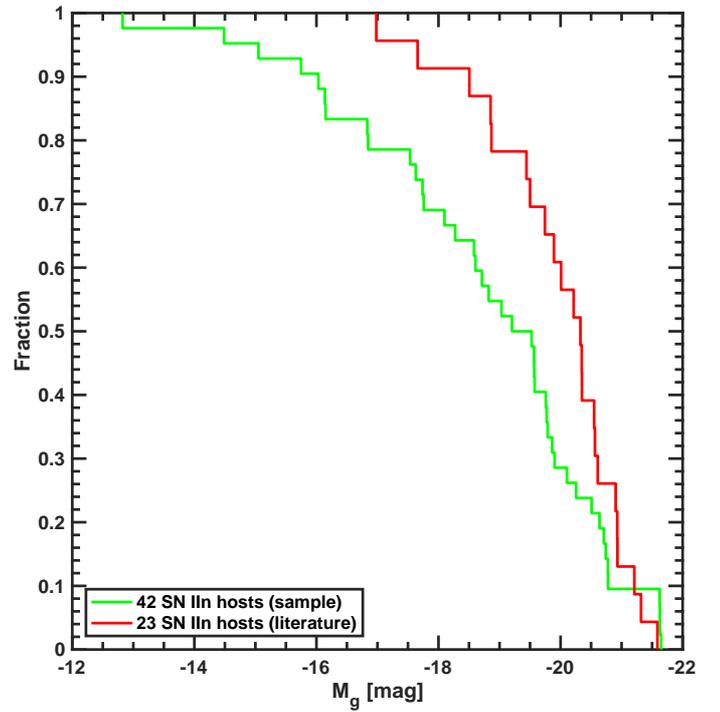}
\caption{Cumulative distribution plot of host-galaxy SDSS/Pan-STARRS $g$-band absolute magnitudes (from Schulze et al., in prep.) for the 42 hosts of our sample SNe and for 23 SN~IIn hosts in the SDSS field from the \citet{taddia15} sample.\label{fig:galmag}}
\end{center}
\end{figure}

We limited our comparison to the \citet{taddia15} SN~IIn sample hosts also in the SDSS field, obtaining their Petrosian magnitudes under the \textit{PhotoTag} section of the SDSS DR15\footnote{\protect\url{http://skyserver.sdss.org/dr15/en/tools/}} Object Explorer web pages, considering galaxies with \texttt{petroMagErr\_g} $< 0.2$\,mag. This gave 23 galaxies. We used \citet{SF11} to get the foreground extinctions and the redshifts from \citet{taddia15} with our infall-compensation method (Sect.~\ref{sec:sample}) to calculate absolute magnitudes. For the redshifts ($\lesssim 0.02$) of these literature SN hosts, K-corrections are generally small ($\lesssim 0.2$\,mag; \citealp{chilingarian10}) and can be neglected. The cumulative distribution functions of the host galaxy $g$-band absolute magnitudes are shown in Fig.~\ref{fig:galmag}. For completeness, we note that two SNe~IIn from the \citet{taddia15} sample (SNe 1997ab and 2007va) were hosted by dwarf galaxies; these galaxies did not pass our \texttt{petroMagErr\_g} requirement.

The 42 sample SN host galaxies have $-21.6 < M_g < -12.8$\,mag (8.8\,mag interval). For the 23 SN host galaxies from the \citet{taddia15} sample, the absolute magnitudes are $-21.6 < M_g < -17.0$\ (4.6\,mag interval). The host galaxies of the SNe from our untargeted sample thus cover a range of absolute magnitudes about 4\,mag wider (towards the fainter end) than the SN host galaxies from the targeted searches for which we have reliable SDSS photometry, also having a different distribution. Schulze et al. (in prep.) shows that the mass distribution of the SN~IIn hosts in our sample is consistent with the mass distribution of the hosts of the entire PTF/iPTF yield of SNe~IIn. Schulze et al. (in prep.) also show it to be consistent with the distribution of masses for PTF/iPTF hosts of SNe~Ib, Ic, IIb and II. These galaxy mass distributions are also consistent with mass distributions of CANDELS \citep[Cosmic Assembly Near-IR Deep Extragalactic Legacy Survey,][]{grogin11} galaxies weighted by star-formation rate \citep[][their Fig.~12]{schulze18}. This confirms that SNe~IIn arise from massive stars and indicates that there is nothing peculiar with the SN~IIn hosts in the sample.

The luminosity-metallicity relation by \citet[][their Fig.~5]{tremonti04} shows that the spread in $M_g$ for our sample hosts in Fig.~\ref{fig:galmag} corresponds to a wide range of global galaxy metallicities, from above solar metallicity (up to approximately $\rm  12 + \log{(\frac{O}{H})} \approx 9$ on the scale of \citealp{kobulnicky04}), to $12 + \log{(\frac{O}{H})} \lesssim 8$ (i.e., subsolar) for the fainter end of our distribution ($M_g \gtrsim -16$\,mag). Earlier work on SN~IIn environments \citep{kelly12,habergham14,taddia15,anderson15,galbany18,kuncarayakti18} suggests that studies of global properties of host galaxies are insufficient when detailed conclusions are to be drawn about SNe~IIn and their progenitor channels\footnote{Spearman correlation tests (unweighted) for SN peak absolute magnitudes, rise times, and decline rates versus $M_g$ of the host galaxies (c.f. Sect.~\ref{sec:corr}) give significance levels of $\sigma \lesssim 1$, i.e., no correlations can be seen.}, but it is clear that an untargeted search like the one used to build our SN sample has the potential to probe a more representative part of the galaxy population.

\subsection{Interpretation using models\label{sec:model}}
As discussed in Sect.~\ref{sec:corr}, \citet{ofek14_10jl} claim the possible existence of a correlation between the rise time and the peak luminosity of SNe~IIn, assuming that the CSM close to the progenitor star is dense enough to allow shock breakout (SBO) to happen in the CSM. In Sect.~\ref{sec:corr}, we saw that such a correlation hardly exists.

As an alternative to studying the relations between peak absolute magnitude and rise time or decline rates (Sect.~\ref{sec:corr}), the relation between peak absolute magnitude and overall duration can also be considered. The location of the SNe in this duration-luminosity phase space (DLPS) can give insights into the mechanisms of the SNe. Using our spline fits (Sect.~\ref{sec:peak}) we can measure the duration of the light-curves as the time it takes for the light-curve to rise and decline 1\,mag relative to its peak \citep[as in][]{villar17}. This duration measurement is possible for 31 of our spline-fitted SN light-curves in Fig.~\ref{fig:lcfit_allIIn}.

\begin{figure}
   \centering
    \includegraphics[width=\linewidth]{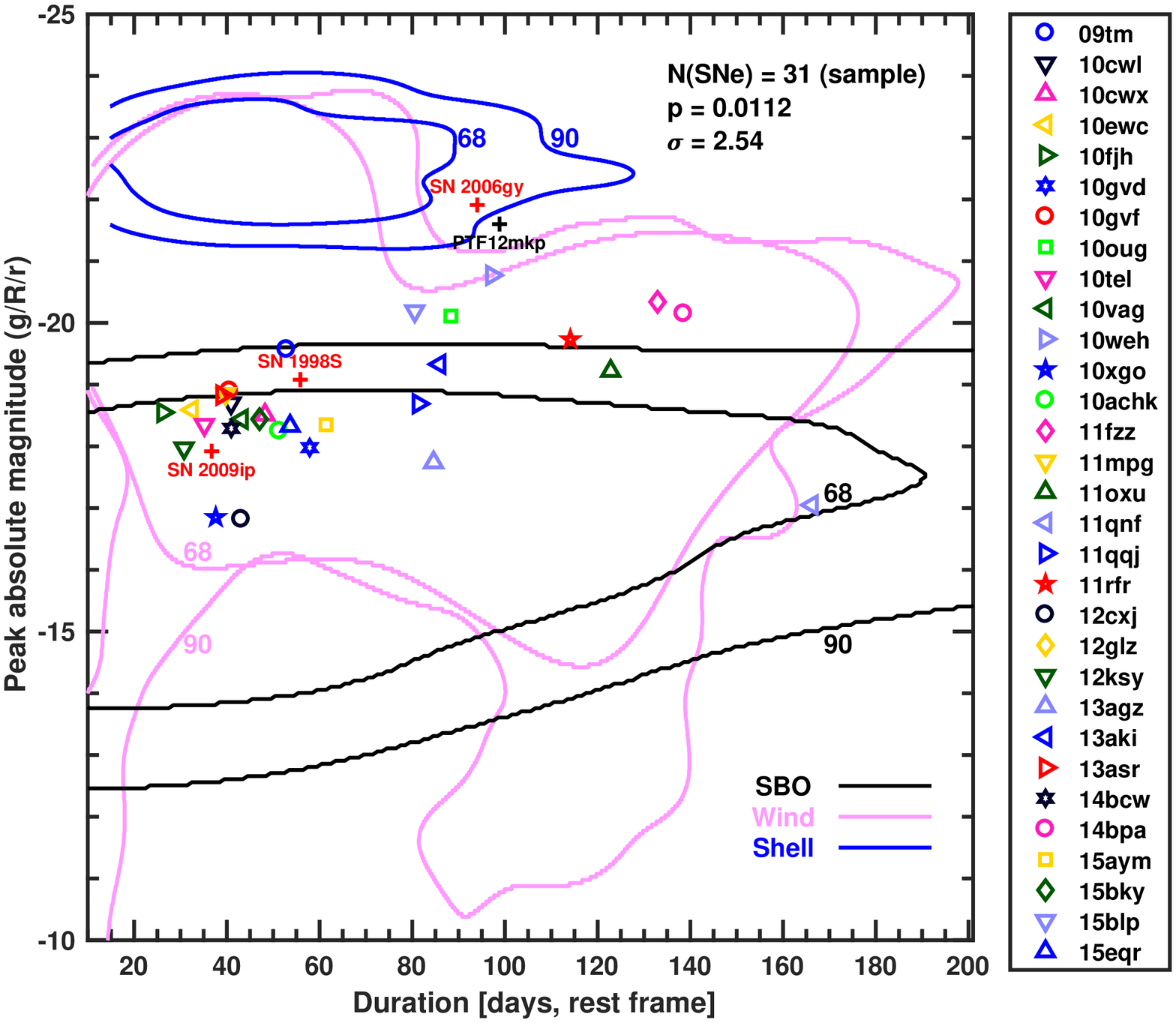}
    \caption{Percentile contours for the DLPS of CSI-driven SNe as explored by \citet{villar17} using the MOSFiT code \citep{guillochon18}. SN-like events are shown in pink (wind-like CSM) and blue (shell-like CSM) whereas shock-breakout (SBO) transients are shown in black. For each contour colour, the inner contour represents the 68$\rm ^{th}$ percentile and the outer contour the 90$\rm ^{th}$ percentile. From our sample, 31 SNe are plotted, along with the SLSN IIn PTF12mkp. For comparison, the much-studied SNe 1998S \citep{liu00}, 2006gy \citep{smith07phot06gy}, and 2009ip (2012B event, \citealp{graham14}) are included as well. For the $\rm N(SNe) = 31$ sample SNe, the Spearman correlation coefficient $p$ and the corresponding significance $\sigma$ are shown.\label{fig:mosfit1}}
\end{figure}

A comprehensive exploration of the DLPS for different astronomical transients was done by \citet{villar17}, who used the MOSFiT (Modular Open Source Fitter for Transients) code \citep{guillochon18} to explore the locations in the DLPS occupied by transients driven by different mechanisms (such as CSI) occurring under wide but realistic parameter ranges. Using the analytical models based on \citet{chatzopoulos12} for SN-like CSI transients (with both wind-like and shell-like CSM profiles) and the model by \citet{ofek14_10jl} for SBO interacting transients, \citet{villar17} calculated 68$\rm ^{th}$ and 90$\rm ^{th}$ percentile contours for the DLPS populated by CSI-driven transients. The CSM mass in all models was picked from the range 0.1--10$~M_{\odot}$, sampled log-uniformly. In Fig.~\ref{fig:mosfit1} we compare our sample SNe to the DLPS percentile contours by \citet{villar17}. The slight overlap of the contours in some places for the SN-like transients is due to the contours being generated based on only 1000 MC realisations (Ashley Villar, personal communication). We also plot SLSN IIn PTF12mkp (Sect.~\ref{sec:sample}). We also did spline measurements (as in Sect.~\ref{sec:peak}) for photometry of SNe 1998S \citep{liu00}, 2006gy \citep{smith07phot06gy}, and 2009ip (2012B event, \citealp{graham14}) to include them in Fig.~\ref{fig:mosfit1} for comparison. A unweighted Spearman test for the 31 sample SNe~IIn plotted in Fig.~\ref{fig:mosfit1} gives $p = 0.0112$, corresponding to a significance of $\sim 2.5\sigma$. A correlation like this (more luminous SNe~IIn generally evolving more slowly) could be expected, based on the investigation presented in Fig.~\ref{fig:corr}. It is noteworthy that in Sect.~\ref{sec:corr}, we found a correlation (at $2.7 \sigma$) between peak brightness and decline rate, but no correlation ($\sim 1\sigma$) between peak brightness and rise time. The overall slower evolution of more luminous SNe~IIn therefore seems to be dictated by the decline rates of the SNe.

In Fig.~\ref{fig:mosfit1} we see that 19 of the sample SNe~IIn reside within both the SBO 68$\rm ^{th}$ percentile and the wind-like 68$\rm ^{th}$ percentile, suggesting that either model could explain these SNe. At $M \lesssim -20$\,mag, we see 5 SNe outside the SBO 90$\rm ^{th}$ percentile but still inside the wind-like 68$\rm ^{th}$ percentile, making it less likely that they are SBO-driven SNe~IIn. Two of the SNe plotted (comparison events PTF12mkp and SN 2006gy) are found in a region of the DLPS where both wind-like and shell-like CSM could offer a satisfactory explanation of their light curves.

The DLPS plot (Fig.~\ref{fig:mosfit1}) shows a noticeable clustering of events with durations between 20\,d and 60\,d and peak magnitudes between $\sim -18$ and $\sim -19$, whereas the more luminous ($M < -20$\,mag) and long-lasting events are rarer. Observationally, more long-lasting and luminous events should be easier to find and monitor, also at higher redshifts. This might indicate that the less luminous and less long-lasting SNe~IIn are intrinsically more common. This can be interpreted such that progenitors with a relatively lower mass-loss rate towards the end of their existence dominate the zoo of SN~IIn progenitors. A rapidly declining SN~IIn likely has a less dense CSM, whereas a slowly declining SN~IIn should have an extended dense CSM. The differences in extent and density of the CSM suggest a difference in the mass-loss histories of SN~IIn progenitors, with the progenitors of more long-lasting and luminous SNe~IIn having enhanced mass loss for decades or longer \citep[c.f.][]{moriya14} before the SN explosions, while the fainter and briefer SNe~IIn possibly only have stronger mass loss during the years before the SN.

\section{Discussion\label{sec:discu}}
In this section, we discuss the contents of our sample. We examine what could be contaminating our sample in terms of unrelated astronomical transients, and we evaluate sources of errors. Further, we discuss some of our individual SNe and compare them to SNe in the literature with respect to SN impostors, light-curve bumps, and flat light-curve maxima.

\subsection{Possible contaminants in the sample\label{sec:contam}}
Transients showing SN~IIn spectral signatures are usually understood to involve a supergiant star having an outburst (destructive or not) inside a substantial CSM. There are other mechanisms that can mimic the SN~IIn signature and which could possibly contaminate our sample. Here, we discuss two such mechanisms.

Under some circumstances, an active galactic nucleus (AGN) can appear like a SN~IIn both spectroscopically and photometrically. For example, an early label put on SNe 1987F and 1988I (later classified as SNe~IIn) was "Seyfert 1-like" \citep{filippenko89}. The erratic photometric behaviour of an AGN (caused e.g. by changes in accretion rate of its supermassive black hole) can lead to variability episodes \citep[see e.g.][]{graham17} with duration and absolute magnitude comparable to some SNe~IIn. Most of the SNe in our sample are located off-centre in their host galaxies. Vetting The Million Quasars catalogue (version 6.3, 2019 June) compiled by \citet{flesch15} shows that no known AGN is found in the catalogue within 5$\arcsec$ of our SN positions. This suggests that the transients in our sample are likely not AGN in outburst mimicking SNe~IIn.

Some thermonuclear SNe have been observed showing spectra with SN~IIn-like Balmer lines superimposed on ordinary SN~Ia spectra \citep[e.g.][]{hamuyetal03}. These SNe~Ia-CSM have been proposed to happen when white dwarfs explode inside the CSM of its non-degenerate binary companion \citep[see][and references therein]{silverman13}. In simulations by \citet{leloudas15} it was shown that SNe~Ia-CSM can be mistaken for genuine SNe~IIn under some circumstances. \citet{leloudas15} shows that a SN~Ia (of typical peak magnitude $M \approx -19$) can be hidden under an apparent SN~IIn spectrum if the CSI is strong and the peak magnitude of the SN~IIn is $M \approx -20.9$. In this interval, our sample contains PTF12frn (Table~\ref{tab:peakmag_time}). We caution that this SN might have an overestimated host extinction (Sect.~\ref{sec:sample}). PTF12frn was a part of the feeder sample for the \citet{silverman13} study, but its spectra made them conclude that it was not a SN~Ia-CSM. It therefore seems likely that none of our sample SNe are SNe~Ia-CSM.

\subsection{Sources of errors\label{sec:errors}}
Some of the assumptions made when measuring the light-curve properties could affect our analysis and conclusions. Sources of error important for absolute magnitudes and colours are the extinction corrections, distances, and the bolometric correction for the SNe. The values we obtain for SN decline rates and rise times are not affected by these. 

No comprehensive study of bolometric corrections ($BC$) for SNe~IIn has yet been done. The Balmer emission lines in their spectra as well as the infrared contributions from dust \citep{fox13} makes their $BC$ hard to estimate. For blackbody temperatures between 7500 and 11,000\,K, the bolometric correction of our Mould $R$-band filter is $-0.6 < BC < -0.06$\,mag \citep[][footnote 22]{ofek14_10jl}. These $BC$ values for SNe~IIn are adopted by \citet{moriyamaeda14}. For practical reasons (and following \citealp{ofek14prec} and \citealp{nyholm17}), we have been using $BC = 0$\,mag in our analysis with the $R/r$ and $g$ bands, and this likely introduces an error.

Several samples of CC~SNe \citep[e.g.][]{valenti16,taddia15,taddia19} show host colour excess in the range $0 \lesssim E(B-V) \lesssim 0.4$\,mag. In the literature compilation for SN~Ia host galaxies by \citet{cikota16}, the absorption-to-reddening ratios are in the range $1 \lesssim R_V \lesssim 3.5$. Owing to the low sensitivity of our spectra to the \ion{Na}{I\,D} absorption diagnostic (Sect.~\ref{sec:sample}) it appears likely that we could underestimate the host extinction for some of the SNe in our sample, and the distribution of peak absolute magnitudes is systematically brighter than reported. As indicated by the small spread (a few tenths of a magnitude around light-curve peak epoch) in the colour evolution displayed in Fig.~\ref{fig:colours}, in practice it seems unlikely that we have large extinction correction errors.

A 5\,km\,s$^{-1}$\,Mpc$^{-1}$ uncertainty in $H_0$ corresponds to an uncertainty  in the distance moduli of 0.15\,mag. This is a systematic error, applying equally to all SNe in the sample, and thus introduces a random uncertainty in the derived luminosity function.

Concerning the selection of the sample, the $+40$\,d (after discovery) lower limit for having detections (Sect.~\ref{sec:sample}) affects SNe with fast declines and means that the sample is not inclusive of such events. During the course of this work, it became clear that the post-peak P60 detections initially used to make PTF09bcl a member of the sample were actually upper limits. Owing to the good coverage of the PTF09bcl rise and peak, we did keep PTF09bcl in the sample despite that.

\subsection{Possible SN impostor PTF11qnf\label{sec:impost}}
\begin{figure}
    \includegraphics[width=\linewidth]{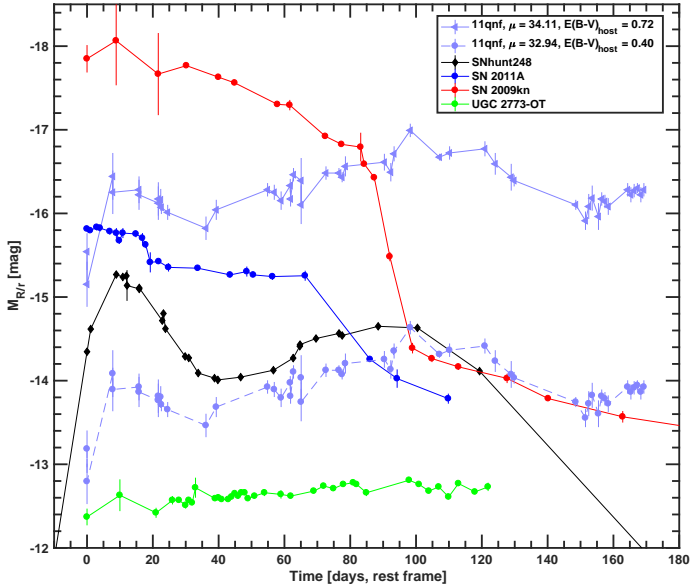}
    \caption{Comparison between SN impostor candidate PTF11qnf and events from the literature ($R/r$-band absolute magnitudes). For PTF11qnf, we show the light-curve assuming two different combinations of distance modulus and host extinction: $\mu = 34.11$\,mag given in Table~\ref{tab:IInsummary} and the $E(B-V)_{\rm host}$ from Sect.~\ref{sec:sample}, as well as assuming the Tully-Fisher based $\mu = 32.94$\,mag \citep{lagattuta13} and the $E(B-V)_{\rm host}$ estimated in Sect.~\ref{sec:impost}. Using the latter combination of $\mu$ and $E(B-V)_{\rm host}$ a similarity between PTF11qnf and SNhunt248 can be discerned. SNhunt248 \citep{kankare15}, SN~2011A \citep{dejaeger15}, SN~2009kn \citep{kankare12}, and U2773-OT \citep[2009 outburst;][]{smith10} are shown with distance moduli and extinction values are taken from the cited papers. Epoch 0\,d is set at the first detection in the light-curve (for PTF11qnf, SN~2009kn, and SN~2011A) or the observed start of the outburst episodes (for SNhunt248 and U2773-OT). Lines are drawn to guide the eye.\label{fig:11qnf}}
\end{figure}

The question of SN impostors \citep[e.g.][]{vandyk00, vandyk12} is important to address. Such transients are generally less luminous than SNe~IIn, but can still display the multi-component Balmer emission lines characteristic of SNe~IIn, thus mimicking some of the SN~IIn behaviour and earning them the label "impostors". No CC is assumed to happen in an SN impostor. The compilation by \citet{smith11_lbv} shows that SN impostors typically have peak magnitudes in the range $-16 < M < -10$\,mag and usually have erratic light-curve shapes. Outbursts of LBV stars is one common mechanism invoked to explain SN impostors \citep{smith11_lbv}. The faint end of the peak absolute magnitude distribution for our sample (Fig.~\ref{fig:peakmags}) lies at $M \approx -17$\,mag, and in this respect the majority of our sample SNe would not be regarded as SN impostors. The case of PTF11qnf, with a plateau-like and erratic light-curve peaking at $M \approx -17$\,mag, calls for special attention as a possible SN impostor candidate. We attempt to evaluate the nature of PTF11qnf by comparing it to other transients showing similar behaviour (Fig.~\ref{fig:11qnf}).

An interesting comparison can be made to the transient \object{SNHunt248}, which showed a fast rise to a double-peaked maximum lasting $\gtrsim 100$\,d at $M \approx -14$\,mag \citep{mauerhan15, kankare15}, interpreted by those authors as the nondestructive outburst of a hypergiant star. The deepest upper limits in the photometry from the P48 suggest that the absolute magnitude of PTF11qnf was $M_R \gtrsim -15$ between $-960$ to $-10$\,d and between $+260$ to $+1950$\,d, relative to the first detection (JD 2,455,866.9). This is not a deep limit, and still allows for pre-discovery variability such as that observed in SNHunt248 \citep{kankare15}, but demonstrates that the $\sim 180$\,d plateau episode we observe in PTF11qnf was preceded and followed by fainter periods. \object{SN 2011A} \citep{dejaeger15} had a SN~IIn spectrum, a maximum observed luminosity in the range between those of SNe and SN impostors, and a light-curve having two plateaus. \citet{dejaeger15} favoured a SN impostor scenario for this transient. The 2009 outburst of the transient \object{U2773-OT} \citep{smith10} exhibited a light-curve plateau of $M_R \approx -12$\,mag lasting $\sim 120$\,d, as well as H$\alpha$ emission with a narrow component. \citet{smith10} concluded that U2773-OT was a LBV in outburst, and later studies of its $\gtrsim 15$\,yr light-curve led \citet{smith16u2773} to compare U2773-OT to the LBV star $\eta$~Car. While there is a spread of $\sim 4$\,mag in maximum brightness, the duration of the light-curves of SNHunt248, SN~2011A, and U2773-OT are similar. For comparison, Fig.~\ref{fig:11qnf} also shows \object{SN 2009kn} \citep{kankare12}, a SN~IIn which had a plateau-like behaviour on a time scale similar to that of PTF11qnf, but was initially more luminous and most likely was a CC~SN \citep{kankare12}.

The single spectrum we have of PTF11qnf (Fig.~\ref{fig:class_spec}) was taken during its initial hump and shows a flat continuum with conspicuous H$\alpha$ and H$\beta$ emission. A Gaussian fit to its H$\alpha$ narrow component gives a full width at half-maximum of $\sim 400$\,km\,s$^{-1}$. This spectrum is comparable to some of the LBV outburst spectra compiled by \citet{smith11_lbv}, as well as to the early spectra of SN~2011A \citep{dejaeger15}. The \ion{Fe}{II} lines reported by \citet{kankare15} for SNhunt248 were not seen in PTF11qnf, but otherwise their spectra are comparable.

The redshift of PTF11qnf is the lowest in our sample and its host (\object{UGC 3344}) is located in a group of galaxies (LDC 410; \citealp{crook07}). This makes the distance estimate based on redshift particularly uncertain. Using the Tully-Fisher method, \citet{lagattuta13} obtained $\mu = 32.94$\,mag for the host of PTF11qnf. In Fig.~\ref{fig:colours}, $E(B-V)_{\rm host} = 0.72$\,mag was assumed based on \ion{Na}{I\,D} absorption, but this gives a conspicuous offset in $g-i$ towards the blue, compared to all other entries in this figure. The slope of the spectrum of PTF11qnf as well as the redder colours reported for SNHunt248 by \citet{kankare15} suggests that we might have overestimated the colour excess from the PTF11qnf host. Instead assuming $E(B-V)_{\rm host} = 0.4$\,mag would put PTF11qnf within the $g-i$ range of the other entries in Fig.~\ref{fig:colours}. Using the Tully-Fisher $\mu$ and $E(B-V)_{\rm host} = 0.4$\,mag, Fig.~\ref{fig:11qnf} shows that PTF11qnf might be more similar in luminosity to SNHunt248, having a late-time evolution similar to the second bump of SNHunt248.

The spectroscopic and photometric similarities of PTF11qnf to a number of transients interpreted as nondestructive outbursts of massive stars points towards this transient not being a SN, but rather an impostor. For this reason we removed PTF11qnf from the analysis in Sects.~\ref{sec:risetimes} and \ref{sec:corr}.

\subsection{Bumpy SN~IIn light-curves\label{sec:bumps}}
The light-curves of some SNe~IIn have breaks in their post-peak decline, when the SNe temporarily re-brightens before the decline in brightness resumes. Such episodes appear as a bump in the light-curve and have been seen in, for example, SNe~IIn 2006jd \citep{stritzinger12}, 2009ip \citep{margutti14,graham14,martin15}, and \object{iPTF13z} \citep{nyholm17}. The bumps have been proposed \citep[e.g.][]{graham14, nyholm17} to happen when SN ejecta encounter regions of increased density in the CSM.

Figure \ref{fig:LCs} indicates that some SNe in our sample have light-curve bumps. Photometry of the PTF10tel bump was presented by \citet{ofek13} while some of the other bumps have not been discussed before. The possible bump of PTF10weh ($\sim 0.2$\,mag break in the $r$-band decline, $\sim 116$\,d after peak) seems to occur also in the $Bgi$ bands, while the start of the possible bump of iPTF13aki ($\sim 0.4$\,mag rise in the $r$-band, 120\,d after peak) is possibly also captured in the $i$-band. Whereas the PTF10weh bump could be a true, (pseudo)bolometric bump, the nature of the possible iPTF13aki bump is harder to evaluate. The difficulty in judging the nature of light-curve bumps is illustrated by the light-curve of PTF11fzz.

For PTF11fzz, the quite conspicuous bump ($\sim 0.5$\,mag amplitude) seen around 197\,d after peak is likely a consequence of the H$\alpha$ line centre being shifted to 7097\,\AA\ at the redshift of PTF11fzz ($z = 0.081$). This puts the H$\alpha$ emission inside the Mould $R$ filter, but outside the SDSS $r$ filter. We lack spectra of PTF11fzz around this epoch, but a comparison to PTF12glz (at the similar $z = 0.079$) is instructive. A large offset ($R_{\rm Mould} - r_{\rm SDSS} \approx -0.8$\,mag) can be seen (Fig.~\ref{fig:LCs}) in our $R/r$ photometry of PTF12glz around 220\,d after its estimated peak epoch. We have a spectrum of PTF12glz taken about 235\,d after its estimated peak epoch \citep[the JD 2,456,422 Keck-I/LRIS spectrum from][]{soumagnac19}, which is comparable to the epoch where we see the apparent bump in PTF11fzz. Synthetic photometry on this PTF12glz spectrum using \texttt{synphot} by \citet{ofek14code} gives a $R_{\rm Mould} - r_{\rm SDSS} \approx -0.6$\,mag difference. This is comparable to the offset we see in the observed photometry. The large offsets ($R_{\rm Mould} - r_{\rm SDSS}$) seen in two different SNe~IIn both at $z \approx 0.08$, and confirmed via synthetic photometry for one of them (PTF12glz), makes it likely that the "bump" we see in PTF11fzz is a photometric filter effect.

The precursor activity as well as the bump of both PTF10tel ($\sim 40$\,d past peak) and SN~2009ip ($\sim 40$\,d past peak of the 2012B event) has been noted before \citep{ofek13, graham14, smith14_09ip}. Adding to this, in the data release by \cite{hicken17} we see that the Type~IIn \object{SN 2008aj} has a post-peak decline and a bump comparable to bumps seen in PTF10tel and SN~2009ip. \citet{pastorello18} point out the similarity between SN~2009ip and a number of events with similar rise and decline rates, as well as comparable peak brightness. Here, we add SN~2008aj to that comparison and emphasise the presence of light-curve bumps.

\begin{figure}
\begin{center}
\includegraphics[width=\linewidth]{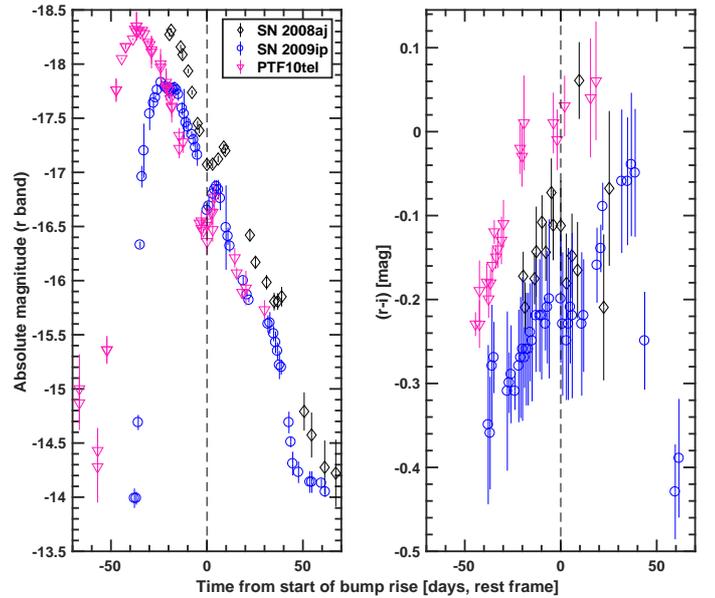}
\caption{Comparison of the $r$-band absolute magnitudes and $r-i$ colours for the bumps seen in Type~IIn SNe 2008aj, 2009ip (2012B event), and PTF10tel. Photometry and extinction corrections are taken from \citet{hicken17}, \citet{graham14}, and this paper, respectively. For clarity, colour data points with uncertainties $> 0.1$\,mag are not shown. The vertical, dashed black line indicates the estimated start epoch of the light curve bumps. \label{fig:bumpcomp}}
\end{center}
\end{figure}

For all three SNe, we plot the $r$-band light-curves and the $r-i$ colour in Fig.~\ref{fig:bumpcomp}. A common property of the three bumps is that their rises last $\sim 10$\,d and the height of the rise is $\sim 0.3$\,mag. We also note that the $r$-band decline rates pre-peak and post-peak are comparable. In $r-i$ colour, SNe 2008aj and 2009ip display a somewhat more similar behaviour pre-peak compared to PTF10tel, but all three SNe show a general change to redder colour before the bump start. This change slows down around the start of the bump, to resume afterwards. For comparison, the $r-i$ colour of the most long-lasting bump of iPTF13z (called B$_3$ by \citealp{nyholm17}) showed a reddening before the bump, but turned bluer as the bump ended. As in iPTF13z, the bumps shown in Fig.~\ref{fig:bumpcomp} all appear in multiple optical bands, suggesting that they are driven by the continuum of the SNe rather than by (for example) H$\alpha$ evolution \citep[as in the bump of SN~2006jd; see][]{stritzinger12}. For SN~2009ip, an LBV star has been proposed as the progenitor \citep[e.g.][]{mauerhan13_09ip,smith17lbv}. The remarkable similarity between the evolution of SN~2008aj, SN~2009ip and PTF10tel around main peak as shown in Fig.~\ref{fig:bumpcomp} might indicate similar progenitors (e.g. LBV stars) for all three events.

\subsection{The frequency of SN~IIn light-curve bumps\label{sec:freqbumps}}
Bumps in SN~IIn light-curves are apparently rare. Conspicuous light-curve bumps have been observed in $\lesssim 10$ SNe~IIn, that is in $\lesssim 2$\% of the $\approx 550$ public SNe~IIn listed in the Open Supernova Catalog (OSC; \citealp{guillochon17}) as of 2019 September. We used our observations of SNe~IIn from PTF/iPTF and MC simulations of detection probability to constrain the frequency of SN~IIn bumps.

The bumps are intrinsically dissimilar, in terms of amplitudes, durations, shapes, and time of occurrence in the light curve. For simplicity, we consider bump B$_1$ of iPTF13z \citep{nyholm17} to be our model bump. Fitting a second-degree polynomial to the $R$-band photometry of iPTF13z between 125 and 162 rest-frame days after discovery gives $m(t) = 0.0016 (t - t_0)^2 - 0.51 + \Delta m$\,mag as a simple model of bump B$_1$. The model is valid within 0.51\,mag of bump peak at $t_0$. We draw a magnitude offset $\Delta m$ uniformly between 1 and 3\,mag fainter than the SN peak and a time offset uniformly between 50 and 200\,d after SN peak for the bump to start. This random offset relative to the peak apparent magnitude of a SN in our sample introduces our model bump in that SN in a way resembling how the bump appeared in iPTF13z. We use the SN peak apparent magnitudes, but do not take any other light-curve property of the sample SNe into account.
 
To represent how SN~IIn photometry is obtained, for each SN in our sample we consider all epochs when the SN was observed (after discovery) with the P48 and P60 in the photometric bands listed in Table~\ref{tab:peakmag_time}. Such sets of epochs are representative for how PTF/iPTF operated, also capturing interruptions caused by for example cloudy nights or solar conjunction gaps. When the model bump has been introduced at a random place relative to the peak of a sample SN, the bump (stretched in time to the observer's frame) is sampled using the photometric epochs for that SN. The bump is considered to be detected if the following criteria are fulfilled: $\geq 3$ epochs fall on the bump, with a time of $\geq 10$\,d separating the first and the last such epoch, and the faintest of the epochs giving $m \leq 20.5$\,mag.

Introducing 100 such model bumps randomly (as described above) in each of the 39 SNe of our sample with measured peak epoch (Table~\ref{tab:peakmag_time}) gives an acceptance fraction of $\sim 8.5\%$, that is the bump being detected in $\sim 8.5\%$ of the cases. From our observations, we know that no bump like B$_1$ of iPTF13z was seen in these 39 SNe in the sample. A smaller bump in PTF10tel was indeed seen, but in this MC experiment we consider only our model bump from iPTF13z. Knowing that 1 bump (B$_1$ itself) was detected among the 69 remaining SNe~IIn from PTF/iPTF ($\sim 1.5\%$; see Table~\ref{tab:IInselection}) allows us to make a quantitative estimate of the fraction of SNe~IIn displaying light-curve bumps.

We model the observation as independent observations of $N_\mathrm{a} = 39$ SNe~IIn, calling this sample "a", with a probability $p_\mathrm{a} \cdot \alpha$ for a bump to be detected in one SN~IIn. Here, $\alpha$ is the probability of SNe to feature a bump, and $p_{\mathrm{a}} = 8.5\%$ is the probability for this bump to be detected. In addition to this sample, we have sample "b", with $N_\mathrm{b} = 69$ SNe~IIn with a bump probability $p_{\mathrm{b}} \cdot \alpha$. For this sample, we do not know the bump detection probability, and conservatively allow it to take any value between 0 and 1\footnote{Constraining $p_{\mathrm{b}} < p_{\mathrm{a}}$, which could be a reasonable assumption as this sample is sampled less, will move the lower edge of our confidence level up slightly.}. We model the observations as two independent samplings of SNe with a different probability to detect a bump, disregarding possible correlations due to selecting the samples. The total likelihood for the observation of $n_\mathrm{a} = 0$ and $n_\mathrm{b} = 1$ in the two samples may be written as the product of two binomial distributions, with $\alpha$ and $p_{\mathrm{b}}$ as parameters:

\begin{equation}
\label{eq:binom}
L(\alpha,p_\mathrm{b}) = Bin(n_\mathrm{a}|N_\mathrm{a},p_\mathrm{a}\cdot\alpha)\cdot Bin(n_\mathrm{b}|N_\mathrm{b},p_\mathrm{b}\cdot\alpha)  .
\end{equation}

The maximum likelihood estimate of $\alpha$ is $0.014$. Using the profile likelihood ratio and the assumption of an asymptotically distributed test statistic (Wilks' theorem), the $1\sigma$ confidence interval for $\alpha$ is $[0.0043,0.16]$. A closer examination of the likelihood surface showed that the upper constraint on $\alpha$ is dominated by the term due to sample "a", while the lower edge is set by the observation of one bump in sample "b". Improved knowledge of $p_\mathrm{b}$ may improve the lower constraint. With the current unconstrained probability, the profiled likelihood raises $p_\mathrm{b}$ to 1 for low $\alpha$.

From our 39 sample SNe~IIn considered here and our MC experiment, we therefore estimate that bumps like the B$_1$ bump of iPTF13z happen in approximately $1.4^{+14.6}_{-1.0} \%$ of all SNe~IIn. This interval, for our specific case of B$_1$, is also consistent with our rough initial estimate based on SN~IIn bumps in the literature and the number of SNe~IIn from the OSC.

Given the assumption that SN~IIn light-curve bumps are caused by SN ejecta encountering denser regions of CSM, the rarity of such bumps is noteworthy. Modelling \citep[e.g.][]{kochanek09} and observations \citep[][but see also \citealp{kochanek18}]{sana12,dunstall15} suggest that at least $50\%$ of massive stars reside in binary or multiple systems. In such binaries, the mass lost by the constituent stars can be shaped into spiral patterns \citep{tuthill08} which could cause light-curve bumps if one of the stars explodes as a CSI-dominated SN \citep[as invoked by][for SN~1979C]{schwarz96}. The LBV stars, often proposed as SN~IIn progenitors, are known to have structured CSM \citep[e.g.][]{weis11} which could also lead to bumpy light-curves if swept up by SN ejecta. Selection effects due to geometry and viewing angle might be at work, preventing light-curve bumps from occurring or being seen in the majority of SNe~IIn. Distinctly non-spherical geometries have been demonstrated using spectropolarimetry for SNe~IIn 1997eg \citep{hoffman08}, 1998S \citep{leonard00}, 2009ip \citep{reilly17}, and 2010jl \citep{patat11}. The relative rarity of light-curve bumps might be a separate indicator of non-spherical geometries in SNe~IIn, or that CSM density changes big enough to cause conspicuous bumps somehow are hard to make.

A simpler explanation might be the bias caused by the short ($\sim 100$\,d) time window after the SN explosion during which we can discover the bumps. For typical CSM and SN ejecta velocities, a SN progenitor outburst must occur shortly before the SN in order to produce a suitably located density change causing a bump we can actually observe.

\subsection{SNe~IIn with plateaus and slow declines\label{sec:flatmax}}
Plateaus have been seen in the light-curves of some SNe~IIn, earning them the designation Type~IIn-P (where "P" stands for plateau). Examples are SNe 1994W \citep{sollerman98}, 2009kn \citep{kankare12} and 2011ht \citep{fraser13, mauerhan13}, which all exhibited plateau-like maxima lasting $\sim 100$ days followed by a rapid decline ($\sim 0.25$\,mag\,d$^{-1}$) from the plateau. Different models have been proposed to explain the Type~IIn-P SNe, for example colliding CSM shells \citep[see][and references therein]{dessart16}. 

In our sample, we see a version of the SNe~IIn-P with a behaviour different from the examples above. Instead of dropping quickly in brightness after the plateau, these SNe instead showed a slower, linear decline after their plateaus. Among our sample SNe, we see that PTF11oxu and PTF11rfr have a rather fast rise to the plateau (rise time to plateau start $\lesssim 25$\,d) and a plateau lasting $\gtrsim 50$\,d, followed by a linear decline. The plateaus occur at $M_r \approx -19$\,mag and $M_r \approx -19.6$\,mag for PTF11oxu and PTF11rfr, respectively. The spectral sequence of PTF11rfr opens a possibility that the narrow component of its H$\alpha$ line used to classify it as a SN~Type~IIn might come from the host galaxy. The plateau magnitude of $M_r \approx -19.6$ mag for PTF11rfr is significantly more luminous than the average SN~Type~IIP peak magnitude of $M_B \approx -16.75\pm0.98$ mag \citep{richardson14} making CSI a plausible scenario for PTF11rfr.

\begin{figure}
\begin{center}
\includegraphics[width=\linewidth]{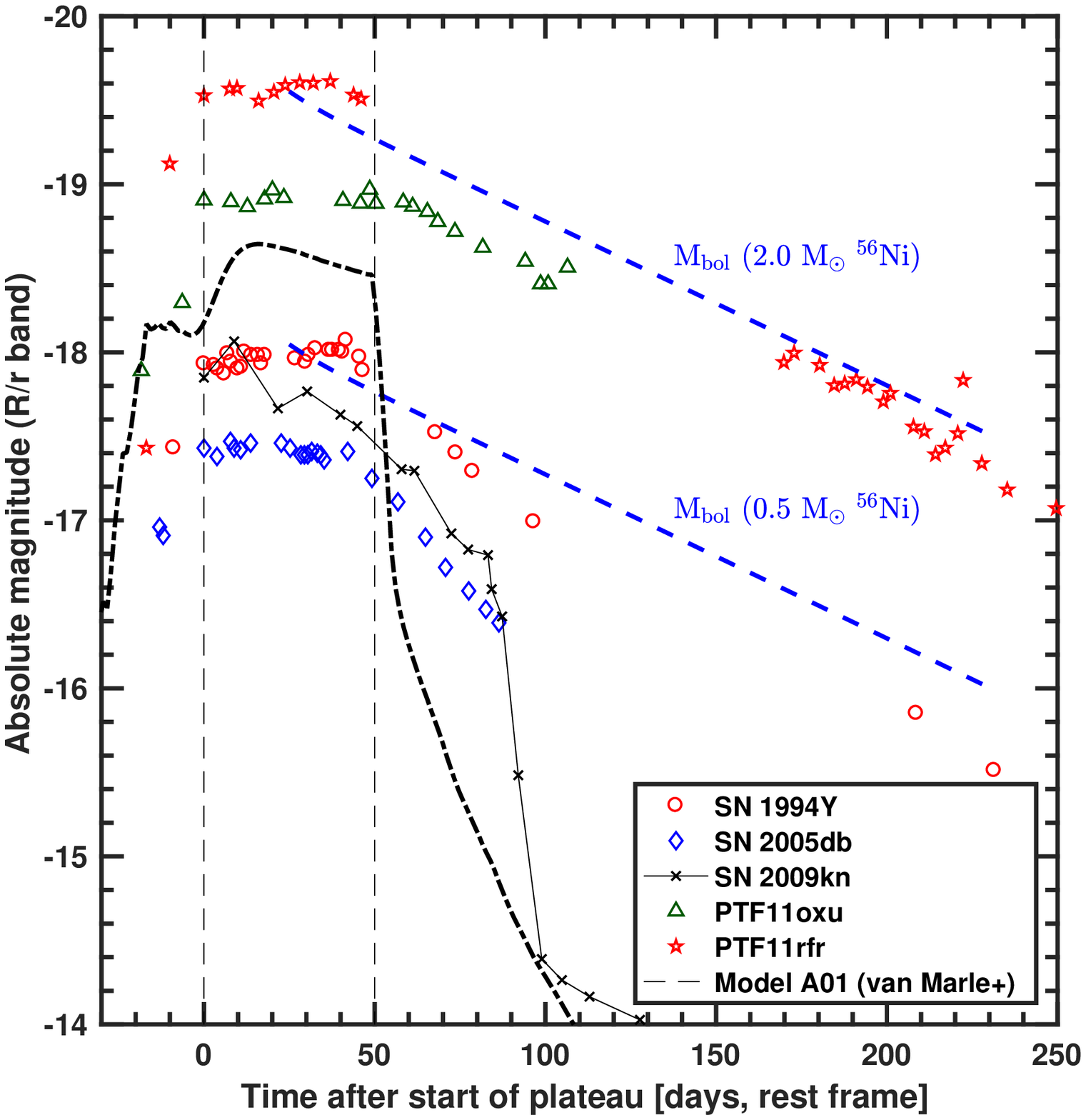}
\caption{Compilation of SNe~IIn with rapid rises, plateau-like maxima, and linear declines. Shown here are the sample SNe PTF11oxu and PTF11rfr (photometry binned in 3 day bins for clarity) along with SNe 1994Y \citep{ho01} and 2005db \citep{kiewe12}. For comparison, the Type~IIn-P SN~2009kn \citep{kankare12} is shown. Dashed vertical lines highlight 50\,d plateau duration. The bolometric decay rates \citep{nadyozhin94} for 0.5 and 2.0 M$_{\odot}$ of synthesised $\rm ^{56}Ni$ are shown as dashed blue lines (assuming a rise time to plateau start of 20\,d) for comparison. They are not fits to the photometry. Model A01 (bolometric luminosity) by \citet{vanmarle10} is also shown.\label{fig:flatmax}}
\end{center}
\end{figure}

Vetting the literature, we find that photometry of SNe 1994Y \citep{ho01} and 2005db \citep{kiewe12} show comparable, but intrinsically fainter, $R/r$-band light-curves. These light-curves are plotted in Fig.~\ref{fig:flatmax}. The $I$-band light-curve of Type~IIn SN \object{OGLE 2013-SN-016} \citep[][raw photometry by the Optical Gravitational Lensing Experiment, OGLE]{wyrzykowski13, nicholl13} appears to have had a similar plateau duration and decline phase. Figure~\ref{fig:flatmax} shows that the spread in plateau absolute magnitude is $\gtrsim 2$ mag and that the decline rates differ after the end of the plateau. The models by \citet{vanmarle10} of SNe~IIn, where the SN ejecta collide with a CSM shell, can produce light-curves with plateaus of approximately the right duration and luminosity; see \citet[their Figs. 11 and 12]{vanmarle10}. As an example, in Fig.~\ref{fig:flatmax} we plot model A01 by \citet{vanmarle10}, where SN ejecta hit a 1~M$_{\odot}$ CSM shell. The steep rise and the duration of the model A01 plateau, as well as the plateau luminosity of this model, is comparable to the SNe shown. However, model A01 (and the other \citealp{vanmarle10} models) gives a too steep decline after the plateau compared to the SNe of our proposed sub-type. The slow decline of the SNe~IIn here could be due to an extended CSM outside the shell. While Type~IIn-P (like \object{SN 1994W}, with fast decline) have been studied by several authors, the SNe~IIn with plateaus and slow declines warrants further investigation to determine if they constitute a distinct sub-type of SNe~IIn.

\section{Conclusions\label{sec:concl}}
We have studied a sample of 42 SNe~IIn from the untargeted PTF/iPTF survey. Our main conclusions are summarised below.

The luminosity function of SNe~IIn, after compensating for Malmquist bias and K-corrections, is $M_{\rm peak} = -18.72\pm1.32$\,mag (if applying the $M > -21$\,mag cut to exclude ostensibly "superluminous" SNe) or $M_{\rm peak} = -19.18\pm1.32$\,mag (if that cut is not applied, and all SNe with Type~IIn spectral signatures are included). The typical rise times to peak for SNe~IIn are $20\pm6$\,d and $50\pm11$\,d, dividing the SNe~IIn into fast and slow risers. The typical decline rates of SNe~IIn for the initial 50\,d after light-curve peak can be described with the Gaussian distribution $0.027\pm0.016$\,mag\,d$^{-1}$. A bimodal distribution possibly exists, with slower and faster declines (with slopes $0.013\pm0.006$\,mag\,d$^{-1}$ and $0.040\pm0.010$\,mag\,d$^{-1}$).

We find no strong correlation between the peak luminosity of SNe~IIn and their rise times. A correlation between the peak luminosity and the duration within 1\,mag of peak was found. Correlations between rise time and decline rate, and peak absolute magnitude and decline rate, can also be discerned.

The colour evolution of SNe~IIn is rather uniform. The colour indices at peak of the sample SNe~IIn are $g-r = 0.06\pm0.21$\,mag and $g-i = 0.13\pm0.46$\,mag. K-corrections for SNe~IIn at light-curve peak are found to be within 0.2\,mag for the observer's frame $r$-band, for SNe~IIn at $z < 0.25$.

The SNe in the sample were hosted by galaxies of absolute magnitude $-22 \lesssim M_g \lesssim -13$\,mag, that is with global metallicities ranging from supersolar to subsolar. No correlations between host global metallicity and SN~IIn rise times, decline rates, or peak absolute magnitude were found.

We suggest a possible new variety of "SNe~IIn-P", with a light-curve showing a quick rise to a $\gtrsim 50$\,d plateau, followed by a slow, linear decline. Example events include PTF11oxu and PTF11rfr, as well as SNe 1994Y and 2005db. They are thus unlike the prototypical SNe~IIn-P events such as SNe 1994W and 2009kn, which dropped much faster from their plateaus. Light-curve bumps like the B$_1$ bump of iPTF13z are rare, and happen in an estimated $1.4^{+14.6}_{-1.0} \%$ of the SNe~IIn.

\begin{acknowledgements}
A.N. gratefully acknowledges a grant from Stiftelsen Gustaf och Ellen Kobbs stipendiefond, and wishes to thank Ashley Villar for providing MOSFiT contours in table format for Fig.~\ref{fig:mosfit1}, Axel Runnholm for help with Python, Ofer Yaron for help with WISeREP, Timothy E. Holy for making the \texttt{distinguishable\_colors} function, Maayane Soumagnac and Peter Lundqvist for discussions, and opponent Philip James for questions and comments at the PhD thesis defence in Stockholm 2019 September 23.

We gratefully acknowledge support from the Knut and Alice Wallenberg Foundation. The Oskar Klein Centre is funded by the Swedish Research Council. The intermediate Palomar Transient Factory project is a scientific collaboration among the California Institute of Technology, Los Alamos National Laboratory, the University of Wisconsin (Milwaukee), the Oskar Klein Centre, the Weizmann Institute of Science, the TANGO Program of the University System of Taiwan, and the Kavli Institute for the Physics and Mathematics of the Universe. This work was supported by the GROWTH project funded by the National Science Foundation under Grant No 1545949. A.V.F. is grateful for financial assistance from National Science Foundation (NSF) grant AST-1211916, the TABASGO Foundation, the Christopher R. Redlich Fund, and the Miller Institute for Basic Research in Science (U.C. Berkeley). E.O.O. is grateful for support by grants from the Israeli Ministry of Science, ISF, Minerva, BSF, BSF transformative program, Weizmann-UK, and the I-CORE Program of the Planning and Budgeting Committee and the Israel Science Foundation (grant No. 1829/12). G.L. was supported by research grant 19054 from VILLUM FONDEN. Part of this research was carried out at the Jet Propulsion Laboratory, California Institute of Technology, under a contract with the National Aeronautics and Space Administration (NASA). SED Machine is based upon work supported by the National Science Foundation under Grant No. 1106171. Based in part on observations obtained at the Gemini Observatory, under programs: GN-2010B-Q13 and GN-2010A-Q20 (PI: Howell), which is operated by the Association of Universities for Research in Astronomy, Inc., under a cooperative agreement with the NSF on behalf of the Gemini partnership: the NSF (United States), National Research Council (Canada), CONICYT (Chile), Ministerio de Ciencia, Tecnolog\'{i}a e Innovaci\'{o}n Productiva (Argentina), Minist\'{e}rio da Ci\^{e}ncia, Tecnologia e Inova\c{c}\~{a}o (Brazil), and Korea Astronomy and Space Science Institute (Republic of Korea). Partly based on observations obtained with the Apache Point Observatory 3.5-m telescope, which is owned and operated by the Astrophysical Research Consortium. Some of the data presented herein were obtained at the W. M. Keck Observatory, which is operated as a scientific partnership among the California Institute of Technology, the University of California, and NASA. The Observatory was made possible by the generous financial support of the W. M. Keck Foundation. The authors wish to recognise and acknowledge the very significant cultural role and reverence that the summit of Maunakea has always had within the indigenous Hawaiian community. We are most fortunate to have the opportunity to conduct observations from this mountain. Partly based on observations at Kitt Peak National Observatory, National Optical Astronomy Observatory, which is operated by the Association of Universities for Research in Astronomy (AURA) under cooperative agreement with the NSF. The authors are honoured to be permitted to conduct astronomical research on Iolkam Du'ag (Kitt Peak), a mountain with particular significance to the Tohono O'odham. Partly based on observations made with the Nordic Optical Telescope, operated by the Nordic Optical Telescope Scientific Association at the Observatorio del Roque de los Muchachos, La Palma, Spain, of the Instituto de Astrofisica de Canarias. Partly based on observations made with the University of Hawaii’s 2.2-m telescope, at Maunakea Observatory, Hawaii, USA. Partly based on observations made with the William Herschel Telescope operated on the island of La Palma by the Isaac Newton Group of Telescopes in the Spanish Observatorio del Roque de los Muchachos of the Instituto de Astrof\'{i}sica de Canarias. Partly based on observations made with the 5-m Hale Telescope (P200), at Palomar Observatory, California, USA. Partly based on observations made with the Kast spectrograph on the Shane 3-m telescope at Lick Observatory, Mount Hamilton, California, USA. Research at Lick Observatory is partially supported by a generous gift from Google. We thank S. Bradley Cenko, Kelsey I. Clubb, Melissa L. Graham, Michael T. Kandrashoff, Patrick L. Kelly, Alekzandir Morton, Peter E. Nugent, and Jeffrey M. Silverman for help with some of the observations and reductions. This research used the Latest Supernovae web page (maintained by D. Bishop) and the Open Supernova Catalog (maintained by J. Guillochon and J. Parrent). This research used the SIMBAD and VizieR databases operated at CDS, Strasbourg, France, as well as the NASA/IPAC Extragalactic Database (NED) which is operated by the Jet Propulsion Laboratory, California Institute of Technology, under contract with NASA, and NASA’s Astrophysics Data System. {\sc STSDAS} and {\sc PyRAF} are products of the Space Telescope Science Institute, which is operated by AURA for NASA. We acknowledge the whole PTF and iPTF collaborations for finding and following the SNe~IIn studied in this sample.
\end{acknowledgements}

\onecolumn
\clearpage

\begin{deluxetable}{lllllll}
\center
\tabletypesize{\scriptsize}
\tablewidth{0pt}
\tablecolumns{7}
\tablecaption{Sample studies of SNe~IIn based on optical observations.\label{tab:IInsamples}}
\tablehead{
\colhead{Reference} &
\colhead{Sample} &
\colhead{Lowest $z$} &
\colhead{Median $z$} &
\colhead{Maximum $z$} &
\colhead{Mode of sample} &
\colhead{Main area}\\
\colhead{} &
\colhead{size} &
\colhead{in sample} &
\colhead{in sample} &
\colhead{in sample} &
\colhead{source survey(s)} &
\colhead{of study}}
\startdata
\citet{li02} & 7 & 0.0022 & 0.0047 & 0.018 & Targeted & Late-time photometry \\ 
\citet{richardson02} & 9 & 0.0028 & 0.0082 & 0.043 & Targeted & Luminosity function \\ 
\citet{li11} & 7 & 0.0024 & 0.0088 & 0.012 & Targeted & Luminosity function \\ 
\citet{kiewe12} & 4 & 0.015 & 0.024 & 0.031 & Targeted & Light-curves and spectra \\ 
\citet{taddia13} & 7 & 0.0071 & 0.017 & 0.029 & Mainly targeted & Light-curves and spectra \\ 
\citet{ofek14rise} & 19 & 0.031 & 0.073 & 0.28 & Untargeted & Rise times \\ 
\citet{ofek14prec} & 16 & 0.004 & 0.031 & 0.14 & Mainly untargeted & Precursor outbursts \\ 
\citet{richardson14} & 31 & 0.0021 & 0.016 & 0.35 & Mainly targeted & Luminosity function \\ 
\citet{bilinski15} & 6 & 0.0061 & 0.0095 & 0.011 & Targeted & Precursor outbursts \\ 
\citet{delarosa16} & 8 & 0.004 & 0.01 & 0.017 & Mainly targeted & Light-curves \\ \hline
This work & 42 & 0.014 & 0.078 & 0.313 & Untargeted & Light-curves
\enddata
\tablecomments{The works included in the table all discuss photometric properties of SNe~IIn. For the sample size counts, the SNe labelled as SNe~IIn by the respective authors were used. If available, the redshifts ($z$) of the SNe in each sample were taken from the respective papers. Otherwise, distance moduli or luminosity distances given in the papers were converted to redshifts using cosmological parameters from Sect.~\ref{sec:sample} without infall correction. \citet{li02} gives no distance information for their sample SNe; here, NED redshifts for their respective SN host galaxies were used.}
\end{deluxetable}

\begin{deluxetable}{lccccccccc}
\tabletypesize{\scriptsize}
\tablewidth{0pt}
\tablecaption{Photometry of the 42 SNe~IIn in the sample.\label{tab:phot}}
\tablehead{
\colhead{SN} &
\colhead{Epoch} &
\colhead{Magnitude} &
\colhead{Magnitude} &
\colhead{Limiting} &
\colhead{Telescope} &
\colhead{Camera} &
\colhead{Filter} &
\colhead{Reference} &
\colhead{Pipeline}\\
\colhead{} &
\colhead{(JD)} &
\colhead{(AB)} &
\colhead{error} &
\colhead{magnitude} &
\colhead{} &
\colhead{} &
\colhead{} &
\colhead{image} &
\colhead{}}
\startdata
PTF09tm   & $2454959.6605$ & $99.00$ & $99.0$ & $20.02$ & P48 & CFH12K & $R$ & P48ref & PTFIDE \\
PTF09tm   & $2454959.7523$ & $99.00$ & $99.0$ & $20.26$ & P48 & CFH12K & $R$ & P48ref & PTFIDE \\
... & ... & ... & ... & ... & ... & ... & ... & ... & ... \\
PTF09tm   & $2455005.8071$ & $18.87$ & $0.09$ & $99.00$ & P48 & CFH12K & $R$ & P48ref & PTFIDE \\
PTF09tm   & $2455005.8816$ & $18.83$ & $0.09$ & $99.00$ & P48 & CFH12K & $R$ & P48ref & PTFIDE \\
... & ... & ... & ... & ... & ... & ... & ... & ... & ... \\
PTF10weh  & $2455724.9677$ & $21.46$ & $0.13$ & $99.00$ & P60 & GRBCam & $r$ & SDSSref & FPipe \\
PTF10weh  & $2455481.6718$ & $18.69$ & $0.01$ & $99.00$ & P60 & GRBCam & $i$ & SDSSref & FPipe \\
... & ... & ... & ... & ... & ... & ... & ... & ... & ... \\
PTF11qnf  & $2456037.7320$ & $19.99$ & $0.09$ & $99.00$ & P60 & GRBCam & $r$ & PS1ref & FPipe \\
... & ... & ... & ... & ... & ... & ... & ... & ... & ... \\
iPTF15aym & $2457480.9090$ & $21.31$ & $0.06$ & $99.00$ & P60 & SEDM & $g$ & SDSSref & FPipe \\
... & ... & ... & ... & ... & ... & ... & ... & ... & ...
\enddata
\tablecomments{All magnitudes reported are AB magnitudes from PSF photometry on reference subtracted images. The columns give the name of the SN, the Julian date (JD) of the photometry, the magnitude and its 1$\sigma$ error (if SN detected, otherwise given as "99") or the upper limit (magnitudes, if SN not detected, otherwise given as "99"), the telescope used (P48 or P60), the camera used (CFH12K, GRBCam or SEDM), the filter used (Mould $R$ or SDSS $g$ for P48, SDSS $griz$ or Johnson $B$ for P60), the source of the reference image (P48, P60, SDSS or PS1 i.e. Pan-STARRS) and the pipeline used for the photometry (PTFIDE or FPipe). FPipe uses SDSS stars as reference stars, or PS1 stars when outside the SDSS footprint. Details on instruments etc. are given in Sect.~\ref{sec:obs}. Photometric upper limits from PTFIDE are given at 5$\sigma$ level. Upper limits from FPipe are given at 3$\sigma$ for GRBCam and at 5$\sigma$ for SEDM, respectively. Selected parts of the table are shown here, as a guide to its content. The full table is available at the CDS.}
\end{deluxetable}

\begin{deluxetable}{lcclcl}
\centering
\tabletypesize{\scriptsize}
\tablewidth{0pt}
\tablecaption{Log of the classification spectra for the 42 SNe~IIn in our sample.\label{tab:speclog}}
\tablehead{
\colhead{SN} &
\colhead{Epoch} &
\colhead{SN age, relative to}&
\colhead{Telescope} &
\colhead{Instrument}\\
\colhead{} &
\colhead{(MJD)} &
\colhead{peak (days, rest frame)}&
\colhead{} &
\colhead{}}
\startdata
PTF09tm & 55037 & 8 & Lick & Kast \\
PTF09uy & 55008 & -23 & Keck I & LRIS \\
PTF09bcl & 55055 & 9 & WHT & ISIS \\
PTF10cwl & 55270 & 4 & Keck I & LRIS \\
PTF10cwx & 55273 & 1 & P200 & DBSP \\
PTF10ewc & 55285 & -14 & WHT & ISIS \\
PTF10fjh & 55300 & -8 & Gemini North & GMOS \\
PTF10flx & 55299 & -6 & Gemini North & GMOS \\
PTF10gvd & 55322 & 9 & Keck I & LRIS \\
PTF10gvf & 55323 & -13 & Keck I & LRIS \\
PTF10oug & 55396 & -30 & P200 & DBSP \\
PTF10qwu & 55421 & -- & P200 & DBSP \\
PTF10tel & 55434 & -11 & Gemini North & GMOS \\
PTF10tyd & 55443 & -23 & P200 & DBSP \\
PTF10vag & 55454 & -10 & Lick & Kast \\
PTF10weh & 55502 & -2 & Lick & Kast \\
PTF10xgo & 55477 & -2 & P200 & DBSP \\
PTF10abui & 55536 & -10 & P200 & DBSP \\
PTF10achk & 55547 & -5 & UH & SNIFS \\
PTF10acsq & 55574 & -- & P200 & DBSP \\
PTF11fzz & 55736 & -51 & P200 & DBSP \\
PTF11mpg & 55833 & 6 & Keck II & DEIMOS \\
PTF11oxu & 55872 & -37 & WHT& ISIS \\
PTF11qnf & 55892 & -76 & Keck I & LRIS \\
PTF11qqj & 55895 & -7 & P200 & DBSP \\
PTF11rfr & 55916 & -30 & P200 & DBSP \\
PTF11rlv & 55953 & 20 & Kitt Peak & RC Spec \\
PTF12cxj & 56035 & -14 & Gemini North & GMOS \\
PTF12frn & 56124 & 4 & Keck II & DEIMOS \\
PTF12glz & 56123 & -58 & P200 & DBSP \\
PTF12ksy & 56238 & -14 & P200 & DBSP \\
iPTF13agz & 56442 & 26 & Apache Point & DIS \\
iPTF13aki & 56395 & -7 & P200 & DBSP \\
iPTF13asr & 56421 & 2 & Keck I & LRIS \\
iPTF13cuf & 56545 & -- & Keck II & DEIMOS \\
iPTF14bcw & 56803 & -16 & Keck II & DEIMOS \\
iPTF14bpa & 56832 & -28 & P200 & DBSP \\
iPTF15aym & 57176 & -13 & NOT & ALFOSC \\
iPTF15bky & 57196 & -8 & NOT & ALFOSC \\
iPTF15blp & 57226 & 17 & P200 & DBSP \\
iPTF15eqr & 57362 & -10 & Keck I & LRIS \\
iPTF16fb & 57445 & 4 & P200 & DBSP
\enddata
\tablecomments{Epochs are given in MJD ($= JD - 2,400,000.5$). The spectra were taken at epochs $-11 \pm 20$\,d ($1\sigma$ Gaussian spread) relative to light-curve peak (Table~\ref{tab:peakmag_time}). For our spectroscopy, we used the following telescopes (with spectrographs given in parenthesis): Apache Point 3.5 m (DIS, Dual Imaging Spectrograph), Gemini North 8.1 m (GMOS, Gemini Multi-Object Spectrograph), Keck I 10 m (LRIS, Low-Resolution Imaging Spectrometer), Keck II 10 m (DEIMOS, DEep Imaging Multi-Object Spectrograph), Kitt Peak 4 m (Ritchey-Chretien Spectrograph), Lick 3 m Shane telescope (Kast Double Spectrograph), Nordic Optical Telescope (NOT) 2.5 m (ALFOSC, Alhambra Faint Object Spectrograph and Camera), P200 5.1 m Hale telescope (DBSP, Double Spectrograph), University of Hawaii (UH) 2.2 m (SNIFS, Supernova Integral Field Spectrograph) \& William Herschel Telescope (WHT) 4.2 m (ISIS, Intermediate-dispersion Spectrograph and Imaging System). For SNe with age \textit{"--"} when the spectrum was taken, no peak epoch was determined.}
\end{deluxetable}

\begin{deluxetable}{lll}
\center
\tabletypesize{\scriptsize}
\tablewidth{0pt}
\tablecolumns{5}
\tablecaption{Selection of the SN~IIn sample.\label{tab:IInselection}}
\tablehead{
\colhead{Selection of SNe} &
\colhead{Number of SNe}}
\startdata
All SNe found and classified by PTF/iPTF 2009--2017& 3018  \\
Spectroscopically classified SNe~II & 692 \\
Spectroscopically classified SNe~IIn & 111    \\ 
SNe~IIn having upper limits less than 40\,d before discovery & 55   \\ 
SNe~IIn also having detections past 40\,d after discovery & 42
\enddata
\tablecomments{The selection of SNe~II(n) in each row is a subset of the SNe in the row above. The 692 "SNe~II" include SNe~IIP/L/b/n. The 40\,d mentioned is in the observer's frame. The 15 SLSNe~II found by PTF/iPTF are not included in the SNe~II row and below.}
\end{deluxetable}

\onecolumn

\begin{deluxetable}{lcccccccccl}
\rotate
\tabletypesize{\scriptsize}
\tablewidth{0pt}
\tablecaption{Locations and discovery circumstances of the 42 SNe~IIn in the sample.\label{tab:IInsummary}}
\tablehead{
\colhead{SN} &
\colhead{$\alpha$ (J2000.0)} &
\colhead{$\delta$ (J2000.0)} &
\colhead{Redshift} &
\colhead{$\mu$} &
\colhead{Discovery date} &
\colhead{$E(B-V)_{\rm MW}$} &
\colhead{$M^{host}_{g}$} &
\colhead{Alternative} &
\colhead{Discovery} &
\colhead{Inclusion in}\\
\colhead{} &
\colhead{(h:m:s)} &
\colhead{(\degr:\arcmin:\arcsec)} &
\colhead{(heliocentric)} &
\colhead{(mag)} &
\colhead{(MJD)} &
\colhead{(mag)} &
\colhead{(mag)} &
\colhead{designation(s)} &
\colhead{announcement(s)} &
\colhead{other sample(s)}}
\startdata
\object{PTF09tm} & $13$:$46$:$55.94$ & $+61$:$33$:$15.6$ & $0.0349$ & $36.01$ & $55005.31$ & $0.016$ & $-20.1$ & -- & -- & \citetalias{arcavi10}, \citet{stoll13} \\
\object{PTF09uy} & $12$:$43$:$55.80$ & $+74$:$41$:$08.1$ & $0.3135$ & $41.08$ & $55005.33$ & $0.020$ & $-18.1$ & -- & -- & \citetalias{ackermann15} \\
\object{PTF09bcl} & $18$:$06$:$26.78$ & $+17$:$51$:$43.0$ & $0.0621$ & $37.27$ & $55034.20$ & $0.081$ & $-21.63$ & -- & -- & \citetalias{arcavi10,ackermann15}, \citet{stoll13} \\
\object{PTF10cwl} & $12$:$36$:$22.06$ & $+07$:$47$:$38.0$ & $0.0849$ & $37.99$ & $55268.23$ & $0.019$ & $-17.74$ & CSS100320:123622+074737 & \citet{drake10_PTF10cwl} & \citetalias{ofek14rise,ackermann15} \\
\object{PTF10cwx} & $12$:$33$:$16.53$ & $-00$:$03$:$10.6$ & $0.0731$ & $37.66$ & $55268.24$ & $0.022$ & $-18.71$ & -- & -- & \citetalias{arcavi10,ofek14rise,ackermann15}, \citet{stoll13} \\
\object{PTF10ewc} & $14$:$01$:$59.08$ & $+33$:$50$:$11.6$ & $0.0542$ & $36.99$ & $55284.33$ & $0.013$ & $-17.76$ & -- & -- & \citetalias{ackermann15} \\
\object{PTF10fjh} & $16$:$46$:$55.36$ & $+34$:$09$:$34.7$ & $0.0321$ & $35.86$ & $55296.26$ & $0.021$ & $-21.65$ & SN~2010bq & \citet{duszanowicz10_PTF10fjh, challis10_PTF10fjh} & \citetalias{ofek14prec,ackermann15} \\
\object{PTF10flx} & $16$:$46$:$58.28$ & $+64$:$26$:$48.5$ & $0.0674$ & $37.44$ & $55297.43$ & $0.026$ & $-19.86$ & -- & -- & \citetalias{ackermann15} \\
\object{PTF10gvd} & $16$:$53$:$02.12$ & $+67$:$00$:$08.9$ & $0.0693$ & $37.50$ & $55322.27$ & $0.036$ & $-16.14$ & -- & -- & \citetalias{ackermann15} \\
\object{PTF10gvf} & $11$:$13$:$45.24$ & $+53$:$37$:$44.9$ & $0.0809$ & $37.85$ & $55322.22$ & $0.010$ & $-19.77$ & -- & -- & \citetalias{ofek14rise,ofek14prec,ackermann15} \\
\object{PTF10oug} & $17$:$20$:$44.79$ & $+29$:$04$:$25.6$ & $0.1501$ & $39.29$ & $55391.28$ & $0.037$ & $-18.61$ & -- & -- & \citetalias{ofek14rise,ackermann15} \\
\object{PTF10qwu} & $16$:$51$:$10.36$ & $+28$:$18$:$06.2$ & $0.2258$ & $40.27$ & $55417.21$ & $0.040$ & $-16.03$ & -- & -- & \citetalias{ackermann15} \\
\object{PTF10tel} & $17$:$21$:$30.68$ & $+48$:$07$:$47.4$ & $0.0349$ & $36.01$ & $55433.18$ & $0.015$ & $-16.83$ & SN~2010mc & \citet{ofek12_PTF10tel, howell12_PTF10tel} & \citetalias{ofek14rise,ofek14prec,ackermann15} \\
\object{PTF10tyd} & $17$:$09$:$19.41$ & $+27$:$49$:$08.6$ & $0.0633$ & $37.32$ & $55440.15$ & $0.058$ & $-20.51$ & -- & -- & \citetalias{ofek14rise,ackermann15} \\
\object{PTF10vag} & $21$:$47$:$18.48$ & $+18$:$07$:$51.5$ & $0.0516$ & $36.81$ & $55451.16$ & $0.106$ & $-15.75$ & -- & -- & \citetalias{ofek14rise,ackermann15} \\
\object{PTF10weh} & $17$:$26$:$50.46$ & $+58$:$51$:$07.4$ & $0.1379$ & $39.08$ & $55461.18$ & $0.028$ & $-12.82$ & -- & \citet{benami10_PTF10weh} & \citetalias{ofek14rise,ofek14prec,ackermann15} \\
\object{PTF10xgo} & $21$:$55$:$57.38$ & $+01$:$19$:$14.1$ & $0.0336$ & $35.85$ & $55472.29$ & $0.048$ & $-17.54$ & -- & -- & \citetalias{ackermann15} \\
\object{PTF10abui} & $06$:$12$:$18.46$ & $-22$:$46$:$15.6$ & $0.0516$ & $36.80$ & $55535.29$ & $0.062$ & $-19.57$ & -- & \citet{arcavi10_PTF10abui} & \citetalias{ackermann15} \\
\object{PTF10achk} & $03$:$05$:$57.54$ & $-10$:$31$:$21.0$ & $0.0325$ & $35.72$ & $55545.10$ & $0.056$ & $-20.74$ & -- & -- & \citetalias{ofek14rise,ofek14prec,ackermann15} \\
\object{PTF10acsq} & $08$:$01$:$33.17$ & $+46$:$45$:$52.5$ & $0.1730$ & $39.61$ & $55558.36$ & $0.063$ & $-16.85$ & -- & -- & \citetalias{ackermann15} \\
\object{PTF11fzz} & $11$:$10$:$46.68$ & $+54$:$06$:$18.8$ & $0.0813$ & $37.86$ & $55730.20$ & $0.009$ & $-14.49$ & -- & -- & \citetalias{ofek14rise,ofek14prec,ackermann15} \\
\object{PTF11mpg} & $22$:$17$:$36.66$ & $+00$:$36$:$48.4$ & $0.0933$ & $38.15$ & $55824.26$ & $0.050$ & $-18.58$ & PS1-11aqj & Pan-STARRS Alerts & \citetalias{ackermann15} \\
\object{PTF11oxu} & $03$:$38$:$34.38$ & $+22$:$32$:$59.4$ & $0.0878$ & $37.99$ & $55853.30$ & $0.176$ & $-21.62$ & SN~2011jc & \citet{drake11_PTF11oxu} & \citetalias{ackermann15} \\
\object{PTF11qnf} & $05$:$44$:$54.14$ & $+69$:$09$:$06.9$ & $0.0148$ & $34.11$ & $55888.40$ & $0.106$ & $-20.64$ & -- & -- & \citetalias{ackermann15} \\
\object{PTF11qqj} & $09$:$58$:$01.64$ & $+00$:$43$:$14.7$ & $0.0931$ & $38.18$ & $55891.44$ & $0.025$ & $-19.03$ & PS1-12y & -- & \citetalias{ackermann15} \\
\object{PTF11rfr} & $01$:$42$:$16.98$ & $+29$:$16$:$25.7$ & $0.0675$ & $37.38$ & $55906.08$ & $0.042$ & $-18.82$ & -- & -- & \citetalias{ofek14prec,ackermann15} \\
\object{PTF11rlv} & $12$:$49$:$34.04$ & $-09$:$20$:$40.5$ & $0.1323$ & $39.00$ & $55922.50$ & $0.037$ & $-19.79$ & -- & -- & \citetalias{ackermann15} \\
\object{PTF12cxj} & $13$:$12$:$38.68$ & $+46$:$29$:$06.3$ & $0.0356$ & $36.07$ & $56033.29$ & $0.010$ & $-19.57$ & -- & -- & \citetalias{ofek14rise,ofek14prec,ackermann15} \\
\object{PTF12frn} & $16$:$22$:$00.16$ & $+32$:$09$:$38.9$ & $0.1365$ & $39.07$ & $56096.43$ & $0.019$ & $-20.78$ & -- & -- & \citet{ganot16} \\
\object{PTF12glz} & $15$:$54$:$53.04$ & $+03$:$32$:$07.5$ & $0.0793$ & $37.84$ & $56114.22$ & $0.130$ & $-19.76$ & -- & \citet{galyam12_PTF12glz} & \citetalias{ofek14rise}, \citet{ganot16} \\
\object{PTF12ksy} & $04$:$11$:$46.09$ & $-12$:$28$:$00.8$ & $0.0314$ & $35.66$ & $56237.29$ & $0.037$ & $-20.71$ & -- & -- & \citetalias{ofek14rise} \\
\object{iPTF13agz} & $14$:$34$:$32.12$ & $+25$:$09$:$43.6$ & $0.0571$ & $37.11$ & $56386.36$ & $0.028$ & $-19.9$ & -- & -- & \citetalias{ofek14rise} \\
\object{iPTF13aki} & $14$:$35$:$34.35$ & $+38$:$38$:$31.0$ & $0.1700$ & $39.59$ & $56392.25$ & $0.010$ & $-18.27$ & -- & This work & -- \\
\object{iPTF13asr} & $12$:$47$:$28.61$ & $+27$:$04$:$03.6$ & $0.1543$ & $39.36$ & $56415.27$ & $0.011$ & $-19.57$ & -- & This work & -- \\
\object{iPTF13cuf} & $02$:$04$:$52.97$ & $+14$:$37$:$59.7$ & $0.2199$ & $40.18$ & $56516.44$ & $0.041$ & $-16.15$ & -- & \citet{arcavi13_iPTF13cuf} & -- \\
\object{iPTF14bcw} & $13$:$48$:$41.18$ & $+35$:$52$:$17.1$ & $0.1206$ & $38.78$ & $56803.29$ & $0.009$ & $-15.05$ & -- & This work & -- \\
\object{iPTF14bpa} & $15$:$26$:$59.96$ & $+24$:$41$:$17.5$ & $0.1220$ & $38.81$ & $56820.21$ & $0.037$ & $-20.78$ & PS15aod & PS1 Object List & -- \\
\object{iPTF15aym} & $13$:$26$:$26.67$ & $+55$:$23$:$43.4$ & $0.0334$ & $35.93$ & $57168.24$ & $0.012$ & $-19.2$ & -- & This work & -- \\
\object{iPTF15bky} & $15$:$04$:$40.80$ & $+12$:$37$:$43.4$ & $0.0288$ & $35.66$ & $57192.35$ & $0.032$ & $-21.63$ & SN~2015Z & \citet{arbour15, leonard15_iPTF15bky} & -- \\
\object{iPTF15blp} & $16$:$27$:$15.21$ & $+41$:$08$:$58.1$ & $0.1949$ & $39.91$ & $57193.32$ & $0.006$ & $-17.63$ & PS15axx & PS1 Object List & -- \\
\object{iPTF15eqr} & $04$:$01$:$15.67$ & $+33$:$16$:$58.3$ & $0.0467$ & $36.55$ & $57361.22$ & $0.354$ & $-20.25$ & AT 2016oy, PS16qg & \citet{young16_iPTF15eqr} & -- \\
\object{iPTF16fb} & $10$:$22$:$09.25$ & $+15$:$28$:$19.2$ & $0.0811$ & $37.87$ & $57427.46$ & $0.045$ & $-19.52$ & SN~2016afj, PS16ago & \citet{faran16_iPTF16fb} & --
\enddata
\tablecomments{The redshift, distance modulus $\mu$ and MW colour excess $E(B-V)_{\rm MW}$ for each SN was obtained as described in Sect.~\ref{sec:sample}. The mean statistical error of the heliocentric redshifts $z$ is $\sigma_{z} = 0.00025$, but the resolution of our spectra (typically a few $\times 10^2$\,km\,s$^{-1}$) suggests that $\sigma_{z} \approx 0.001$ is a more realistic error estimate. The given discovery date ($MJD = JD - 2,400,000.5$) refers to the time of PTF/iPTF discovery, i.e., the epoch of the P48 image which prompted a human scanner in the PTF/iPTF project to flag the transient as a candidate SN. The host galaxy $g$-band absolute magnitudes ($M^{host}_{g}$) were obtained as described in Sect.~\ref{sec:host}. The majority of SNe in our sample have been included in some (or all) of the following three earlier SN~IIn samples in the literature: \citet[][O14c in the table]{ofek14rise}, \citet[][O14a in the table]{ofek14prec}, and \citet[][A15 in the table]{ackermann15}. The general SN~II sample by \citet[][A10 in the table]{arcavi10} and the works by \citet{ganot16} and \citet{stoll13} also included some of the SNe in our sample.}
\end{deluxetable}

\begin{deluxetable}{lclclllcl}
\rotate
\tabletypesize{\scriptsize}
\tablewidth{0pt}
\tablecaption{Peak magnitudes, rise times, explosion epochs, peak epochs, and decline rates for the 42 SNe~IIn in the sample.\label{tab:peakmag_time}}
\tablehead{
\colhead{SN} &
\colhead{Peak absolute} &
\colhead{Peak apparent} &
\colhead{Rise time (rest frame)$\pm 1\sigma$}&
\colhead{Explosion MJD $\pm 1\sigma$}&
\colhead{Peak MJD $\pm 1\sigma$}&
\colhead{0--50 days decline rate}&
\colhead{CSS spline smoothing}&
\colhead{Band}\\
\colhead{} &
\colhead{magnitude $\pm 1\sigma$} &
\colhead{magnitude $\pm 1\sigma$} &
\colhead{(days)} &
\colhead{(obs. frame)} &
\colhead{(obs. frame)} &
\colhead{$\pm 1\sigma$ (mag\,d$^{-1}$)} &
\colhead{parameter $\times 100$} &
\colhead{}}
\startdata
PTF09tm & $-19.57\pm0.03$ & $16.93\pm0.03$ & $23.90\pm1.97$ & $55003.91\pm1.22$ & $55028.65\pm1.63$ & $--$ & 1.000 & R/r \\
PTF09uy & $-21.48\pm0.09$ & $19.65\pm0.09$ & $43.76\pm15.25$ & $54980.45\pm19.98$ & $55037.92\pm1.50$ & $0.0288\pm0.0040$ & 0.500 & R/r \\
PTF09bcl & $-18.81\pm0.05$ & $18.67\pm0.05$ & $12.80\pm1.33$ & $55031.96\pm0.93$ & $55045.55\pm1.06$ & $--$ & 1.000 & R \\
PTF10cwl & $-18.70\pm0.03$ & $19.34\pm0.03$ & $12.61\pm1.95$ & $55251.09\pm1.83$ & $55264.77\pm1.06$ & $0.0382\pm0.0027$ & 0.400 & R \\
PTF10cwx & $-18.51\pm0.04$ & $19.20\pm0.04$ & $26.49\pm2.11$ & $55243.69\pm1.95$ & $55272.11\pm1.15$ & $0.0518\pm0.0060$ & 4.000 & R \\
PTF10ewc & $-18.59\pm0.04$ & $18.43\pm0.04$ & $28.08\pm4.98$ & $55270.45\pm5.24$ & $55300.06\pm0.36$ & $0.0400\pm0.0018$ & 6.000 & R \\
PTF10fjh & $-18.55\pm0.05$ & $17.36\pm0.05$ & $14.41\pm1.13$ & $55293.18\pm0.89$ & $55308.06\pm0.76$ & $0.0570\pm0.0034$ & 3.000 & R/r \\
PTF10flx & $-18.84\pm0.04$ & $18.66\pm0.03$ & $--$ & $--$ & $55304.92\pm1.01$ & $0.0362\pm0.0018$ & 1.000 & R \\
PTF10gvd & $-17.98\pm0.05$ & $19.62\pm0.05$ & $11.65\pm3.47$ & $55299.99\pm3.22$ & $55312.44\pm1.85$ & $0.0210\pm0.0019$ & 7.000 & R \\
PTF10gvf & $-18.90\pm0.03$ & $18.98\pm0.03$ & $14.78\pm1.34$ & $55320.63\pm1.06$ & $55336.61\pm0.99$ & $--$ & 4.000 & R \\
PTF10oug & $-20.11\pm0.02$ & $19.28\pm0.02$ & $51.05\pm3.95$ & $55371.29\pm4.33$ & $55430.00\pm1.39$ & $0.0196\pm0.0011$ & 0.030 & R \\
PTF10qwu & $-20.81\pm0.05$ & $19.57\pm0.05$ & $--$ & $--$ & $--$ & $--$ & 4.000 & R/r \\
PTF10tel & $-18.35\pm0.02$ & $17.70\pm0.02$ & $16.28\pm0.47$ & $55428.22\pm0.00$ & $55445.07\pm0.49$ & $0.0422\pm0.0015$ & 1.000 & R/r \\
PTF10tyd & $-18.23\pm0.03$ & $19.23\pm0.03$ & $45.10\pm4.10$ & $55419.09\pm3.37$ & $55467.04\pm2.76$ & $--$ & 0.100 & R \\
PTF10vag & $-18.44\pm0.06$ & $18.65\pm0.04$ & $17.78\pm1.27$ & $55445.63\pm0.58$ & $55464.33\pm1.20$ & $0.0542\pm0.0051$ & 2.000 & R/r \\
PTF10weh & $-20.77\pm0.01$ & $18.39\pm0.01$ & $49.27\pm1.63$ & $55448.58\pm1.61$ & $55504.64\pm0.93$ & $0.0148\pm0.0003$ & 0.100 & R/r \\
PTF10xgo & $-16.85\pm0.06$ & $19.14\pm0.03$ & $26.36\pm2.57$ & $55452.09\pm2.40$ & $55479.34\pm1.14$ & $0.0314\pm0.0019$ & 3.000 & R/r \\
PTF10abui & $-17.85\pm0.03$ & $19.10\pm0.03$ & $--$ & $--$ & $55546.94\pm1.44$ & $--$ & 0.090 & R \\
PTF10achk & $-18.25\pm0.13$ & $17.63\pm0.13$ & $29.56\pm3.94$ & $55521.39\pm3.07$ & $55551.91\pm2.67$ & $0.0310\pm0.0048$ & 1.000 & R/r \\
PTF10acsq & $-20.13\pm0.06$ & $19.65\pm0.06$ & $--$ & $--$ & $--$ & $--$ & 5.000 & R/r \\
PTF11fzz & $-20.34\pm0.04$ & $17.55\pm0.03$ & $62.36\pm3.91$ & $55723.86\pm1.74$ & $55791.29\pm3.85$ & $0.0064\pm0.0003$ & 0.800 & R/r \\
PTF11mpg & $-18.85\pm0.04$ & $19.44\pm0.04$ & $15.81\pm2.76$ & $55809.05\pm2.44$ & $55826.34\pm1.78$ & $0.0396\pm0.0057$ & 8.000 & R/r \\
PTF11oxu & $-19.22\pm0.02$ & $19.21\pm0.02$ & $68.93\pm5.48$ & $55837.33\pm4.16$ & $55912.31\pm4.26$ & $0.0080\pm0.0003$ & 0.010 & R \\
PTF11qnf & $-17.05\pm0.05$ & $19.32\pm0.05$ & $134.86\pm4.70$ & $55831.70\pm2.23$ & $55968.55\pm4.22$ & $0.0166\pm0.0010$ & 4.000 & R/r \\
PTF11qqj & $-18.69\pm0.08$ & $19.55\pm0.08$ & $--$ & $--$ & $55903.18\pm1.83$ & $0.0148\pm0.0016$ & 0.100 & R \\
PTF11rfr & $-19.73\pm0.01$ & $17.76\pm0.01$ & $49.70\pm2.60$ & $55895.39\pm0.07$ & $55948.44\pm2.78$ & $0.0134\pm0.0015$ & 1.000 & R \\
PTF11rlv & $-18.88\pm0.05$ & $20.22\pm0.05$ & $--$ & $--$ & $55930.38\pm3.02$ & $--$ & 0.006 & R \\
PTF12cxj & $-16.82\pm0.04$ & $19.28\pm0.04$ & $22.21\pm2.00$ & $56026.30\pm1.95$ & $56049.30\pm0.70$ & $0.0378\pm0.0032$ & 0.200 & R \\
PTF12frn & $-21.16\pm0.04$ & $19.42\pm0.04$ & $37.85\pm2.82$ & $56076.92\pm3.00$ & $56119.94\pm1.12$ & $--$ & 0.500 & R \\
PTF12glz & $-20.44\pm0.02$ & $17.75\pm0.01$ & $62.75\pm1.03$ & $56118.16\pm0.97$ & $56185.88\pm0.55$ & $0.0052\pm0.0001$ & 0.002 & R/r \\
PTF12ksy & $-17.97\pm0.02$ & $17.78\pm0.01$ & $26.04\pm1.18$ & $56225.84\pm1.18$ & $56252.69\pm0.29$ & $--$ & 2.000 & R \\
iPTF13agz & $-17.74\pm0.01$ & $19.44\pm0.01$ & $52.65\pm1.05$ & $56359.31\pm0.49$ & $56414.97\pm1.00$ & $0.0170\pm0.0003$ & 0.100 & R \\
iPTF13aki & $-19.33\pm0.03$ & $20.28\pm0.03$ & $13.64\pm2.91$ & $56387.61\pm0.93$ & $56403.57\pm3.27$ & $0.0102\pm0.0005$ & 5.000 & R \\
iPTF13asr & $-18.83\pm0.04$ & $20.56\pm0.04$ & $20.27\pm1.22$ & $56395.26\pm0.93$ & $56418.66\pm1.06$ & $--$ & 1.000 & R \\
iPTF13cuf & $-19.28\pm0.08$ & $21.00\pm0.08$ & $--$ & $--$ & $--$ & $--$ & 0.300 & R \\
iPTF14bcw & $-18.29\pm0.06$ & $20.52\pm0.06$ & $--$ & $--$ & $56821.01\pm1.71$ & $--$ & 1.000 & R/r \\
iPTF14bpa & $-20.15\pm0.06$ & $18.76\pm0.06$ & $36.05\pm7.92$ & $56822.87\pm1.09$ & $56863.31\pm8.82$ & $0.0056\pm0.0003$ & 0.700 & R/r \\
iPTF15aym & $-18.35\pm0.02$ & $17.62\pm0.01$ & $24.51\pm1.63$ & $57164.31\pm1.65$ & $57189.64\pm0.35$ & $0.0242\pm0.0004$ & 1.000 & g \\
iPTF15bky & $-18.44\pm0.01$ & $17.34\pm0.01$ & $22.73\pm0.78$ & $57180.86\pm0.76$ & $57204.24\pm0.26$ & $0.0456\pm0.0014$ & 1.000 & g \\
iPTF15blp & $-20.19\pm0.02$ & $19.74\pm0.02$ & $20.76\pm3.06$ & $57180.51\pm3.39$ & $57205.31\pm1.38$ & $0.0136\pm0.0005$ & 0.100 & g \\
iPTF15eqr & $-18.33\pm0.02$ & $19.18\pm0.02$ & $18.71\pm1.18$ & $57353.31\pm0.80$ & $57372.90\pm0.94$ & $--$ & 0.500 & R/r \\
iPTF16fb & $-18.47\pm0.05$ & $19.57\pm0.05$ & $--$ & $--$ & $57441.06\pm3.64$ & $--$ & 0.100 & g 
\enddata
\tablecomments{The absolute magnitudes are corrected for MW and host galaxy extinction (Sect.~\ref{sec:sample}). The magnitude 1$\sigma$ uncertainties are taken from the spline fits (Sect.~\ref{sec:peak}), with the CSS spline smoothing parameters given in the table. The total uncertainty of the rise time is estimated as the uncertainty of the rise time (Sect.~\ref{sec:risetimes}) from the MC experiment and the uncertainty of the peak epoch, added in quadrature. The decline rates are determined as described in Sect.~\ref{sec:decline}. The photometric bands are called \textit{R} (Mould $R$), \textit{r} (SDSS $r$) and \textit{g} (SDSS $g$) in the table. Explosion and peak epochs are given as $MJD = JD - 2,400,000.5$.}
\end{deluxetable} 

\begin{appendix}
\onecolumn
\section{Classification spectra and light-curves\label{sec:bigfig}}

\begin{landscape}
\begin{figure*}
   \centering
    \includegraphics[width=\linewidth,keepaspectratio]{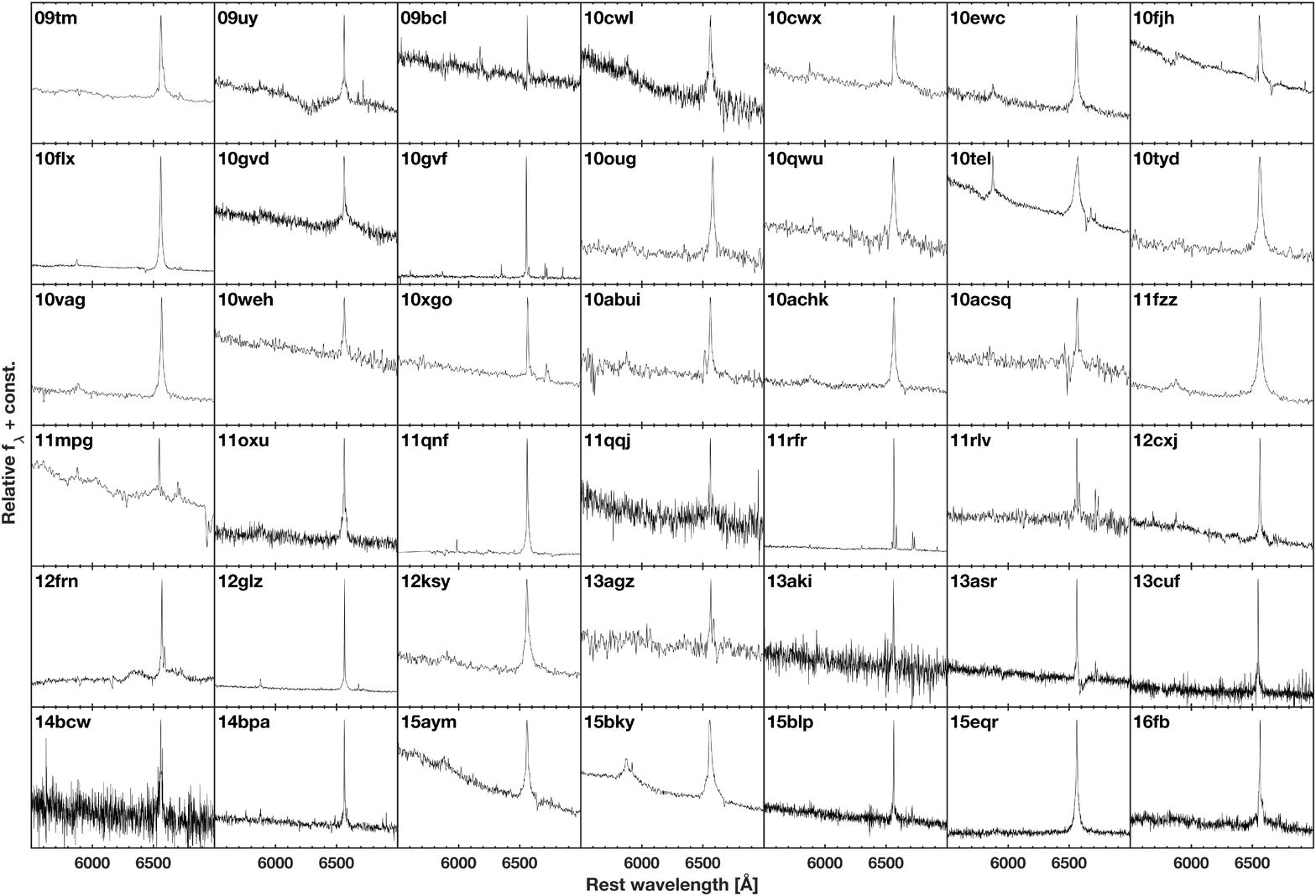}
    \caption{Classification spectra of the 42 SNe~IIn included in our study, each showing the H$\alpha$ narrow or intermediate-width emission characteristic of SNe~IIn. Details of the spectra are given in Table~\ref{tab:speclog}. In some spectra, telluric B-band absorption (or residual, incompletely removed absorption) is visible at 6860\,\AA\ in the observed frame of reference. Spectra of PTF10weh and PTF12cxj were presented by \citet{ofek14prec}, spectra of PTF10gvf by \citet{khazov16}, of PTF10tel by \citet{ofek13} and \citet{khazov16}, and of PTF12glz by \citet{soumagnac19}.\label{fig:class_spec}}
\end{figure*}
\end{landscape}

\clearpage
\begin{landscape}
\begin{figure*}
   \centering
    \includegraphics[width=\linewidth,keepaspectratio]
    {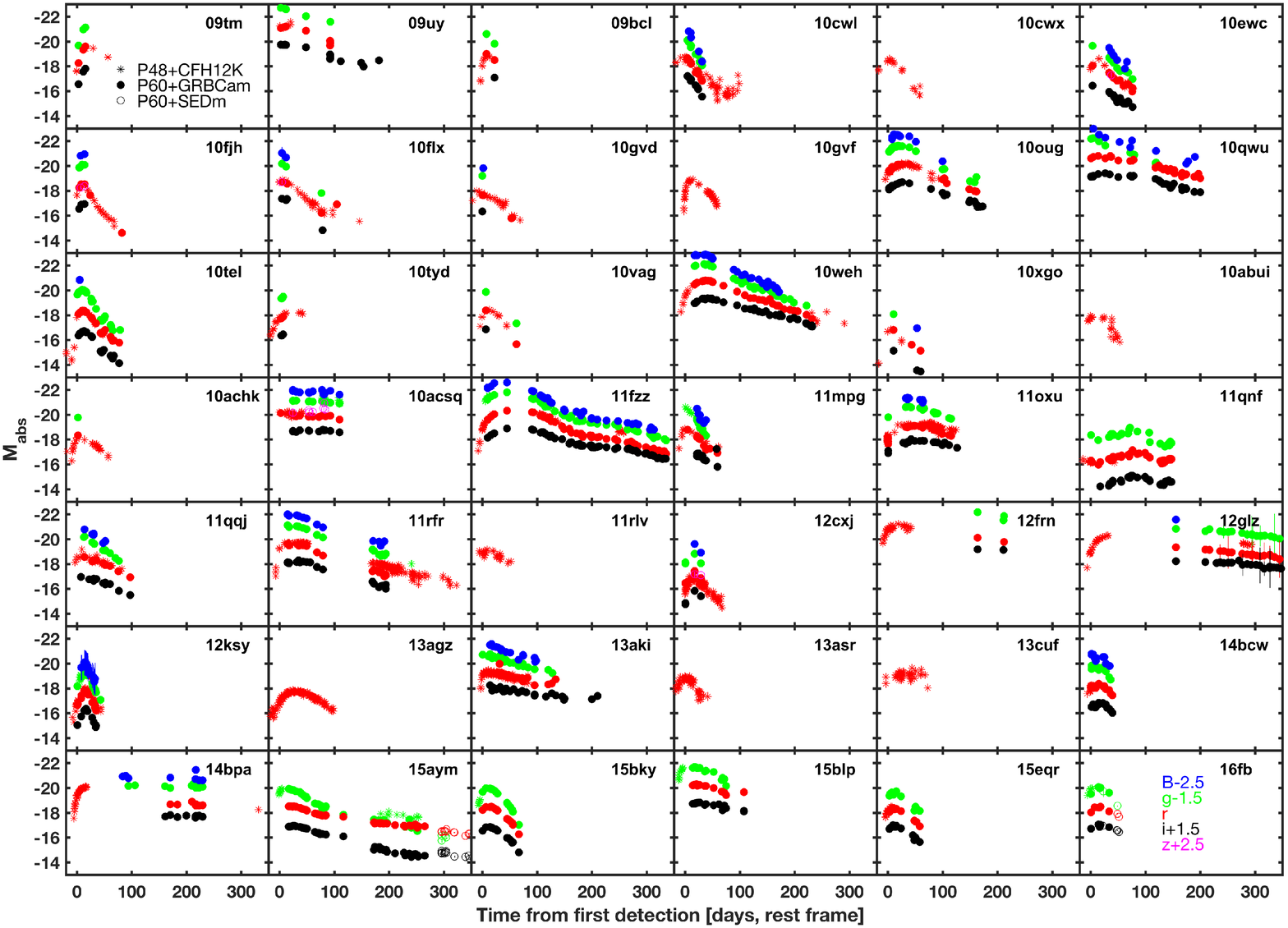}
    \caption{Light curves of the 42 SNe~IIn in the sample, with time shown in the rest frame. In these plots, for clarity, we do not show photometric upper limits.\label{fig:LCs}}
\end{figure*}
\end{landscape}

\clearpage
\begin{landscape}
\begin{figure*}
   \centering
    \includegraphics[width=\linewidth,keepaspectratio]{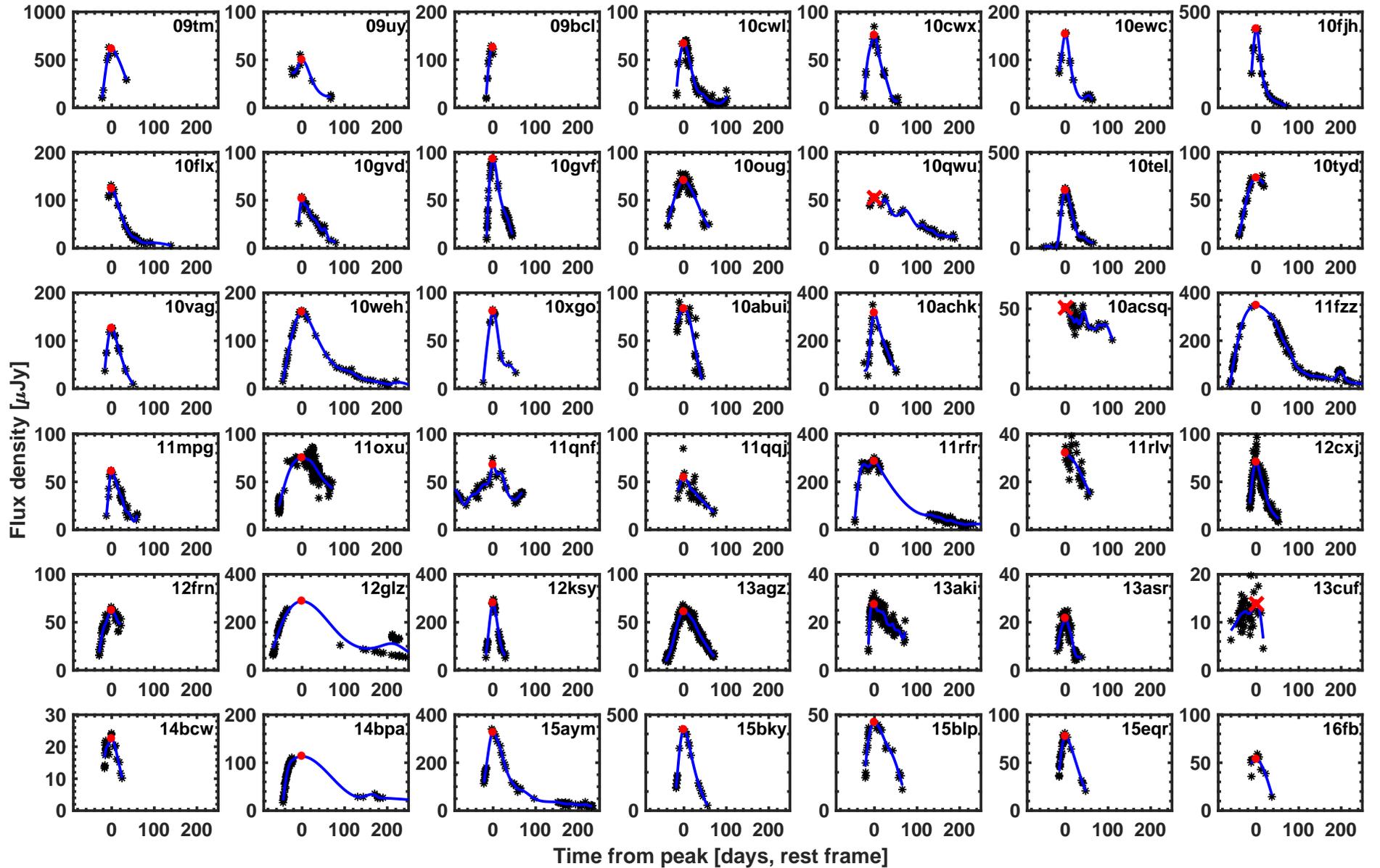} 
    \caption{Cubic smoothing splines fit to our SN~IIn light-curves, to determine the times and magnitudes of the light-curve peaks (Sect.~\ref{sec:peak}). Photometric bands are specified in Table~\ref{tab:peakmag_time}. The CSS fit is indicated by a blue curve and the peak of the light-curve is marked by a red dot. The smoothing parameter $s$ for each SN is reported in Table~\ref{tab:peakmag_time}. The SNe with peak-time error $> 10$\,d (PTF10qwu, PTF10acsq, iPTF13cuf) have their estimated peaks marked with red crosses.\label{fig:lcfit_allIIn}}
\end{figure*}
\end{landscape}

\clearpage
\begin{landscape}
\begin{figure*}
   \centering
    \includegraphics[width=\linewidth,keepaspectratio]{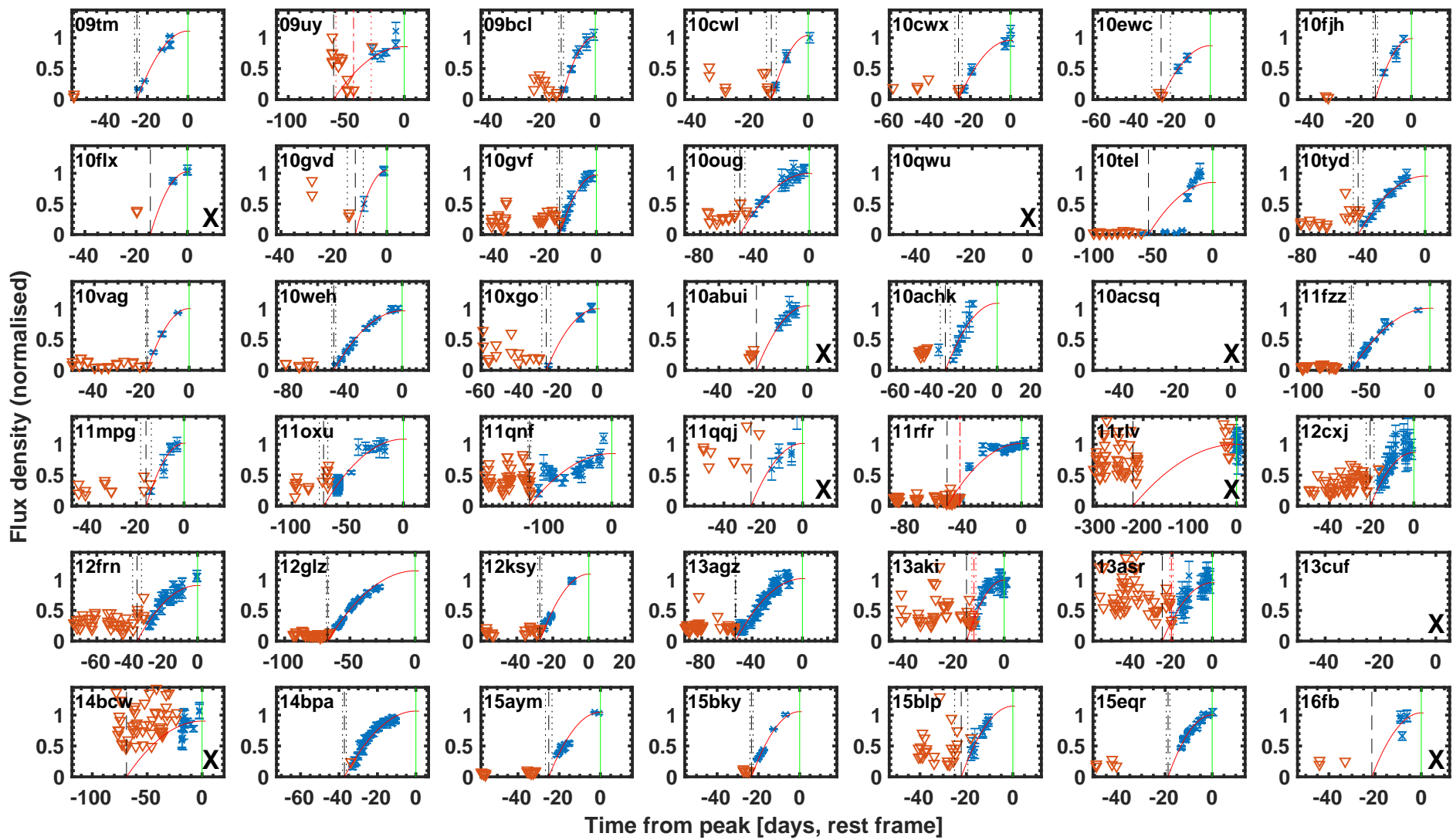}
    \caption{Fits of the $\propto t^2$ template to determine the rise times of the SNe in our sample. Photometric bands are as specified in Table~\ref{tab:peakmag_time}. The black dashed vertical line shows where the $\propto t^2$ template reaches 0 flux density (with 1$\sigma$ uncertainties given as black dotted lines). The red dashed vertical line shows where the explosion epoch estimate is based on the occurrence of an upper limit inconsistent with the $\propto t^2$ template (with uncertainties given as red dotted lines). The SNe with "X" in their plots are not used for the rise-time measurements, according to criteria given in Sect.~\ref{sec:risetimes}. For the SNe with peak-epoch error $> 10$\,d (PTF10qwu, PTF10acsq, iPTF13cuf), our method does not allow any rise-time determination; these SNe are not shown.\label{fig:fitrise_all}}
\end{figure*}
\end{landscape}
\twocolumn

\section{K-corrections for SNe~IIn\label{sec:kcor}}
K-corrections were determined in the $g$, $r$, and $i$ bands, while phases refer to the peak of each SN\footnote{To allow better sampling of the K-correction at higher redshifts, the SNe PTF10qwu, PTF10acsq and iPTF13cuf (excluded in the main analysis, c.f. Sect.~\ref{sec:peak}) are included here, with peak epochs approximated as shown by the 'X' symbols in Fig.~\ref{fig:lcfit_allIIn})} (see Sect.~\ref{sec:peak} and Table~\ref{tab:peakmag_time}) and are given in the rest frame, hence corrected by a factor $(1+z)$. Owing to the lack of detailed individual spectral sequences, the inferred K-corrections are based on the spectroscopic evolution of the entire sample as a whole, with each spectrum corresponding to a specific phase.

For each event, we take the derived redshift as a reference, shifting all the rest-frame spectra of the sample to that particular $z$ using the {\sc iraf} task {\sc dopcor} (called through {\sc PyRAF}\footnote{\protect\url{https://github.com/spacetelescope/pyraf}}). We then compute synthetic magnitudes on the shifted spectra ($m_{z,t}$) using the {\sc iraf} task {\sc calcphot} included in the software package Space Telescope Science Data Analysis System (STSDAS) and K-corrections ($m_{z,t}-m_{z_0,t}$, where $m_{z_0,t}$ are the synthetic magnitudes computed on the rest-frame spectra). All spectra were previously corrected for the foreground MW extinction along the line of sight and, for each filter, we did not consider spectra without the sufficient spectral coverage (i.e., not covering the entire filter bandwidth) after applying the specific redshift correction.

\begin{figure*}
\begin{center}
\includegraphics[width=12cm,angle=0]{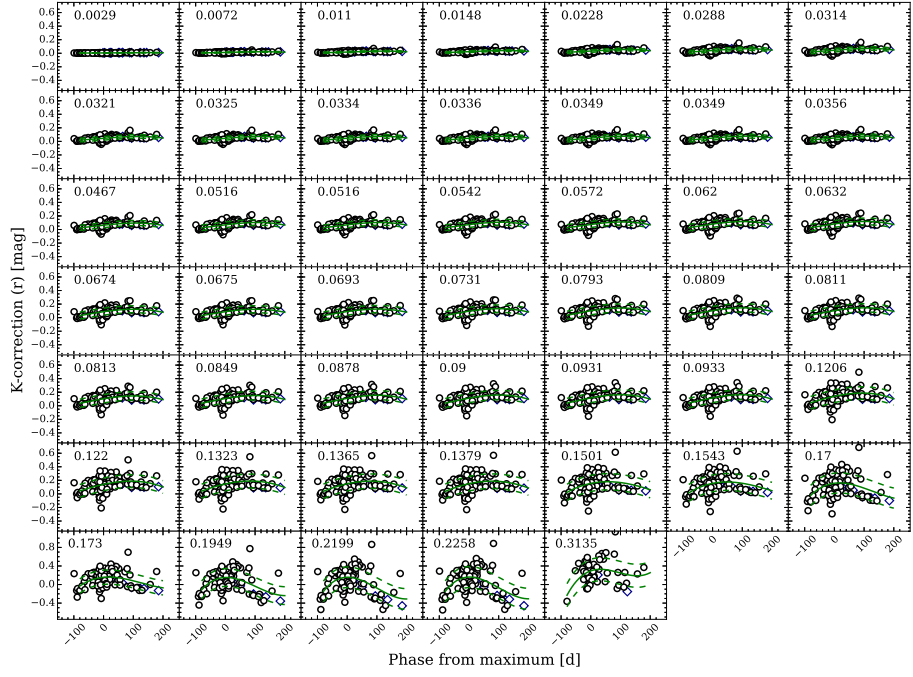}
\caption{K-corrections in the $r$-band for SNe~IIn at different redshifts. Blue points are the corrections computed for well-studied SNe~IIn available in the literature (see the text for more details). Green lines are the best-fit models obtained using third-order polynomials, while dashed lines are the $1\sigma$ uncertainties of each fit. Phases (rest frame) refer to the estimated light-curve peak (Sect.~\ref{sec:peak} and Table~\ref{tab:peakmag_time}) of each SN in the sample.\label{fig:rkcor}}
\end{center}
\end{figure*}

In Fig.~\ref{fig:rkcor} we show the derived K-corrections for each redshift, also including those obtained from other SNe~IIn with well-sampled spectroscopic evolutions available in the literature  \citep[SNe~1988Z, 1998S, 2010jl, and 2015da;][]{turatto93,fassia01,fransson14,tartaglia20}. Final values were inferred fitting third-order polynomials to the K-corrections computed at each epoch and $1\sigma$ errors were assumed as uncertainties. We find corrections of $\lesssim0.2\,\rm{mag}$ at all phases up to $z \approx 0.1$, while at higher redshifts we start to see a significant scatter in the data points.

\begin{figure*}
\begin{center}
\includegraphics[width=9cm,angle=0]{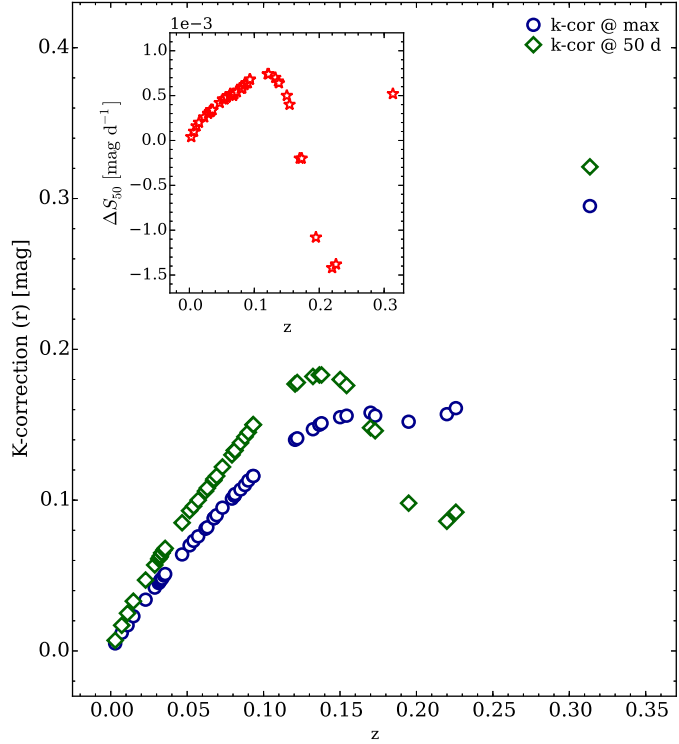}
\caption{Effect of the K-correction on photometric properties of our SN~IIn sample. The required correction is smaller than $\sim 0.2$\,mag in both peak magnitudes and mag$_{50\,\rm{d}}$ at redshifts $\lesssim0.15$, although it can be as high as 0.3\,mag at $z \gtrsim 0.3$. The inset shows the effects of the K-correction on the light-curve slopes measured from peak to +50\,d, which is smaller than $1.5\times10^{-3}$\,mag for all the redshifts of our sample.\label{fig:deltakcor}}
\end{center}
\end{figure*}

Interpolating the derived polynomials at peak brightness and $+50$\,d, we can estimate the effects of the K-corrections on the photometric properties of our sample (see Fig.~\ref{fig:deltakcor}). A LSQ fourth-order fit to the $r$-band K-correction at light-curve peak gives us Eq.~\ref{eq:kcorr_r}, and a LSQ first-order fit to the $r$-band K-correction errors at light-curve peak gives us Eq.~\ref{eq:kcorr_r_err}:

\begin{equation}
\label{eq:kcorr_r}
K_r(z) = 184.16 z^4 -86.69 z^3 + 8.96 z^2 + 0.95 z + 0.01{\rm~mag},
\end{equation}

\begin{equation}
\label{eq:kcorr_r_err}
{\rm and~} K^{\rm Err}_r(z) = 0.89 z - 0.01{\rm~mag}. 
\end{equation}

Equations \ref{eq:kcorr_r} and \ref{eq:kcorr_r_err} are valid for $z \leq 0.3135$. The errors used for Eq.~\ref{eq:kcorr_r_err} are the errors of the fits in Fig.~\ref{fig:rkcor}. The data used to make the fits for Eqs.~\ref{eq:kcorr_r}~and~\ref{eq:kcorr_r_err} are given in Table~\ref{tab:Kcorr}.

At maximum, we see an increase in the K-correction up to $z \approx 0.15$, followed by a flattening around $z \approx 0.2$, while at higher redshifts we see a further increase up to $\sim 0.3$\,mag (although the derived correction at these redshifts is affected by the large scatter in the data points and the lower number of available spectra, owing to the limited spectral coverage). We find a similar evolution in the K-corrections estimated at +50\,d, although we do not notice the flattening observed at maximum, while we see a steep decline from $\sim 0.18$ to $\sim 0.09$\,mag in the $0.15\lesssim z\lesssim 0.2$ range. In Fig.~\ref{fig:deltakcor} (inset) we also show the effect of the derived K-corrections on the slope of light-curves in the first 50\,d after maximum brightness. We note a maximum deviation of $-1.5\times10^{-3}$\,mag at $z \approx 0.25$ in the derived slope, suggesting a negligible contribution of the K-correction up to $z \approx 0.3$, while it starts to play an important role in determining magnitudes at peak at $z \gtrsim 0.15$.

\newpage
\onecolumn

\begin{deluxetable}{ccccc}
\centering
\tabletypesize{\scriptsize}
\tablewidth{0pt}
\tablecaption{K-corrections ($r$-band) for SNe~IIn.\label{tab:Kcorr}}
\tablehead{
\colhead{Redshift} &
\colhead{$K_{\rm 0~d}$} &
\colhead{Error (${\rm 0 ~ d}$)}&
\colhead{$K_{\rm 50~d}$}&
\colhead{Error (${\rm 50 ~ d}$)}\\
\colhead{} &
\colhead{(mag)} &
\colhead{(mag)}&
\colhead{(mag)}&
\colhead{(mag)}}
\startdata
0.0029 & 0.005 & 0.003 & 0.007 & 0.003 \\
0.0072 & 0.012 & 0.007 & 0.017 & 0.007 \\
0.0110 & 0.017 & 0.010 & 0.025 & 0.010 \\
0.0148 & 0.023 & 0.013 & 0.033 & 0.013 \\
0.0228 & 0.034 & 0.019 & 0.047 & 0.019 \\
0.0288 & 0.042 & 0.024 & 0.057 & 0.024 \\
0.0314 & 0.045 & 0.026 & 0.061 & 0.026 \\
0.0321 & 0.046 & 0.026 & 0.062 & 0.026 \\
0.0325 & 0.047 & 0.027 & 0.063 & 0.027 \\
0.0334 & 0.048 & 0.027 & 0.065 & 0.027 \\
0.0336 & 0.048 & 0.028 & 0.065 & 0.028 \\
0.0349 & 0.050 & 0.029 & 0.067 & 0.029 \\
0.0356 & 0.051 & 0.029 & 0.068 & 0.029 \\
0.0467 & 0.064 & 0.037 & 0.085 & 0.037 \\
0.0516 & 0.070 & 0.041 & 0.093 & 0.041 \\
0.0542 & 0.073 & 0.043 & 0.096 & 0.043 \\
0.0572 & 0.076 & 0.045 & 0.100 & 0.045 \\
0.0620 & 0.081 & 0.049 & 0.106 & 0.049 \\
0.0632 & 0.082 & 0.049 & 0.108 & 0.049 \\
0.0674 & 0.088 & 0.052 & 0.113 & 0.052 \\
0.0675 & 0.088 & 0.052 & 0.114 & 0.052 \\
0.0693 & 0.090 & 0.053 & 0.116 & 0.053 \\
0.0731 & 0.095 & 0.055 & 0.122 & 0.055 \\
0.0793 & 0.101 & 0.060 & 0.130 & 0.060 \\
0.0809 & 0.103 & 0.061 & 0.133 & 0.061 \\
0.0811 & 0.103 & 0.061 & 0.133 & 0.061 \\
0.0813 & 0.104 & 0.061 & 0.133 & 0.061 \\
0.0849 & 0.107 & 0.064 & 0.138 & 0.064 \\
0.0878 & 0.110 & 0.066 & 0.142 & 0.066 \\
0.0900 & 0.113 & 0.068 & 0.145 & 0.068 \\
0.0931 & 0.116 & 0.070 & 0.150 & 0.070 \\
0.0933 & 0.116 & 0.070 & 0.150 & 0.070 \\
0.1206 & 0.140 & 0.093 & 0.177 & 0.093 \\
0.1220 & 0.141 & 0.094 & 0.178 & 0.094 \\
0.1323 & 0.147 & 0.103 & 0.182 & 0.103 \\
0.1365 & 0.150 & 0.106 & 0.183 & 0.106 \\
0.1379 & 0.151 & 0.107 & 0.183 & 0.107 \\
0.1501 & 0.155 & 0.119 & 0.180 & 0.119 \\
0.1543 & 0.156 & 0.124 & 0.176 & 0.124 \\
0.1700 & 0.158 & 0.148 & 0.148 & 0.148 \\
0.1730 & 0.156 & 0.154 & 0.146 & 0.154 \\
0.1949 & 0.152 & 0.192 & 0.098 & 0.192 \\
0.2199 & 0.157 & 0.223 & 0.086 & 0.223 \\
0.2258 & 0.161 & 0.227 & 0.092 & 0.227 \\
0.3135 & 0.295 & 0.248 & 0.321 & 0.248
\enddata
\tablecomments{The K-correction at light-curve peak is called $K_{\rm 0~d}$, at 50\,d (rest frame) after peak it is called $K_{\rm 50~d}$. The data in this table are plotted in Fig.~\ref{fig:deltakcor}.}
\end{deluxetable}

\end{appendix}

\end{document}